\PassOptionsToPackage{unicode}{hyperref}
\PassOptionsToPackage{hyphens}{url}
\PassOptionsToPackage{dvipsnames,svgnames,x11names}{xcolor}
\documentclass[
  english,
  11pt]{article}
\usepackage{xcolor}
\usepackage[margin=1in]{geometry}
\usepackage{amsmath,amssymb}
\usepackage{iftex}
\ifPDFTeX
  \usepackage[T1]{fontenc}
  \usepackage[utf8]{inputenc}
  \usepackage{textcomp} 
\else 
  \usepackage{unicode-math} 
  \defaultfontfeatures{Scale=MatchLowercase}
  \defaultfontfeatures[\rmfamily]{Ligatures=TeX,Scale=1}
\fi
\usepackage{lmodern}
\ifPDFTeX\else
\fi
\IfFileExists{upquote.sty}{\usepackage{upquote}}{}
\IfFileExists{microtype.sty}{
  \usepackage[]{microtype}
  \UseMicrotypeSet[protrusion]{basicmath} 
}{}
\makeatletter
\@ifundefined{KOMAClassName}{
  \IfFileExists{parskip.sty}{%
    \usepackage{parskip}
  }{
    \setlength{\parindent}{0pt}
    \setlength{\parskip}{6pt plus 2pt minus 1pt}}
}{
  \KOMAoptions{parskip=half}}
\makeatother
\makeatletter
\ifx\paragraph\undefined\else
  \let\oldparagraph\paragraph
  \renewcommand{\paragraph}{
    \@ifstar
      \xxxParagraphStar
      \xxxParagraphNoStar
  }
  \newcommand{\xxxParagraphStar}[1]{\oldparagraph*{#1}\mbox{}}
  \newcommand{\xxxParagraphNoStar}[1]{\oldparagraph{#1}\mbox{}}
\fi
\ifx\subparagraph\undefined\else
  \let\oldsubparagraph\subparagraph
  \renewcommand{\subparagraph}{
    \@ifstar
      \xxxSubParagraphStar
      \xxxSubParagraphNoStar
  }
  \newcommand{\xxxSubParagraphStar}[1]{\oldsubparagraph*{#1}\mbox{}}
  \newcommand{\xxxSubParagraphNoStar}[1]{\oldsubparagraph{#1}\mbox{}}
\fi
\makeatother

\usepackage{longtable,booktabs,array}
\usepackage{calc} 
\usepackage{etoolbox}
\makeatletter
\patchcmd\longtable{\par}{\if@noskipsec\mbox{}\fi\par}{}{}
\makeatother
\IfFileExists{footnotehyper.sty}{\usepackage{footnotehyper}}{\usepackage{footnote}}
\makesavenoteenv{longtable}
\usepackage{graphicx}
\makeatletter
\newsavebox\pandoc@box
\newcommand*\pandocbounded[1]{
  \sbox\pandoc@box{#1}%
  \Gscale@div\@tempa{\textheight}{\dimexpr\ht\pandoc@box+\dp\pandoc@box\relax}%
  \Gscale@div\@tempb{\linewidth}{\wd\pandoc@box}%
  \ifdim\@tempb\p@<\@tempa\p@\let\@tempa\@tempb\fi
  \ifdim\@tempa\p@<\p@\scalebox{\@tempa}{\usebox\pandoc@box}%
  \else\usebox{\pandoc@box}%
  \fi%
}
\def\fps@figure{htbp}
\makeatother

\ifLuaTeX
\usepackage[bidi=basic,shorthands=off]{babel}
\else
\usepackage[bidi=default,shorthands=off]{babel}
\fi
\ifLuaTeX
  \usepackage{selnolig} 
\fi

\providecommand{\tightlist}{%
  \setlength{\itemsep}{0pt}\setlength{\parskip}{0pt}}

\usepackage[]{natbib}
\usepackage{amsmath,amssymb,bm,booktabs}

\makeatletter
\@ifpackageloaded{caption}{}{\usepackage{caption}}
\AtBeginDocument{%
\ifdefined\contentsname
  \renewcommand*\contentsname{Table of contents}
\else
  \newcommand\contentsname{Table of contents}
\fi
\ifdefined\listfigurename
  \renewcommand*\listfigurename{List of Figures}
\else
  \newcommand\listfigurename{List of Figures}
\fi
\ifdefined\listtablename
  \renewcommand*\listtablename{List of Tables}
\else
  \newcommand\listtablename{List of Tables}
\fi
\ifdefined\figurename
  \renewcommand*\figurename{Figure}
\else
  \newcommand\figurename{Figure}
\fi
\ifdefined\tablename
  \renewcommand*\tablename{Table}
\else
  \newcommand\tablename{Table}
\fi
}
\@ifpackageloaded{float}{}{\usepackage{float}}
\floatstyle{ruled}
\@ifundefined{c@chapter}{\newfloat{codelisting}{h}{lop}}{\newfloat{codelisting}{h}{lop}[chapter]}
\floatname{codelisting}{Listing}

\usepackage{amsthm}
\theoremstyle{plain}
\newtheorem{proposition}{Proposition}[section]
\theoremstyle{plain}
\newtheorem{theorem}{Theorem}[section]
\theoremstyle{plain}
\newtheorem{lemma}{Lemma}[section]
\theoremstyle{plain}
\newtheorem{corollary}{Corollary}[section]
\theoremstyle{remark}
\AtBeginDocument{}

\newtheorem{refremark}{Remark}[section]

\makeatother
\makeatletter
\@ifpackageloaded{caption}{}{\usepackage{caption}}
\@ifpackageloaded{subcaption}{}{\usepackage{subcaption}}
\makeatother
\usepackage{bookmark}
\IfFileExists{xurl.sty}{\usepackage{xurl}}{} 
\hypersetup{
  pdftitle={The response scale as a tuning parameter in robust Rayleigh regression},
  pdfauthor={Nirian Martín},
  pdflang={en},
  pdfkeywords={density power divergence, equivariance, extreme value
distribution, influence function, Rayleigh regression, SAR imagery},
  colorlinks=true,
  linkcolor={blue},
  filecolor={Maroon},
  citecolor={Blue},
  urlcolor={Blue},
  pdfcreator={LaTeX via pandoc}}

\title{The response scale as a tuning parameter in robust Rayleigh
regression}
\author{Nirian Martín\\[0.5em] \normalsize Department of Statistics and Operations Research and Interdisciplinary Mathematics Institute (IMI),\\ \normalsize Complutense University of Madrid, Spain}
\date{November 25, 2026}
\begin{document}
\maketitle
\begin{abstract}
Synthetic Aperture Radar (SAR) magnitudes are usually modelled with the
Rayleigh distribution, whose square is exponential and whose logarithm
follows a type I extreme value distribution. The three formulations lead
to the same maximum likelihood estimator of the regression coefficients,
but to three different minimum density power divergence estimators,
since these estimators are not equivariant under nonlinear
transformations of the observations. This paper studies the consequences
of that fact. The three models are embedded in a family of power
transformations of the exponential response, indexed by a real number
\(c\), and the asymptotic distribution and the influence function of the
minimum density power divergence estimator are obtained in closed form
for the whole family. The asymptotic variance-covariance matrix
factorizes into a scalar function of \(c\) and the tuning parameter, and
a matrix which collects the effect of the heterogeneity of the sample.
The matrix factor is proven to be minimal for the logarithmic
transformation, the only member of the family for which the asymptotic
variance-covariance matrix does not depend on the unknown parameters and
the influence function redescends to zero. The exponent \(c\) acts as a
second tuning parameter, which trades efficiency for gross error
sensitivity. For the logarithmic transformation, Wald-type, Rao-type and
divergence-based test statistics with a chi-square distribution under
the null hypothesis are derived in explicit form. A simulation study and
an application to ground type detection in SAR imagery illustrate the
results.


\end{abstract}

\section{Introduction}\label{sec-intro}

Synthetic Aperture Radar (SAR) imagery is an efficient way to monitor
the surface of the Earth. Under the fully developed speckle assumption,
the magnitude of the SAR return follows a Rayleigh distribution, and
\citet{palm2019} proposed a Rayleigh regression model for ground type
detection, in which a function of the mean of the magnitude is linear in
a set of explanatory variables. A regression model for the Rayleigh
distribution had been previously considered in \citet{aminzadeh1993},
and bias-corrected versions of the maximum likelihood estimator for
small windows of pixels were obtained in \citet{palm2022}. An extension
which admits pixel values equal to zero has been recently proposed in
\citet{palm2024}. Outlying pixels are very common in SAR data, and two
ways of dealing with them have been followed in the literature of
Rayleigh regression. The first one enlarges the model with a parameter
which makes the tail heavier, as in the slashed Rayleigh regression
model of \citet{gallardo2020}. The second one keeps the model, which has
a physical justification, and replaces the maximum likelihood estimator
(MLE) by a robust estimator. This is the approach of \citet{palm2021},
who use the weighted maximum likelihood estimator of \citet{field1994},
and it agrees with the recommendation of \citet{frery2022} for SAR data,
to compute the MLE together with a robust estimator and to rely on the
second one when they differ. This paper follows the second way.

The square of a Rayleigh random variable is exponential, and the
logarithm of an exponential random variable follows a type I extreme
value distribution in which the scale parameter of the exponential
distribution becomes a location parameter. The last formulation is the
usual one for lifetime regression models
\citep{kalbfleisch2002, meeker2021}. Hence, the same regression model
for the SAR magnitudes can be written in terms of three different
response variables. For the MLE the choice is irrelevant, since the MLE
is equivariant under one-to-one transformations of the observations.

The minimum density power divergence (DPD) estimators \citep{basu1998},
extended to independent and non-identically distributed observations in
\citet{ghosh2013} and to generalized linear models in
\citet{ghosh2016glm}, are a well-established way of obtaining robust
estimators with a small loss of efficiency. It is known that they are
not equivariant under transformations of the observations unless the
Jacobian is constant \citetext{\citealp[Section
3.4]{basu1998}; \citealp[Theorem 4.1.4]{broniatowski2012}}. This
property is usually mentioned as a limitation. Equivalently, the DPD
depends on the measure with respect to which the density functions are
taken, and the choice of this reference measure is identified in
\citet{eguchi2024} as a relevant issue which requires further study. A
power transformation of the response amounts to replacing the Lebesgue
measure by a measure whose density function is a power of the original
variable. The log-linear exponential regression model, with censored
observations and stochastic covariates, has been studied with minimum
DPD estimators in \citet{ghosh2017}, where the logarithmic form of the
model is mentioned as an accelerated failure time model but the
divergence is always computed with the density of the lifetime. Hence,
as far as we know, the consequences of the choice of the scale of the
response on the efficiency, the robustness and the test statistics have
not been quantified. The purpose of this paper is to do so, and to show
that the lack of equivariance can be exploited, the scale of the
response being a second tuning parameter of the procedure. A preliminary
version of this work, in which the asymptotic distributions of the
estimators for the three models were stated without proofs and the
connection of the type I extreme value model with the maximum
\(L_{q}\)-likelihood estimators was pointed out, was presented in
\citet{martin2026iwsm}.

The same question arises for the test statistics. It is desirable that a
test does not depend on choices which are irrelevant for the hypothesis,
such as the units of measurement or the parametrization of the model,
and it is known that many asymptotic tests do not have this property.
Wald tests are not invariant to the rescaling of the variables in
regression models with Box-Cox transformations
\citep{dagenais1991, dagenais1992}, and, for models estimated through
M-estimators or estimating functions, none of the usual Wald-type,
score-type and likelihood ratio-type statistics is invariant to general
reparametrizations, so that equivalent hypotheses in equivalent models
may lead to different conclusions \citep{dufour2017}. The family of
models considered here is a family of equivalent models, related through
Box-Cox transformations of the response, and the test statistics are
studied from this point of view.

The contributions of the paper concern the estimators, the test
statistics and their application. With respect to the estimators, the
three models are embedded in a family of power transformations of the
exponential response, \(T^{c}\), with the logarithm for \(c=0\), and the
asymptotic distribution of the minimum DPD estimator is obtained in
closed form for the whole family (Theorem~\ref{thm-asymptotic}). Its
asymptotic variance-covariance matrix is the product of a scalar factor
and a matrix factor. The matrix factor is minimal for \(c=0\), and for
any other member of the family the heterogeneity of the sample produces
an additional loss of efficiency (Theorem~\ref{thm-penalty}), while for
small values of the tuning parameter the scalar factor is an explicit
quadratic function of \(c\) which is increasing on \([0,1]\)
(Proposition~\ref{prp-small}), so that for the same tuning parameter the
type I extreme value model is more efficient than the Rayleigh model,
and this one more efficient than the exponential model. The influence
function is also obtained for the whole family. It is bounded if and
only if \(c\leq 1\) and it redescends to zero only for \(c=0\)
(Proposition~\ref{prp-influence}), although at equal efficiency the
gross error sensitivity is not minimal for \(c=0\), which shows that
\(c\) trades efficiency for gross error sensitivity. Two further
properties distinguish the logarithmic scale. Observations equal to
zero, which are frequent in SAR images, do not produce any asymptotic
bias in the estimator (Proposition~\ref{prp-zeros}), and the minimum DPD
estimator coincides with the maximum \(L_{q}\)-likelihood estimator of
\citet{ferrari2010} only in that scale.

With respect to the test statistics, for \(c=0\) the Wald-type, Rao-type
and divergence-based statistics have explicit expressions which do not
require the estimation of any matrix. Their asymptotic distribution is
chi-square under the null hypothesis and non-central chi-square under
contiguous alternative hypotheses (Theorem~\ref{thm-tests}). They are
invariant to changes of the units of measurement and to
reparametrizations of the linear predictor, and they are not invariant
to the scale of the response, the asymptotic relative efficiency between
two scales being the ratio of the corresponding scalar factors
(Proposition~\ref{prp-invariance}). Finally, the ground type detection
experiment of \citet{palm2021} is analysed again with the proposed
estimators and test statistics, taking into account the dependence
between neighbouring pixels (Section~\ref{sec-sar}), and the estimators
are compared in a simulation study with the weighted maximum likelihood
estimator used in that paper (Section~\ref{sec-sim-d}).

The rest of the paper is organized as follows. Section~\ref{sec-model}
introduces the model and the family of responses.
Section~\ref{sec-mdpde} defines the minimum DPD estimators and
establishes their asymptotic distribution. Section~\ref{sec-efficiency}
and Section~\ref{sec-robustness} compare the members of the family in
terms of efficiency and robustness. Section~\ref{sec-tests} is devoted
to hypothesis testing, Section~\ref{sec-simulation} to a simulation
study and Section~\ref{sec-sar} to the application to SAR imagery. All
the proofs are given in the Appendix.

\section{The model and the power family of responses}\label{sec-model}

\subsection{Log-linear exponential regression}\label{sec-loglinear}

Let \(T_{1},\ldots,T_{n}\) be independent random variables,
\(T_{i}\sim\mathcal{E}(\theta_{i})\), with
\(\operatorname{E}[T_{i}]=\theta_{i}>0\). For a fixed vector
\(\boldsymbol{x}_{i}^{\top}=(1,x_{i1},\ldots,x_{ik})\) of explanatory
variables and a vector of coefficients
\(\boldsymbol{\beta}=(\beta_{0},\beta_{1},\ldots,\beta_{k})^{\top}\in\mathbb{R}^{k+1}\),
the log-linear exponential regression model is
\begin{equation}\protect\phantomsection\label{eq-model}{
\eta_{i}=\log\theta_{i}=\boldsymbol{x}_{i}^{\top}\boldsymbol{\beta},\quad i=1,\ldots,n,
}\end{equation} or \(\boldsymbol{\eta}=\mathbb{X}\boldsymbol{\beta}\) in
matrix notation, with
\(\mathbb{X}^{\top}=\left[\boldsymbol{x}_{1}\cdots\boldsymbol{x}_{n}\right]\)
and \(\boldsymbol{\theta}=\exp\{\mathbb{X}\boldsymbol{\beta}\}\). It is
assumed throughout that \(\mathbb{X}\) has full column rank.

The same model can be formulated for two other response variables.

\begin{itemize}
\tightlist
\item
  The Rayleigh random variable
  \(R_{i}=T_{i}^{1/2}\sim\mathcal{R}(\vartheta_{i})\), with
  \(\theta_{i}=2\vartheta_{i}^{2}\). Since
  \(\mu_{i}=\operatorname{E}[R_{i}]=\sqrt{\pi/2}\,\vartheta_{i}\), the
  model of \citet{palm2019},
  \(g(\mu_{i})=\boldsymbol{x}_{i}^{\top}\boldsymbol{\beta}\), contains
  (\ref{eq-model}) with \(g(\mu_{i})=\log(\tfrac{4}{\pi}\mu_{i}^{2})\).
  With the logarithmic link function used in \citet{palm2019} and
  \citet{palm2022},
  \(\log\mu_{i}=\boldsymbol{x}_{i}^{\top}\boldsymbol{\delta}\), the two
  parametrizations are related through
  \(\beta_{0}=2\delta_{0}+\log\tfrac{4}{\pi}\) and
  \(\beta_{l}=2\delta_{l}\), \(l=1,\ldots,k\).
\item
  The type I extreme value random variable
  \(Y_{i}=\log T_{i}\sim\mathcal{EV}_{I}(\eta_{i})\), with density
  function
  \(f_{Y_{i}}(y;\boldsymbol{\beta})=\exp\{y-\eta_{i}-e^{y-\eta_{i}}\}\),
  \(y\in\mathbb{R}\), for which \(\eta_{i}\) is a location parameter.
\end{itemize}

\subsection{The power family}\label{sec-family}

The three responses are members of the family indexed by
\(c\in\mathbb{R}\),
\begin{equation}\protect\phantomsection\label{eq-family}{
Z_{i}^{(c)}=\begin{cases}T_{i}^{c}, & c\neq 0,\\ \log T_{i}, & c=0,\end{cases}
}\end{equation} with \(c=1\) for the exponential model,
\(c=\tfrac{1}{2}\) for the Rayleigh model and \(c=0\) for the type I
extreme value model. For \(c\neq 0\) the density function of
\(Z_{i}^{(c)}\) is
\begin{equation}\protect\phantomsection\label{eq-density}{
f_{Z_{i}^{(c)}}(z;\boldsymbol{\beta})=\frac{1}{|c|\,\theta_{i}}z^{\frac{1}{c}-1}\exp\left\{-\frac{z^{1/c}}{\theta_{i}}\right\},\quad z>0,
}\end{equation} which for \(c>0\) is the Weibull density function with
known shape parameter \(1/c\), and for \(c<0\) the density function of
the corresponding inverse Weibull distribution. When the exponent is
treated as an unknown parameter, the family is known as the power
Rayleigh distribution \citep{bhat2020}. Here \(c\) is known and fixed by
the analyst, and it is the effect of this choice on the minimum DPD
estimators that is studied.

\begin{refremark}
The case \(c=0\) is the limit of the family in the sense of
\citet{boxcox1964}. The transformation \((T_{i}^{c}-1)/c\) tends to
\(\log T_{i}\) as \(c\rightarrow 0\) and it is an affine transformation
of \(T_{i}^{c}\) whose coefficients do not depend on \(i\). Since the
minimum DPD estimator defined in Section~\ref{sec-mdpde} is equivariant
under affine transformations which are common to all the observations,
the estimator based on the Box-Cox transformation is the same as the one
based on \(Z_{i}^{(c)}\).

\label{rem-boxcox}

\end{refremark}

\subsection{Maximum likelihood estimation}\label{sec-mle}

Let \(t=z^{1/c}\) for \(c\neq 0\) and \(t=e^{z}\) for \(c=0\). For every
\(c\in\mathbb{R}\) the score function of the \(i\)-th observation is
\begin{equation}\protect\phantomsection\label{eq-score}{
\boldsymbol{s}_{\boldsymbol{\beta}}^{(i)}(z)=\frac{\partial}{\partial\boldsymbol{\beta}}\log f_{Z_{i}^{(c)}}(z;\boldsymbol{\beta})=\boldsymbol{x}_{i}\left(\frac{t}{\theta_{i}}-1\right),
}\end{equation} which does not depend on \(c\). Therefore the MLE of
\(\boldsymbol{\beta}\), \(\widehat{\boldsymbol{\beta}}\), is the same
for all the members of the family, it is the solution of
\(\sum_{i=1}^{n}\boldsymbol{x}_{i}\left(t_{i}/\theta_{i}-1\right)=\boldsymbol{0}_{k+1}\),
the Fisher information matrix of the sample is
\(\mathbb{X}^{\top}\mathbb{X}\) and
\begin{equation}\protect\phantomsection\label{eq-mle}{
\sqrt{n}(\widehat{\boldsymbol{\beta}}-\boldsymbol{\beta})\overset{\mathcal{L}}{\underset{n\rightarrow\infty}{\longrightarrow}}\mathcal{N}\left(\boldsymbol{0}_{k+1},\boldsymbol{G}_{0}^{-1}\right),\qquad \boldsymbol{G}_{0}=\lim_{n\rightarrow\infty}\frac{1}{n}\mathbb{X}^{\top}\mathbb{X}.
}\end{equation}

\begin{refremark}
The exponential distribution is the gamma distribution with unit
dispersion parameter, and hence (\ref{eq-model}) is a generalized linear
model \citep{mccullagh1989} with gamma response, dispersion parameter
fixed to one and logarithmic link function, which is not the canonical
one. The working weight of every observation is equal to one, which is
the reason why the Fisher information matrix does not depend on
\(\boldsymbol{\beta}\), and the logarithm is the variance stabilizing
transformation of the gamma family, which is the reason why
\(\theta_{i}\) becomes a location parameter for \(c=0\). From a
computational point of view, \(\widehat{\boldsymbol{\beta}}\) is
obtained in \texttt{R} with \texttt{glm}, taking
\texttt{family\ =\ Gamma(link\ =\ "log")} for \(t_{i}\), or with
\texttt{survreg} of the \texttt{survival} package \citep{therneau2024},
taking \texttt{dist\ =\ "exponential"} for \(t_{i}\),
\texttt{dist\ =\ "rayleigh"} for \(r_{i}\) (the coefficients are then
\(\boldsymbol{\beta}/2\)) or \texttt{dist\ =\ "extreme"} with
\texttt{scale\ =\ 1} for \(y_{i}\). Any of them provides a starting
value for the minimum DPD estimators.

\label{rem-glm}

\end{refremark}

\section{Minimum density power divergence estimators}\label{sec-mdpde}

\subsection{Definition and estimating equations}\label{sec-definition}

Following \citet{basu1998} and \citet{ghosh2013}, for a tuning parameter
\(\lambda>0\) the minimum DPD estimator of \(\boldsymbol{\beta}\) based
on \(Z_{1}^{(c)},\ldots,Z_{n}^{(c)}\) is
\begin{equation}\protect\phantomsection\label{eq-mdpde}{
\widehat{\boldsymbol{\beta}}_{\lambda}^{(c)}=\arg\min_{\boldsymbol{\beta}\in\mathbb{R}^{k+1}}\frac{1}{n}\sum_{i=1}^{n}\left\{\int f_{Z_{i}^{(c)}}^{1+\lambda}(z;\boldsymbol{\beta})dz-\frac{1+\lambda}{\lambda}f_{Z_{i}^{(c)}}^{\lambda}(z_{i};\boldsymbol{\beta})\right\},
}\end{equation} and the MLE is obtained as \(\lambda\rightarrow 0^{+}\).
Let \(w_{i}=t_{i}/\theta_{i}\), with \(t_{i}=z_{i}^{1/c}\) for
\(c\neq 0\) and \(t_{i}=e^{z_{i}}\) for \(c=0\), and
\begin{equation}\protect\phantomsection\label{eq-je}{
j_{c}(\lambda)=\frac{\Gamma((1-c)\lambda+1)}{(1+\lambda)^{(1-c)\lambda+3}}\left[c^{2}\lambda^{2}+(1-c)\lambda+1\right],\qquad
e_{c}(\lambda)=-c\lambda\frac{\Gamma((1-c)\lambda+1)}{(1+\lambda)^{(1-c)\lambda+2}}.
}\end{equation}

\begin{proposition}[]\protect\hypertarget{prp-equations}{}\label{prp-equations}

For \(\lambda>0\) and \(c<1+\tfrac{1}{\lambda}\), the minimum DPD
estimator \(\widehat{\boldsymbol{\beta}}_{\lambda}^{(c)}\) is a solution
of \begin{equation}\protect\phantomsection\label{eq-estimating}{
\boldsymbol{u}_{\lambda}^{(c)}(\boldsymbol{\beta})=\sum_{i=1}^{n}\theta_{i}^{-c\lambda}\boldsymbol{x}_{i}\,h_{c,\lambda}(w_{i})=\boldsymbol{0}_{k+1},\qquad h_{c,\lambda}(w)=(w-1)w^{(1-c)\lambda}e^{-\lambda w}-e_{c}(\lambda).
}\end{equation}

\end{proposition}

The members of the family differ in three elements of
(\ref{eq-estimating}), the weight \(\theta_{i}^{-c\lambda}\) given to
each individual, the factor \(w_{i}^{(1-c)\lambda}\) which multiplies
the score, and the centring term \(e_{c}(\lambda)\). For \(c=0\) the
first and the third ones disappear, since \(e_{0}(\lambda)=0\), and
(\ref{eq-estimating}) reduces to
\begin{equation}\protect\phantomsection\label{eq-estimatingEV}{
\sum_{i=1}^{n}\boldsymbol{x}_{i}(w_{i}-1)w_{i}^{\lambda}e^{-\lambda w_{i}}=\boldsymbol{0}_{k+1},
}\end{equation} which is the likelihood equation of the model with the
weight \(\varpi_{i}=w_{i}^{\lambda}e^{-\lambda w_{i}}\) for the \(i\)-th
observation. Hence, for \(c=0\) the minimum DPD estimator can be
computed by fitting repeatedly a weighted gamma generalized linear model
with logarithmic link function (Remark~\ref{rem-glm}), the weights being
updated in each iteration, and the weights of the last iteration show
which observations have been discarded.

For \(c=0\) the minimum DPD estimator coincides with the maximum
\(L_{q}\)-likelihood estimator of \citet{ferrari2010} with
\(q=1-\lambda\), which maximizes
\(\sum_{i=1}^{n}f_{Z_{i}^{(c)}}^{\lambda}(z_{i};\boldsymbol{\beta})\).
The function minimized in (\ref{eq-mdpde}) differs from this criterion
in the terms
\(\int f_{Z_{i}^{(c)}}^{1+\lambda}(z;\boldsymbol{\beta})dz\), which are
proportional to \(\theta_{i}^{-c\lambda}\) and therefore do not depend
on \(\boldsymbol{\beta}\) if and only if \(c=0\). In any other scale the
two estimators are different, the estimating equation of the maximum
\(L_{q}\)-likelihood estimator being (\ref{eq-estimating}) without the
centring term \(e_{c}(\lambda)\), and this paper deals only with the
minimum DPD estimators.

The estimating equation can be solved with a Fisher scoring type
algorithm, that is, the Newton method in which the matrix of derivatives
of the estimating function is replaced by its expectation under the
model, \begin{equation}\protect\phantomsection\label{eq-algorithm}{
\boldsymbol{\beta}^{(m+1)}=\boldsymbol{\beta}^{(m)}+\frac{1}{j_{c}(\lambda)}\left(\mathbb{X}^{\top}diag^{-c\lambda}\{\boldsymbol{\theta}^{(m)}\}\mathbb{X}\right)^{-1}\boldsymbol{u}_{\lambda}^{(c)}(\boldsymbol{\beta}^{(m)}),
}\end{equation} with
\(\boldsymbol{\theta}^{(m)}=\exp\{\mathbb{X}\boldsymbol{\beta}^{(m)}\}\),
the matrix
\(j_{c}(\lambda)\mathbb{X}^{\top}diag^{-c\lambda}\{\boldsymbol{\theta}\}\mathbb{X}\)
being the expectation of minus the matrix of derivatives of
\(\boldsymbol{u}_{\lambda}^{(c)}(\boldsymbol{\beta})\). For
\(\lambda=0\) it is the Fisher scoring algorithm of the MLE, since
\(j_{c}(0)=1\) and \(\mathbb{X}^{\top}\mathbb{X}\) is the Fisher
information matrix. For \(c=0\) the matrix to be inverted,
\(\mathbb{X}^{\top}\mathbb{X}\), is the same in all the iterations.

\subsection{Asymptotic distribution}\label{sec-asymptotic}

For \(a\in\mathbb{R}\) let
\begin{equation}\protect\phantomsection\label{eq-G}{
\boldsymbol{G}_{a}(\boldsymbol{\theta})=\lim_{n\rightarrow\infty}\frac{1}{n}\mathbb{X}^{\top}diag^{-a}\{\boldsymbol{\theta}\}\mathbb{X},
}\end{equation} so that
\(\boldsymbol{G}_{0}(\boldsymbol{\theta})=\boldsymbol{G}_{0}\) does not
depend on \(\boldsymbol{\theta}\). The following conditions on the
explanatory variables are considered.

\begin{itemize}
\tightlist
\item
  (C1) There is a constant \(M\) such that
  \(\|\boldsymbol{x}_{i}\|\leq M\) for every \(i\).
\item
  (C2) The limits \(\boldsymbol{G}_{0}\),
  \(\boldsymbol{G}_{c\lambda}(\boldsymbol{\theta})\) and
  \(\boldsymbol{G}_{2c\lambda}(\boldsymbol{\theta})\) exist, and
  \(\boldsymbol{G}_{0}\) is positive definite.
\end{itemize}

Under (C1),
\(e^{-M\|\boldsymbol{\beta}\|}\leq\theta_{i}\leq e^{M\|\boldsymbol{\beta}\|}\)
for every \(i\), so that
\(\boldsymbol{G}_{a}(\boldsymbol{\theta})-e^{-|a|M\|\boldsymbol{\beta}\|}\boldsymbol{G}_{0}\)
is positive semidefinite, and under (C2) the matrices
\(\boldsymbol{G}_{c\lambda}(\boldsymbol{\theta})\) and
\(\boldsymbol{G}_{2c\lambda}(\boldsymbol{\theta})\) are also positive
definite.

\begin{theorem}[]\protect\hypertarget{thm-asymptotic}{}\label{thm-asymptotic}

Let \(\lambda>0\) and \(c<1+\tfrac{1}{2\lambda}\), and assume that (C1)
and (C2) hold and, if \(c>1\), also the regularity conditions of
\citet{ghosh2013}. Then, with probability tending to one as
\(n\rightarrow\infty\), the estimating equation (\ref{eq-estimating})
has a solution \(\widehat{\boldsymbol{\beta}}_{\lambda}^{(c)}\) such
that
\(\sqrt{n}(\widehat{\boldsymbol{\beta}}_{\lambda}^{(c)}-\boldsymbol{\beta})\)
is bounded in probability, and for any sequence of solutions with this
property \begin{equation}\protect\phantomsection\label{eq-asymptotic}{
\sqrt{n}(\widehat{\boldsymbol{\beta}}_{\lambda}^{(c)}-\boldsymbol{\beta})\overset{\mathcal{L}}{\underset{n\rightarrow\infty}{\longrightarrow}}\mathcal{N}\left(\boldsymbol{0}_{k+1},\omega_{c}(\lambda)\boldsymbol{G}_{c\lambda}^{-1}(\boldsymbol{\theta})\boldsymbol{G}_{2c\lambda}(\boldsymbol{\theta})\boldsymbol{G}_{c\lambda}^{-1}(\boldsymbol{\theta})\right),
}\end{equation} where
\begin{equation}\protect\phantomsection\label{eq-omega}{
\omega_{c}(\lambda)=\frac{j_{c}(2\lambda)-e_{c}^{2}(\lambda)}{j_{c}^{2}(\lambda)}.
}\end{equation}

\end{theorem}

For \(c\leq 1\), which includes the three models of interest, the proof
given in the Appendix is self-contained and it does not require more
conditions than (C1) and (C2). For \(1<c<1+\tfrac{1}{2\lambda}\) the
estimating function is not bounded, and the result is obtained from the
general theorem of \citet{ghosh2013} under the conditions established
there. As usual for this kind of estimators, the result refers to a
suitable sequence of solutions of the estimating equation, and in what
follows \(\widehat{\boldsymbol{\beta}}_{\lambda}^{(c)}\) denotes such a
sequence.

\begin{corollary}[]\protect\hypertarget{cor-three}{}\label{cor-three}

The asymptotic variance-covariance matrices of the minimum DPD
estimators of \(\boldsymbol{\beta}\) for the type I extreme value,
Rayleigh and exponential models are, respectively, \[
\varphi(\lambda)\boldsymbol{G}_{0}^{-1},\qquad
\psi(\lambda)\boldsymbol{G}_{\lambda/2}^{-1}(\boldsymbol{\theta})\boldsymbol{G}_{\lambda}(\boldsymbol{\theta})\boldsymbol{G}_{\lambda/2}^{-1}(\boldsymbol{\theta}),\qquad
\phi(\lambda)\boldsymbol{G}_{\lambda}^{-1}(\boldsymbol{\theta})\boldsymbol{G}_{2\lambda}(\boldsymbol{\theta})\boldsymbol{G}_{\lambda}^{-1}(\boldsymbol{\theta}),
\] with \[
\begin{aligned}
\varphi(\lambda)&=\omega_{0}(\lambda)=\frac{(1+\lambda)^{4+2\lambda}}{(1+2\lambda)^{3+2\lambda}}\frac{1}{\mathcal{B}(1+\lambda,1+\lambda)},\\
\psi(\lambda)&=\omega_{1/2}(\lambda)=\frac{4}{(\lambda^{2}+2\lambda+4)^{2}}\left(\frac{4(1+\lambda)^{\lambda+5}}{(1+2\lambda)^{\lambda+3}}\frac{\lambda^{2}+\lambda+1}{\mathcal{B}(\tfrac{\lambda}{2}+1,\tfrac{\lambda}{2}+1)}-\lambda^{2}(1+\lambda)^{2}\right),\\
\phi(\lambda)&=\omega_{1}(\lambda)=\frac{(1+\lambda)^{6}}{(\lambda^{2}+1)^{2}}\left(\frac{4\lambda^{2}+1}{(1+2\lambda)^{3}}-\frac{\lambda^{2}}{(1+\lambda)^{4}}\right),
\end{aligned}
\] \(\mathcal{B}(a,b)\) being the beta function.

\end{corollary}

The three functions of Corollary~\ref{cor-three} are the ones which were
given in \citet{martin2026iwsm}. For \(c=0\) the asymptotic
variance-covariance matrix is proportional to the one of the MLE and it
does not depend on \(\boldsymbol{\theta}\). For \(c\neq 0\) it depends
on the unknown parameters through
\(\boldsymbol{G}_{c\lambda}(\boldsymbol{\theta})\) and
\(\boldsymbol{G}_{2c\lambda}(\boldsymbol{\theta})\).

\section{Comparison of efficiency}\label{sec-efficiency}

According to Theorem~\ref{thm-asymptotic}, the loss of efficiency of
\(\widehat{\boldsymbol{\beta}}_{\lambda}^{(c)}\) with respect to the MLE
has two sources, a matrix factor which depends on the design and on
\(\boldsymbol{\theta}\), and the scalar factor \(\omega_{c}(\lambda)\).
They are studied separately.

\subsection{The matrix factor: effect of the
heterogeneity}\label{sec-matrix}

\begin{theorem}[]\protect\hypertarget{thm-penalty}{}\label{thm-penalty}

Let \(a\neq 0\) and \(\boldsymbol{D}=diag^{-a}\{\boldsymbol{\theta}\}\).
For every \(n\),
\begin{equation}\protect\phantomsection\label{eq-penalty}{
(\mathbb{X}^{\top}\boldsymbol{D}\mathbb{X})^{-1}\mathbb{X}^{\top}\boldsymbol{D}^{2}\mathbb{X}(\mathbb{X}^{\top}\boldsymbol{D}\mathbb{X})^{-1}-(\mathbb{X}^{\top}\mathbb{X})^{-1}
}\end{equation} is a positive semidefinite matrix, and it is the zero
matrix if and only if
\(\boldsymbol{D}\mathbb{X}=\mathbb{X}\boldsymbol{\Delta}\) for some
nonsingular matrix \(\boldsymbol{\Delta}\). As a consequence, whenever
the limits exist, \[
\boldsymbol{G}_{a}^{-1}(\boldsymbol{\theta})\boldsymbol{G}_{2a}(\boldsymbol{\theta})\boldsymbol{G}_{a}^{-1}(\boldsymbol{\theta})-\boldsymbol{G}_{0}^{-1}
\] is positive semidefinite.

\end{theorem}

\begin{corollary}[]\protect\hypertarget{cor-penalty}{}\label{cor-penalty}

Under the conditions of Theorem~\ref{thm-asymptotic}, the asymptotic
variance-covariance matrix of
\(\widehat{\boldsymbol{\beta}}_{\lambda}^{(c)}\) minus
\(\omega_{c}(\lambda)\boldsymbol{G}_{0}^{-1}\) is positive semidefinite.
It is the zero matrix in the following cases.

\begin{enumerate}
\def\labelenumi{\alph{enumi})}
\tightlist
\item
  For \(c=0\), with any design matrix.
\item
  For identically distributed observations.
\item
  For one-way layouts, in which the sample is divided into populations
  and \(\theta_{i}\) is constant inside each population.
\end{enumerate}

\end{corollary}

Hence, for the comparison of populations, as required in ground type
detection, the efficiency of all the members of the family with respect
to the MLE is completely described by \(\omega_{c}(\lambda)\), as in the
i.i.d. case. With explanatory variables which make \(\theta_{i}\) vary
inside the populations, the models with \(c\neq 0\) suffer an additional
loss of efficiency and the model with \(c=0\) does not.

The size of the additional loss depends on the dispersion of
\(\boldsymbol{\theta}\). Table~\ref{tbl-penalty} shows, for a simple
regression with \(\eta_{i}=\beta_{0}+\beta_{1}x_{i}\) and \(x_{i}\)
equally spaced on \([0,1]\), the ratio between the element of the matrix
factor associated with \(\beta_{1}\) and the corresponding element of
\(\boldsymbol{G}_{0}^{-1}\). The loss is negligible when \(\theta_{i}\)
varies in a short range and it is important for the exponential model
when \(\max_{i}\theta_{i}/\min_{i}\theta_{i}\) is large.

\begin{longtable}[]{@{}
  >{\centering\arraybackslash}p{(\linewidth - 10\tabcolsep) * \real{0.1667}}
  >{\centering\arraybackslash}p{(\linewidth - 10\tabcolsep) * \real{0.1667}}
  >{\centering\arraybackslash}p{(\linewidth - 10\tabcolsep) * \real{0.1667}}
  >{\centering\arraybackslash}p{(\linewidth - 10\tabcolsep) * \real{0.1667}}
  >{\centering\arraybackslash}p{(\linewidth - 10\tabcolsep) * \real{0.1667}}
  >{\centering\arraybackslash}p{(\linewidth - 10\tabcolsep) * \real{0.1667}}@{}}
\caption{Matrix factor for the slope of a simple regression, relative to
the model with \(c=0\).}\label{tbl-penalty}\tabularnewline
\toprule\noalign{}
\begin{minipage}[b]{\linewidth}\centering
\(\beta_{1}\)
\end{minipage} & \begin{minipage}[b]{\linewidth}\centering
\(\max_{i}\theta_{i}/\min_{i}\theta_{i}\)
\end{minipage} & \begin{minipage}[b]{\linewidth}\centering
\(c=\tfrac12\), \(\lambda=0.5\)
\end{minipage} & \begin{minipage}[b]{\linewidth}\centering
\(c=\tfrac12\), \(\lambda=1\)
\end{minipage} & \begin{minipage}[b]{\linewidth}\centering
\(c=1\), \(\lambda=0.5\)
\end{minipage} & \begin{minipage}[b]{\linewidth}\centering
\(c=1\), \(\lambda=1\)
\end{minipage} \\
\midrule\noalign{}
\endfirsthead
\toprule\noalign{}
\begin{minipage}[b]{\linewidth}\centering
\(\beta_{1}\)
\end{minipage} & \begin{minipage}[b]{\linewidth}\centering
\(\max_{i}\theta_{i}/\min_{i}\theta_{i}\)
\end{minipage} & \begin{minipage}[b]{\linewidth}\centering
\(c=\tfrac12\), \(\lambda=0.5\)
\end{minipage} & \begin{minipage}[b]{\linewidth}\centering
\(c=\tfrac12\), \(\lambda=1\)
\end{minipage} & \begin{minipage}[b]{\linewidth}\centering
\(c=1\), \(\lambda=0.5\)
\end{minipage} & \begin{minipage}[b]{\linewidth}\centering
\(c=1\), \(\lambda=1\)
\end{minipage} \\
\midrule\noalign{}
\endhead
\bottomrule\noalign{}
\endlastfoot
\(1\) & \(2.7\) & \(1.004\) & \(1.017\) & \(1.017\) & \(1.068\) \\
\(2\) & \(7.4\) & \(1.017\) & \(1.068\) & \(1.068\) & \(1.289\) \\
\(3\) & \(20.1\) & \(1.038\) & \(1.157\) & \(1.157\) & \(1.716\) \\
\end{longtable}

\subsection{The scalar factor}\label{sec-scalar}

The function \(\omega_{c}(\lambda)\) is the ratio between the asymptotic
variances of the minimum DPD estimator and the MLE when the matrix
factor is not present, and its reciprocal is the asymptotic relative
efficiency. The following result describes its behaviour for small
values of the tuning parameter.

\begin{proposition}[]\protect\hypertarget{prp-small}{}\label{prp-small}

For every \(c\in\mathbb{R}\), as \(\lambda\rightarrow 0^{+}\),
\begin{equation}\protect\phantomsection\label{eq-kappa}{
\log\omega_{c}(\lambda)=\kappa(c)\lambda^{2}+O(\lambda^{3}),\qquad \kappa(c)=\left(\frac{\pi^{2}}{6}-1\right)(1-c)^{2}+(1+c)^{2}.
}\end{equation} The function \(\kappa(c)\) is strictly convex, it is
minimized at \(c^{\ast}=1-12/\pi^{2}\simeq-0.216\) and it is strictly
increasing on \([c^{\ast},+\infty)\). In particular,
\(\kappa(0)=\pi^{2}/6\simeq 1.645\), \(\kappa(\tfrac12)\simeq 2.411\)
and \(\kappa(1)=4\).

\end{proposition}

\begin{figure}[!htb]

\centering{

\pandocbounded{\includegraphics[keepaspectratio]{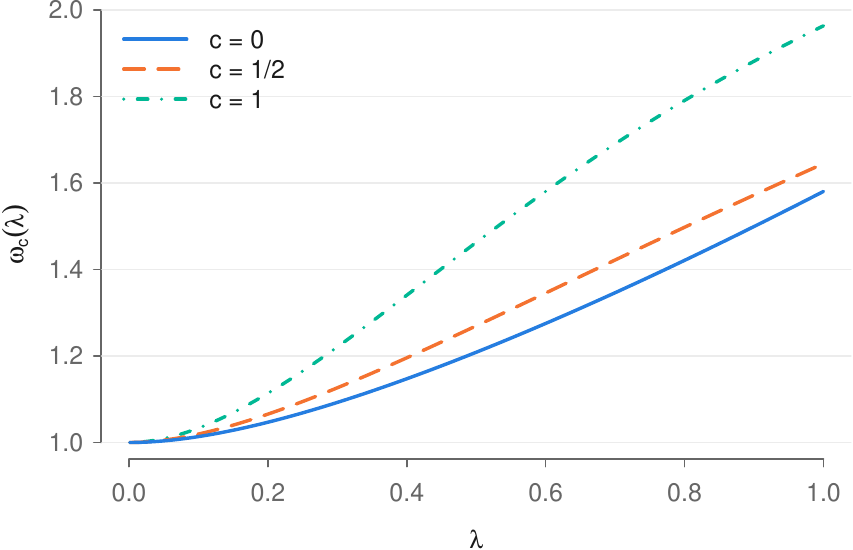}}

}

\caption{\label{fig-omega}Scalar factor \(\omega_{c}(\lambda)\) for the
type I extreme value (\(c=0\)), Rayleigh (\(c=1/2\)) and exponential
(\(c=1\)) models.}

\end{figure}

Therefore, for the same and small enough value of the tuning parameter,
the loss of efficiency of the Rayleigh model is about \(1.5\) times the
one of the type I extreme value model, and the loss of the exponential
model about \(2.4\) times. Table~\ref{tbl-omega} and
Figure~\ref{fig-omega} show that the ordering holds in the whole usual
range of the tuning parameter.

\begin{longtable}[]{@{}
  >{\centering\arraybackslash}p{(\linewidth - 10\tabcolsep) * \real{0.1667}}
  >{\centering\arraybackslash}p{(\linewidth - 10\tabcolsep) * \real{0.1667}}
  >{\centering\arraybackslash}p{(\linewidth - 10\tabcolsep) * \real{0.1667}}
  >{\centering\arraybackslash}p{(\linewidth - 10\tabcolsep) * \real{0.1667}}
  >{\centering\arraybackslash}p{(\linewidth - 10\tabcolsep) * \real{0.1667}}
  >{\centering\arraybackslash}p{(\linewidth - 10\tabcolsep) * \real{0.1667}}@{}}
\caption{Values of
\(\omega_{c}(\lambda)\).}\label{tbl-omega}\tabularnewline
\toprule\noalign{}
\begin{minipage}[b]{\linewidth}\centering
\(\lambda\)
\end{minipage} & \begin{minipage}[b]{\linewidth}\centering
\(c=0\)
\end{minipage} & \begin{minipage}[b]{\linewidth}\centering
\(c=\tfrac14\)
\end{minipage} & \begin{minipage}[b]{\linewidth}\centering
\(c=\tfrac12\)
\end{minipage} & \begin{minipage}[b]{\linewidth}\centering
\(c=\tfrac34\)
\end{minipage} & \begin{minipage}[b]{\linewidth}\centering
\(c=1\)
\end{minipage} \\
\midrule\noalign{}
\endfirsthead
\toprule\noalign{}
\begin{minipage}[b]{\linewidth}\centering
\(\lambda\)
\end{minipage} & \begin{minipage}[b]{\linewidth}\centering
\(c=0\)
\end{minipage} & \begin{minipage}[b]{\linewidth}\centering
\(c=\tfrac14\)
\end{minipage} & \begin{minipage}[b]{\linewidth}\centering
\(c=\tfrac12\)
\end{minipage} & \begin{minipage}[b]{\linewidth}\centering
\(c=\tfrac34\)
\end{minipage} & \begin{minipage}[b]{\linewidth}\centering
\(c=1\)
\end{minipage} \\
\midrule\noalign{}
\endhead
\bottomrule\noalign{}
\endlastfoot
\(0.10\) & \(1.0137\) & \(1.0158\) & \(1.0196\) & \(1.0253\) &
\(1.0333\) \\
\(0.25\) & \(1.0685\) & \(1.0781\) & \(1.0947\) & \(1.1218\) &
\(1.1650\) \\
\(0.50\) & \(1.2086\) & \(1.2328\) & \(1.2696\) & \(1.3348\) &
\(1.4625\) \\
\(0.75\) & \(1.3831\) & \(1.4194\) & \(1.4598\) & \(1.5366\) &
\(1.7415\) \\
\(1.00\) & \(1.5802\) & \(1.6237\) & \(1.6444\) & \(1.6982\) &
\(1.9630\) \\
\end{longtable}

\begin{refremark}
Two numerical facts complete Proposition~\ref{prp-small}. First,
evaluating (\ref{eq-omega}) on a grid, \(c\mapsto\omega_{c}(\lambda)\)
is increasing on \([0,1]\) for every \(\lambda\leq 1.27\), and it is not
for larger values, for which the comparison of the three models changes
(\(\omega_{0}(\lambda)<\omega_{1/2}(\lambda)\) only for
\(\lambda<1.343\) and \(\omega_{0}(\lambda)<\omega_{1}(\lambda)\) only
for \(\lambda<2.011\)). Second, since \(c^{\ast}<0\), the logarithm is
not the member of the family with the smallest scalar factor. Negative
exponents, which correspond to inverse Weibull responses, provide
slightly smaller values, for example \(\omega_{-1}(0.5)=1.1545\) against
\(\omega_{0}(0.5)=1.2086\). They are not considered in the rest of the
paper, since they do not have a physical interpretation for SAR data,
they are affected by the matrix factor of Theorem~\ref{thm-penalty} and,
as shown in Section~\ref{sec-ges}, they have a larger gross error
sensitivity.

\label{rem-range}

\end{refremark}

\section{Robustness}\label{sec-robustness}

\subsection{Influence function}\label{sec-if}

\begin{proposition}[]\protect\hypertarget{prp-influence}{}\label{prp-influence}

Let \(\boldsymbol{U}_{\lambda}^{(c)}\) be the functional associated with
\(\widehat{\boldsymbol{\beta}}_{\lambda}^{(c)}\) and consider a
contamination at the point \(t\), in the scale of the exponential
response, in the distribution of the \(i\)-th observation. Its influence
function is \begin{equation}\protect\phantomsection\label{eq-if}{
\mathcal{IF}_{i}(t,\boldsymbol{U}_{\lambda}^{(c)},\boldsymbol{\beta})=\frac{\theta_{i}^{-c\lambda}}{j_{c}(\lambda)}\left(\frac{1}{n}\mathbb{X}^{\top}diag^{-c\lambda}\{\boldsymbol{\theta}\}\mathbb{X}\right)^{-1}\frac{1}{n}\boldsymbol{x}_{i}\,h_{c,\lambda}\left(\frac{t}{\theta_{i}}\right),
}\end{equation} with \(h_{c,\lambda}\) given in (\ref{eq-estimating}).
For \(\lambda>0\) it is a bounded function of \(t\) if and only if
\(c\leq 1\), and \[
\lim_{t\rightarrow+\infty}h_{c,\lambda}\left(\frac{t}{\theta_{i}}\right)=-e_{c}(\lambda),\qquad
\lim_{t\rightarrow 0^{+}}h_{c,\lambda}\left(\frac{t}{\theta_{i}}\right)=\begin{cases}-e_{c}(\lambda), & c<1,\\ -1-e_{1}(\lambda), & c=1.\end{cases}
\] In particular, the influence function tends to the zero vector as
\(t\rightarrow+\infty\) if and only if \(c=0\).

\end{proposition}

For identically distributed observations and the parameter
\(\eta=\log\theta\), (\ref{eq-if}) reduces to
\(h_{c,\lambda}(t/\theta)/j_{c}(\lambda)\). The three models of interest
have a bounded influence function, but with a different shape. For the
type I extreme value model it is redescending in the sense of
\citet{hampel1986}, and gross outliers are completely discarded. For the
Rayleigh and exponential models it tends to a positive constant, and
observations which are arbitrarily large keep a fixed influence on the
estimator. For the exponential model, in addition, the influence of the
observations which are close to zero is the largest one in absolute
value when \(\lambda\) is not small.

\subsection{Gross error sensitivity at equal efficiency}\label{sec-ges}

The gross error sensitivity for the parameter \(\eta\) is
\(\gamma_{c}^{\ast}(\lambda)=\sup_{w>0}|h_{c,\lambda}(w)|/j_{c}(\lambda)\),
which does not depend on \(\theta\). A comparison of the models with the
same value of \(\lambda\) is not fair, since they do not have the same
efficiency. Table~\ref{tbl-ges} considers, for each \(c\), the value of
\(\lambda\) which provides a given value of \(\omega_{c}(\lambda)\).

\begin{longtable}[]{@{}
  >{\centering\arraybackslash}p{(\linewidth - 12\tabcolsep) * \real{0.1429}}
  >{\centering\arraybackslash}p{(\linewidth - 12\tabcolsep) * \real{0.1429}}
  >{\centering\arraybackslash}p{(\linewidth - 12\tabcolsep) * \real{0.1429}}
  >{\centering\arraybackslash}p{(\linewidth - 12\tabcolsep) * \real{0.1429}}
  >{\centering\arraybackslash}p{(\linewidth - 12\tabcolsep) * \real{0.1429}}
  >{\centering\arraybackslash}p{(\linewidth - 12\tabcolsep) * \real{0.1429}}
  >{\centering\arraybackslash}p{(\linewidth - 12\tabcolsep) * \real{0.1429}}@{}}
\caption{Tuning parameter \(\lambda_{v}\) such that
\(\omega_{c}(\lambda_{v})=v\), and the corresponding gross error
sensitivity.}\label{tbl-ges}\tabularnewline
\toprule\noalign{}
\begin{minipage}[b]{\linewidth}\centering
\(c\)
\end{minipage} & \begin{minipage}[b]{\linewidth}\centering
\(\lambda_{1.1}\)
\end{minipage} & \begin{minipage}[b]{\linewidth}\centering
\(\gamma_{c}^{\ast}(\lambda_{1.1})\)
\end{minipage} & \begin{minipage}[b]{\linewidth}\centering
\(\lambda_{1.2}\)
\end{minipage} & \begin{minipage}[b]{\linewidth}\centering
\(\gamma_{c}^{\ast}(\lambda_{1.2})\)
\end{minipage} & \begin{minipage}[b]{\linewidth}\centering
\(\lambda_{1.4}\)
\end{minipage} & \begin{minipage}[b]{\linewidth}\centering
\(\gamma_{c}^{\ast}(\lambda_{1.4})\)
\end{minipage} \\
\midrule\noalign{}
\endfirsthead
\toprule\noalign{}
\begin{minipage}[b]{\linewidth}\centering
\(c\)
\end{minipage} & \begin{minipage}[b]{\linewidth}\centering
\(\lambda_{1.1}\)
\end{minipage} & \begin{minipage}[b]{\linewidth}\centering
\(\gamma_{c}^{\ast}(\lambda_{1.1})\)
\end{minipage} & \begin{minipage}[b]{\linewidth}\centering
\(\lambda_{1.2}\)
\end{minipage} & \begin{minipage}[b]{\linewidth}\centering
\(\gamma_{c}^{\ast}(\lambda_{1.2})\)
\end{minipage} & \begin{minipage}[b]{\linewidth}\centering
\(\lambda_{1.4}\)
\end{minipage} & \begin{minipage}[b]{\linewidth}\centering
\(\gamma_{c}^{\ast}(\lambda_{1.4})\)
\end{minipage} \\
\midrule\noalign{}
\endhead
\bottomrule\noalign{}
\endlastfoot
\(-1\) & \(0.348\) & \(3.72\) & \(0.628\) & \(3.28\) & \(1.201\) &
\(3.23\) \\
\(0\) & \(0.314\) & \(2.90\) & \(0.487\) & \(2.55\) & \(0.772\) &
\(2.44\) \\
\(\tfrac14\) & \(0.291\) & \(2.78\) & \(0.452\) & \(2.40\) & \(0.725\) &
\(2.24\) \\
\(\tfrac12\) & \(0.259\) & \(2.72\) & \(0.406\) & \(2.28\) & \(0.671\) &
\(2.05\) \\
\(\tfrac34\) & \(0.221\) & \(2.76\) & \(0.345\) & \(2.23\) & \(0.577\) &
\(1.91\) \\
\(1\) & \(0.185\) & \(2.87\) & \(0.282\) & \(2.26\) & \(0.449\) &
\(1.99\) \\
\end{longtable}

\begin{figure}[!htb]

\centering{

\pandocbounded{\includegraphics[keepaspectratio]{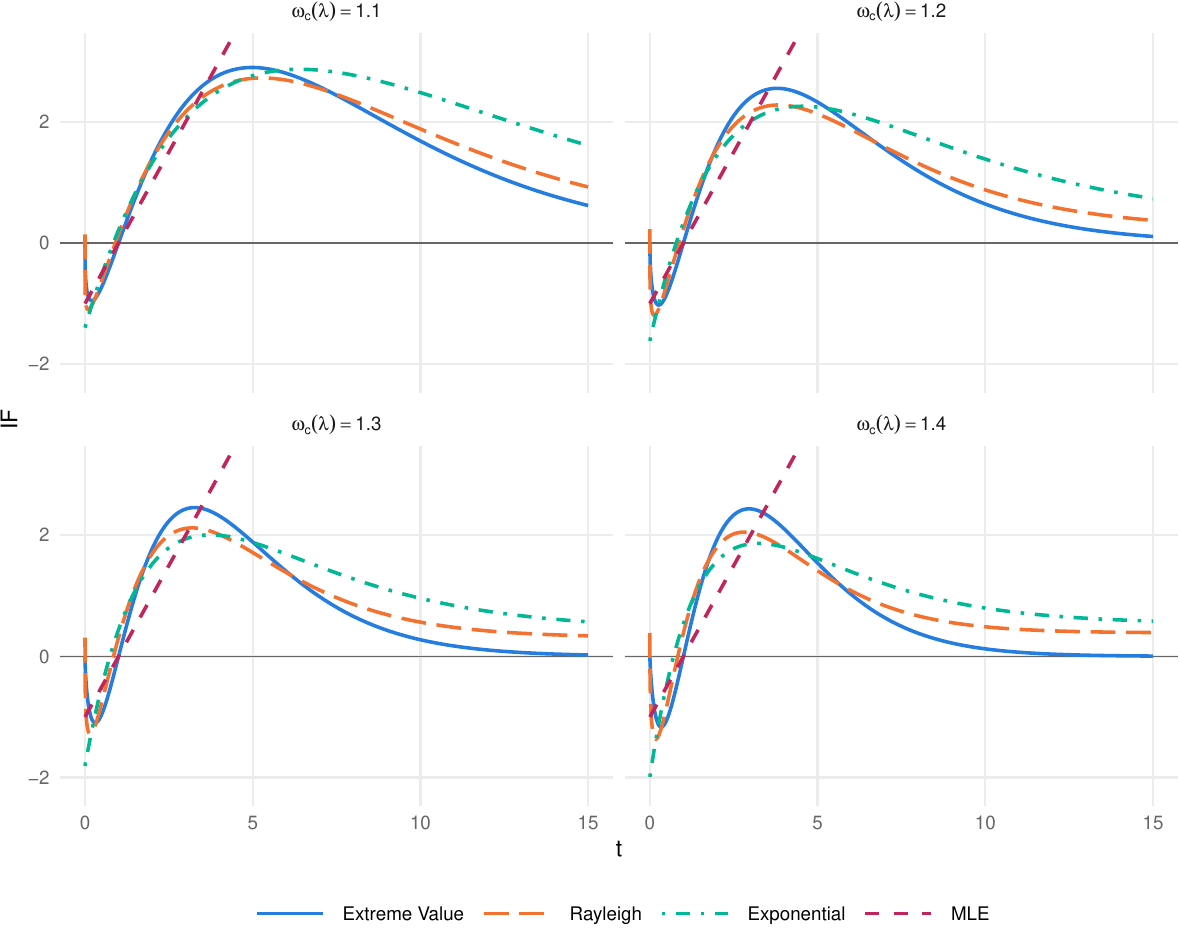}}

}

\caption{\label{fig-influence}Influence functions for the parameter
\(\eta\) at \(\theta=1\) of the type I extreme value, Rayleigh and
exponential models, with the values of \(\lambda\) which provide the
same value of \(\omega_{c}(\lambda)\), together with the one of the
MLE.}

\end{figure}%

Figure~\ref{fig-influence} shows the influence functions of the three
models of interest for four values of \(\omega_{c}(\lambda)\). At equal
efficiency the gross error sensitivity is minimized at an interior point
of \([0,1]\), close to the Rayleigh model for small losses of efficiency
and between the Rayleigh and the exponential models for larger ones, and
the differences among the three models of interest are moderate. The
type I extreme value model has the largest value of the three. Hence
there is not a uniformly best scale. Decreasing \(c\) from \(1\) to
\(0\) improves the efficiency for a given \(\lambda\)
(Proposition~\ref{prp-small}), removes the dependence on the unknown
parameters and the heterogeneity loss (Theorem~\ref{thm-penalty}) and
makes the influence function redescending
(Proposition~\ref{prp-influence}), at the price of a larger maximum
influence of the moderate outliers. The gross error sensitivity is a
worst-case measure, attained at a single point, and the simulation study
of Section~\ref{sec-simulation} shows that for contaminating
distributions which are not concentrated at that point the redescending
shape is the dominant feature.

\subsection{Observations equal to zero}\label{sec-zeros}

Observations which are equal to zero, or very close to it, are common in
practice. In magnitude SAR images they are a consequence of the finite
number of digital levels (Section~\ref{sec-sar}), and they are the
reason why models with a point mass at zero have been proposed
\citep{palm2024}. Let the distribution of \(T_{i}\) be contaminated at
zero, \begin{equation}\protect\phantomsection\label{eq-zeros}{
(1-\varepsilon)\,\mathcal{E}(\theta_{i})+\varepsilon\,\delta_{0},\qquad \varepsilon\in[0,1),
}\end{equation} \(\delta_{0}\) being the degenerate distribution at
zero. The influence function describes the effect of an infinitesimal
value of \(\varepsilon\). For this contamination the type I extreme
value model has a stronger property, which holds for any
\(\varepsilon\).

\begin{proposition}[]\protect\hypertarget{prp-zeros}{}\label{prp-zeros}

Let \(c=0\) and \(\lambda>0\). Under (\ref{eq-zeros}), for every
\(\varepsilon\in[0,1)\) the expectation of the estimating function
(\ref{eq-estimatingEV}) is zero at the true value of
\(\boldsymbol{\beta}\). That is, the functional associated with the
minimum DPD estimator of the type I extreme value model is not modified
by a contamination at zero, whatever its proportion is.

\end{proposition}

The proof is immediate, since for \(\lambda>0\) the function
\((w-1)w^{\lambda}e^{-\lambda w}\) is zero at \(w=0\) and it has zero
expectation for \(W\sim\mathcal{E}(1)\). The property does not hold for
\(\lambda=0\), nor for \(c\neq 0\), since
\(h_{c,\lambda}(0)=-e_{c}(\lambda)\neq 0\) for \(c<1\), \(c\neq 0\), and
\(h_{1,\lambda}(0)=-1-e_{1}(\lambda)\). Table~\ref{tbl-zeros} shows the
asymptotic bias of the estimators of \(\eta\) for identically
distributed observations, obtained by solving
\((1-\varepsilon)\operatorname{E}[h_{c,\lambda}(W/\kappa)]+\varepsilon h_{c,\lambda}(0)=0\)
in \(\kappa\), the bias being \(\log\kappa\), with the values of
\(\lambda\) which provide \(\omega_{c}(\lambda)=1.2\). For the MLE the
bias is \(\log(1-\varepsilon)\).

\begin{longtable}[]{@{}ccccc@{}}
\caption{Asymptotic bias of the estimators of \(\eta\) under a
contamination at zero with proportion \(\varepsilon\), with
\(\omega_{c}(\lambda)=1.2\).}\label{tbl-zeros}\tabularnewline
\toprule\noalign{}
\(\varepsilon\) & MLE & \(c=1\) & \(c=\tfrac12\) & \(c=0\) \\
\midrule\noalign{}
\endfirsthead
\toprule\noalign{}
\(\varepsilon\) & MLE & \(c=1\) & \(c=\tfrac12\) & \(c=0\) \\
\midrule\noalign{}
\endhead
\bottomrule\noalign{}
\endlastfoot
\(0.01\) & \(-0.010\) & \(-0.016\) & \(0.002\) & \(0\) \\
\(0.05\) & \(-0.051\) & \(-0.085\) & \(0.012\) & \(0\) \\
\(0.10\) & \(-0.105\) & \(-0.178\) & \(0.026\) & \(0\) \\
\(0.20\) & \(-0.223\) & \(-0.400\) & \(0.058\) & \(0\) \\
\end{longtable}

The exponential model is more affected than the MLE, because its weight
\(e^{-\lambda w}\) is maximum at \(w=0\), the Rayleigh model has a small
bias, and the type I extreme value model has none. This behaviour has a
counterpart for multinomial data. In \citet{martin2026} it is shown, for
interval-monitored step-stress life tests, that a positive tuning
parameter reduces the influence on the minimum DPD estimator of a cell
whose probability under the model is close to zero, relative to the
other cells, while for cells of moderate probability the comparison with
the MLE may go in the opposite direction. The same two features are
found here. The protection is the strongest one for the observations to
which the model assigns a density close to zero, and the maximum
influence of the moderate outliers is not reduced with respect to the
other scales (Section~\ref{sec-ges}). Proposition~\ref{prp-zeros} and
Table~\ref{tbl-zeros} add that, in the continuous case, the protection
against the observations of small density depends on the scale of the
response, and that it is complete at both ends of the support only for
the logarithmic one.

\section{Hypothesis testing}\label{sec-tests}

\subsection{General linear hypothesis for the type I extreme value
model}\label{sec-tests-general}

Let
\(\boldsymbol{\beta}=(\boldsymbol{\beta}_{\ast}^{\top},\boldsymbol{\beta}^{\ast\top})^{\top}\),
with
\(\boldsymbol{\beta}^{\ast}=(\beta_{k-r+1},\ldots,\beta_{k})^{\top}\),
and consider \begin{equation}\protect\phantomsection\label{eq-H0}{
H_{0}:\boldsymbol{\beta}^{\ast}=\boldsymbol{0}_{r},
}\end{equation}
\(\boldsymbol{\beta}_{\ast}=(\beta_{0},\ldots,\beta_{k-r})^{\top}\)
being a vector of nuisance parameters. With the corresponding partition
of the design matrix,
\(\mathbb{X}=(\mathbb{X}_{\ast},\mathbb{X}^{\ast})\), under
(\ref{eq-H0}) it holds
\(\mathbb{X}\boldsymbol{\beta}=\mathbb{X}_{\ast}\boldsymbol{\beta}_{\ast}\).

In this section \(c=0\),
\(\widehat{\boldsymbol{\beta}}_{\lambda}=\widehat{\boldsymbol{\beta}}_{\lambda}^{(0)}\)
is the minimum DPD estimator and
\(\widetilde{\boldsymbol{\beta}}_{\lambda}=(\widetilde{\boldsymbol{\beta}}_{\ast,\lambda}^{\top},\boldsymbol{0}_{r}^{\top})^{\top}\)
the restricted one, where
\(\widetilde{\boldsymbol{\beta}}_{\ast,\lambda}\) is the minimum DPD
estimator of the model with design matrix \(\mathbb{X}_{\ast}\). Let \[
\boldsymbol{u}_{\lambda}(\boldsymbol{\beta})=\sum_{i=1}^{n}\boldsymbol{x}_{i}(w_{i}-1)w_{i}^{\lambda}e^{-\lambda w_{i}},\qquad
L_{n,\lambda}(\boldsymbol{\beta})=\sum_{i=1}^{n}w_{i}^{\lambda}e^{-\lambda w_{i}},\qquad w_{i}=t_{i}\exp\{-\boldsymbol{x}_{i}^{\top}\boldsymbol{\beta}\},
\] be the estimating function (\ref{eq-estimatingEV}) and the sum of the
\(\lambda\)-th powers of the densities, and \[
j_{0}(\lambda)=\frac{\Gamma(\lambda+1)}{(1+\lambda)^{\lambda+2}},\qquad \varphi(\lambda)=\frac{j_{0}(2\lambda)}{j_{0}^{2}(\lambda)}.
\] The Wald-type, Rao-type and divergence-based test statistics are
defined as \begin{equation}\protect\phantomsection\label{eq-wald}{
W_{n,\lambda}=\frac{1}{\varphi(\lambda)}\,\widehat{\boldsymbol{\beta}}_{\lambda}^{\ast\top}\boldsymbol{S}_{22\cdot 1}\widehat{\boldsymbol{\beta}}_{\lambda}^{\ast},
}\end{equation} \begin{equation}\protect\phantomsection\label{eq-rao}{
R_{n,\lambda}=\frac{1}{j_{0}(2\lambda)}\,\boldsymbol{u}_{\lambda}^{\top}(\widetilde{\boldsymbol{\beta}}_{\lambda})(\mathbb{X}^{\top}\mathbb{X})^{-1}\boldsymbol{u}_{\lambda}(\widetilde{\boldsymbol{\beta}}_{\lambda}),
}\end{equation} \begin{equation}\protect\phantomsection\label{eq-lq}{
S_{n,\lambda}=\frac{2\,j_{0}(\lambda)}{\lambda\,j_{0}(2\lambda)}\left[L_{n,\lambda}(\widehat{\boldsymbol{\beta}}_{\lambda})-L_{n,\lambda}(\widetilde{\boldsymbol{\beta}}_{\lambda})\right],
}\end{equation} where \(\boldsymbol{S}=\mathbb{X}^{\top}\mathbb{X}\) is
partitioned according to
\((\boldsymbol{\beta}_{\ast},\boldsymbol{\beta}^{\ast})\), with blocks
\(\boldsymbol{S}_{11}=\mathbb{X}_{\ast}^{\top}\mathbb{X}_{\ast}\),
\(\boldsymbol{S}_{12}=\mathbb{X}_{\ast}^{\top}\mathbb{X}^{\ast}\),
\(\boldsymbol{S}_{21}=\boldsymbol{S}_{12}^{\top}\) and
\(\boldsymbol{S}_{22}=\mathbb{X}^{\ast\top}\mathbb{X}^{\ast}\), and
\begin{equation}\protect\phantomsection\label{eq-S221}{
\boldsymbol{S}_{22\cdot 1}=\boldsymbol{S}_{22}-\boldsymbol{S}_{21}\boldsymbol{S}_{11}^{-1}\boldsymbol{S}_{12}.
}\end{equation} By the formula of the inverse of a partitioned matrix,
\(\boldsymbol{S}_{22\cdot 1}^{-1}\) is the block of
\((\mathbb{X}^{\top}\mathbb{X})^{-1}\) associated with
\(\boldsymbol{\beta}^{\ast}\). The statistic \(S_{n,\lambda}\) is the
one of \citet{basu2018}, twice the difference between the values of the
empirical DPD at the restricted and unrestricted estimators multiplied
by \(n\), divided by a constant, and it is closely related to the
\(L_{q}\)-likelihood ratio test statistic of \citet{qin2017} with
\(q=1-\lambda\). The statistic \(R_{n,\lambda}\) is of the score type
considered in \citet{martin2023}.

\begin{theorem}[]\protect\hypertarget{thm-tests}{}\label{thm-tests}

Let \(c=0\) and \(\lambda>0\), assume that (C1) holds and that
\(\boldsymbol{G}_{0}\) exists and is positive definite, and let
\(\widehat{\boldsymbol{\beta}}_{\lambda}\) and
\(\widetilde{\boldsymbol{\beta}}_{\ast,\lambda}\) be sequences of
solutions of the unrestricted and restricted estimating equations such
that
\(\sqrt{n}(\widehat{\boldsymbol{\beta}}_{\lambda}-\boldsymbol{\beta})\)
and
\(\sqrt{n}(\widetilde{\boldsymbol{\beta}}_{\ast,\lambda}-\boldsymbol{\beta}_{\ast})\)
are bounded in probability, \(\boldsymbol{\beta}\) being the true value
of the parameter, which exist with probability tending to one in the two
cases considered below. Let \(\boldsymbol{G}_{0}\) be partitioned
according to \((\boldsymbol{\beta}_{\ast},\boldsymbol{\beta}^{\ast})\),
with blocks \(\boldsymbol{G}_{11}\), \(\boldsymbol{G}_{12}\),
\(\boldsymbol{G}_{21}\) and \(\boldsymbol{G}_{22}\), and
\(\boldsymbol{G}_{22\cdot 1}=\boldsymbol{G}_{22}-\boldsymbol{G}_{21}\boldsymbol{G}_{11}^{-1}\boldsymbol{G}_{12}\).

\begin{enumerate}
\def\labelenumi{\alph{enumi})}
\item
  Under the null hypothesis (\ref{eq-H0}), \[
  W_{n,\lambda},\;R_{n,\lambda},\;S_{n,\lambda}\overset{\mathcal{L}}{\underset{n\rightarrow\infty}{\longrightarrow}}\chi_{r}^{2}.
  \]
\item
  Under the contiguous alternative hypotheses
  \(H_{1,n}:\boldsymbol{\beta}^{\ast}=n^{-1/2}\boldsymbol{d}\), with
  \(\boldsymbol{d}\in\mathbb{R}^{r}\) fixed, \[
  W_{n,\lambda},\;R_{n,\lambda},\;S_{n,\lambda}\overset{\mathcal{L}}{\underset{n\rightarrow\infty}{\longrightarrow}}\chi_{r}^{2}(\delta_{\lambda}),\qquad \delta_{\lambda}=\frac{1}{\varphi(\lambda)}\boldsymbol{d}^{\top}\boldsymbol{G}_{22\cdot 1}\boldsymbol{d},
  \] where \(\chi_{r}^{2}(\delta)\) denotes the non-central chi-square
  distribution with \(r\) degrees of freedom and non-centrality
  parameter \(\delta\).
\end{enumerate}

\end{theorem}

Part b) provides an approximation of the power function. For a true
value \(\boldsymbol{\beta}^{\ast}\neq\boldsymbol{0}_{r}\) and a sample
of size \(n\), taking
\(\boldsymbol{d}=n^{1/2}\boldsymbol{\beta}^{\ast}\), the power of the
three tests with significance level \(\alpha\) is approximately
\begin{equation}\protect\phantomsection\label{eq-power}{
P\left(\chi_{r}^{2}\left(\tfrac{1}{\varphi(\lambda)}\boldsymbol{\beta}^{\ast\top}\boldsymbol{S}_{22\cdot 1}\boldsymbol{\beta}^{\ast}\right)>\chi_{r,\alpha}^{2}\right),
}\end{equation} \(\chi_{r,\alpha}^{2}\) being the quantile of order
\(1-\alpha\) of the \(\chi_{r}^{2}\) distribution. For \(\lambda=0\) the
non-centrality parameter is the one of the classical tests, and the
price of the robustness in terms of power is completely described by the
factor \(1/\varphi(\lambda)\). For example, for two populations with
\(m=100\) observations each and \(\theta_{2}/\theta_{1}=1.5\),
(\ref{eq-power}) gives \(0.818\) for \(\lambda=0\), \(0.783\) for
\(\lambda=0.3\) and \(0.742\) for \(\lambda=0.5\), and the proportions
of rejections of the Wald-type test obtained by simulation in
Section~\ref{sec-sim-c} are \(0.820\), \(0.783\) and \(0.742\).

The three statistics are computed without estimating any matrix, the
only quantities which depend on \(\lambda\) being known constants. This
is a consequence of the proportionality of the matrices
\(\boldsymbol{J}_{\lambda}(\boldsymbol{\beta})\) and
\(\boldsymbol{K}_{\lambda}(\boldsymbol{\beta})\) of the Appendix for
\(c=0\). For \(c\neq 0\) the Wald-type and Rao-type statistics require
plug-in estimators of \(\boldsymbol{G}_{c\lambda}(\boldsymbol{\theta})\)
and \(\boldsymbol{G}_{2c\lambda}(\boldsymbol{\theta})\), and, unless the
design is a one-way layout, the asymptotic distribution under the null
hypothesis of the divergence-based statistic is a weighted chi-square
distribution whose weights depend on the unknown parameters. As
\(\lambda\rightarrow 0^{+}\) the three statistics tend to the classical
Wald, Rao and likelihood ratio test statistics of the log-linear
exponential regression model.

\subsection{Invariance}\label{sec-invariance}

Let
\(\boldsymbol{H}=\mathbb{X}(\mathbb{X}^{\top}\mathbb{X})^{-1}\mathbb{X}^{\top}\)
and
\(\boldsymbol{H}_{\ast}=\mathbb{X}_{\ast}(\mathbb{X}_{\ast}^{\top}\mathbb{X}_{\ast})^{-1}\mathbb{X}_{\ast}^{\top}\)
be the Hat matrices of the model and of the model under the null
hypothesis,
\(\widehat{\boldsymbol{\eta}}_{\lambda}=\mathbb{X}\widehat{\boldsymbol{\beta}}_{\lambda}\)
and
\(\widetilde{\boldsymbol{\eta}}_{\lambda}=\mathbb{X}\widetilde{\boldsymbol{\beta}}_{\lambda}\)
the vectors of fitted linear predictors, and
\(\widetilde{\boldsymbol{\psi}}_{\lambda}=(\psi_{\lambda}(t_{1}e^{-\widetilde{\eta}_{1,\lambda}}),\ldots,\psi_{\lambda}(t_{n}e^{-\widetilde{\eta}_{n,\lambda}}))^{\top}\),
with \(\psi_{\lambda}(w)=(w-1)w^{\lambda}e^{-\lambda w}\).

\begin{proposition}[]\protect\hypertarget{prp-invariance}{}\label{prp-invariance}

Let \(c=0\) and \(\lambda>0\).

\begin{enumerate}
\def\labelenumi{\alph{enumi})}
\item
  The statistics (\ref{eq-wald}) and (\ref{eq-rao}) can be written as
  \begin{equation}\protect\phantomsection\label{eq-hat}{
  W_{n,\lambda}=\frac{1}{\varphi(\lambda)}\widehat{\boldsymbol{\eta}}_{\lambda}^{\top}(\boldsymbol{H}-\boldsymbol{H}_{\ast})\widehat{\boldsymbol{\eta}}_{\lambda},\qquad
  R_{n,\lambda}=\frac{1}{j_{0}(2\lambda)}\widetilde{\boldsymbol{\psi}}_{\lambda}^{\top}(\boldsymbol{H}-\boldsymbol{H}_{\ast})\widetilde{\boldsymbol{\psi}}_{\lambda},
  }\end{equation} and \(S_{n,\lambda}\) depends on the data only through
  \(t_{i}e^{-\widehat{\eta}_{i,\lambda}}\) and
  \(t_{i}e^{-\widetilde{\eta}_{i,\lambda}}\), \(i=1,\ldots,n\).
\item
  The three statistics are invariant to the reparametrizations of the
  model which keep the column spaces of \(\mathbb{X}\) and
  \(\mathbb{X}_{\ast}\), in particular to the choice of the baseline
  population in a one-way layout and to linear changes of scale of the
  explanatory variables.
\item
  If the column space of \(\mathbb{X}_{\ast}\) contains the vector
  \(\boldsymbol{1}_{n}\), the three statistics are invariant to a change
  of the units of measurement of the response,
  \(t_{i}\mapsto a\,t_{i}\), \(i=1,\ldots,n\), with \(a>0\).
\end{enumerate}

\end{proposition}

The invariance to the units of measurement also holds for the estimators
and the Wald-type statistics of the other members of the family. If
\(t_{i}\) is replaced by \(a\,t_{i}\) and \(\theta_{i}\) by
\(a\,\theta_{i}\), the left-hand side of (\ref{eq-estimating}) is
multiplied by \(a^{-c\lambda}\), so that only the estimator of the
intercept changes, in \(\log a\), and in the estimator of the
variance-covariance matrix of Theorem~\ref{thm-asymptotic} the factors
\(a^{-c\lambda}\) of
\(\boldsymbol{G}_{c\lambda}(\widehat{\boldsymbol{\theta}})\) and
\(a^{-2c\lambda}\) of
\(\boldsymbol{G}_{2c\lambda}(\widehat{\boldsymbol{\theta}})\) cancel.
Hence the lack of invariance to rescaling which affects the Wald tests
in models with Box-Cox transformations when the exponent is estimated
\citep{dagenais1991, dagenais1992} is not present here, where the
exponent is fixed. In the application of Section~\ref{sec-sar} this
means that the results do not depend on whether the pixels are coded in
\([0,1]\) or in \(\{0,\ldots,255\}\).

On the contrary, the statistics are not invariant to the power
transformation of the response, which is the kind of dependence on an
equivalent formulation of the model described in \citet{dufour2017}. For
one-way layouts, in which by Corollary~\ref{cor-penalty} the asymptotic
variance-covariance matrix of the estimator based on \(Z_{i}^{(c)}\) is
\(\omega_{c}(\lambda)\boldsymbol{G}_{0}^{-1}\), the Wald-type statistic
of the scale \(c\) has under \(H_{1,n}\) the non-centrality parameter
\(\boldsymbol{d}^{\top}\boldsymbol{G}_{22\cdot 1}\boldsymbol{d}/\omega_{c}(\lambda)\).
Therefore the asymptotic relative efficiency, in the sense of Pitman, of
the test based on the scale \(c_{1}\) with respect to the one based on
the scale \(c_{2}\), for the same value of \(\lambda\), is
\begin{equation}\protect\phantomsection\label{eq-pitman}{
\frac{\omega_{c_{2}}(\lambda)}{\omega_{c_{1}}(\lambda)},
}\end{equation} and by Proposition~\ref{prp-small}, for small values of
\(\lambda\) it favours the logarithmic scale with respect to the
Rayleigh and exponential ones. For \(\lambda=0.5\), the efficiency of
the test based on the type I extreme value model is \(1.05\) with
respect to the Rayleigh model and \(1.21\) with respect to the
exponential model (Table~\ref{tbl-omega}).

\subsection{Comparison of populations}\label{sec-tests-oneway}

Let the sample be divided into \(I\) populations with sizes
\(n_{1},\ldots,n_{I}\), \(n=\sum_{j=1}^{I}n_{j}\), and let
\(\eta_{j}=\log\theta_{j}\) be the parameter of the \(j\)-th population.
The estimating equation (\ref{eq-estimatingEV}) is separated into one
equation for each population, the minimum DPD estimator
\(\widehat{\eta}_{j,\lambda}\) being the solution of
\(\sum_{i\in I_{j}}\psi_{\lambda}(t_{i}e^{-\eta})=0\), with
\(\psi_{\lambda}(w)=(w-1)w^{\lambda}e^{-\lambda w}\), and under
\(H_{0}:\eta_{1}=\cdots=\eta_{I}\) the restricted estimator
\(\widetilde{\eta}_{\lambda}\) is the solution of the same equation for
the pooled sample. Then, with
\(\overline{\eta}_{\lambda}=\frac{1}{n}\sum_{j=1}^{I}n_{j}\widehat{\eta}_{j,\lambda}\),
\begin{equation}\protect\phantomsection\label{eq-oneway}{
W_{n,\lambda}=\frac{1}{\varphi(\lambda)}\sum_{j=1}^{I}n_{j}\left(\widehat{\eta}_{j,\lambda}-\overline{\eta}_{\lambda}\right)^{2},\qquad
R_{n,\lambda}=\frac{1}{j_{0}(2\lambda)}\sum_{j=1}^{I}\frac{1}{n_{j}}\left(\sum_{i\in I_{j}}\psi_{\lambda}(t_{i}e^{-\widetilde{\eta}_{\lambda}})\right)^{2},
}\end{equation} and, with
\(\varpi_{\lambda}(w)=w^{\lambda}e^{-\lambda w}\), the divergence-based
statistic is \begin{equation}\protect\phantomsection\label{eq-onewayS}{
S_{n,\lambda}=\frac{2\,j_{0}(\lambda)}{\lambda\,j_{0}(2\lambda)}\sum_{j=1}^{I}\sum_{i\in I_{j}}\left[\varpi_{\lambda}(t_{i}e^{-\widehat{\eta}_{j,\lambda}})-\varpi_{\lambda}(t_{i}e^{-\widetilde{\eta}_{\lambda}})\right].
}\end{equation} The asymptotic distribution of the three statistics
under the null hypothesis is \(\chi_{I-1}^{2}\). As
\(\lambda\rightarrow 0^{+}\), (\ref{eq-onewayS}) tends to the classical
likelihood ratio test statistic,
\(2\sum_{j=1}^{I}n_{j}(\widetilde{\eta}_{0}-\widehat{\eta}_{j,0})\),
where \(\widehat{\eta}_{j,0}\) and \(\widetilde{\eta}_{0}\) are the
logarithms of the sample means of the \(j\)-th population and of the
pooled sample. For two populations,
\begin{equation}\protect\phantomsection\label{eq-two}{
W_{n,\lambda}=\frac{(\widehat{\eta}_{1,\lambda}-\widehat{\eta}_{2,\lambda})^{2}}{\varphi(\lambda)\left(\frac{1}{n_{1}}+\frac{1}{n_{2}}\right)},
}\end{equation} and, when \(n_{1}=n_{2}=m\), the statistic for a
one-sided or two-sided comparison is
\(\sqrt{m}\,(\widehat{\eta}_{1,\lambda}-\widehat{\eta}_{2,\lambda})/\sqrt{2\varphi(\lambda)}\),
whose asymptotic distribution under the null hypothesis is standard
normal.

\section{Simulation study}\label{sec-simulation}

Four experiments are presented. In all of them the minimum DPD
estimators are computed with the algorithm (\ref{eq-algorithm}).

\subsection{Validation of the asymptotic variances with a continuous
covariate}\label{sec-sim-a}

The first experiment considers \(n=200\),
\(\eta_{i}=\beta_{0}+\beta_{1}x_{i}\) with
\((\beta_{0},\beta_{1})=(0,2)\) and \(x_{i}=(i-\tfrac12)/n\), so that
\(\max_{i}\theta_{i}/\min_{i}\theta_{i}\simeq 7.3\), without
contamination, and \(20000\) replications, the same samples being used
for all the estimators. Table~\ref{tbl-simA} compares \(n\) times the
sample variance of the estimator of \(\beta_{1}\) with the value given
by Theorem~\ref{thm-asymptotic}. Since the sample variances of the
different estimators are computed with the same samples, their ratios
with respect to the sample variance of the MLE are more precise than the
variances themselves, and they are compared with the theoretical ratio
and with \(\omega_{c}(\lambda)\), the ratio which would be obtained
ignoring the matrix factor. For the MLE, \(n\) times the sample variance
is \(12.35\) and the theoretical value \(12.00\).

\begin{longtable}[]{@{}
  >{\centering\arraybackslash}p{(\linewidth - 12\tabcolsep) * \real{0.1429}}
  >{\centering\arraybackslash}p{(\linewidth - 12\tabcolsep) * \real{0.1429}}
  >{\centering\arraybackslash}p{(\linewidth - 12\tabcolsep) * \real{0.1429}}
  >{\centering\arraybackslash}p{(\linewidth - 12\tabcolsep) * \real{0.1429}}
  >{\centering\arraybackslash}p{(\linewidth - 12\tabcolsep) * \real{0.1429}}
  >{\centering\arraybackslash}p{(\linewidth - 12\tabcolsep) * \real{0.1429}}
  >{\centering\arraybackslash}p{(\linewidth - 12\tabcolsep) * \real{0.1429}}@{}}
\caption{Values of \(n\) times the variance of the estimator of the
slope and ratios with respect to the MLE,
\(n=200\).}\label{tbl-simA}\tabularnewline
\toprule\noalign{}
\begin{minipage}[b]{\linewidth}\centering
\(\lambda\)
\end{minipage} & \begin{minipage}[b]{\linewidth}\centering
\(c\)
\end{minipage} & \begin{minipage}[b]{\linewidth}\centering
Empirical
\end{minipage} & \begin{minipage}[b]{\linewidth}\centering
Theorem~\ref{thm-asymptotic}
\end{minipage} & \begin{minipage}[b]{\linewidth}\centering
Empirical ratio
\end{minipage} & \begin{minipage}[b]{\linewidth}\centering
Theoretical ratio
\end{minipage} & \begin{minipage}[b]{\linewidth}\centering
\(\omega_{c}(\lambda)\)
\end{minipage} \\
\midrule\noalign{}
\endfirsthead
\toprule\noalign{}
\begin{minipage}[b]{\linewidth}\centering
\(\lambda\)
\end{minipage} & \begin{minipage}[b]{\linewidth}\centering
\(c\)
\end{minipage} & \begin{minipage}[b]{\linewidth}\centering
Empirical
\end{minipage} & \begin{minipage}[b]{\linewidth}\centering
Theorem~\ref{thm-asymptotic}
\end{minipage} & \begin{minipage}[b]{\linewidth}\centering
Empirical ratio
\end{minipage} & \begin{minipage}[b]{\linewidth}\centering
Theoretical ratio
\end{minipage} & \begin{minipage}[b]{\linewidth}\centering
\(\omega_{c}(\lambda)\)
\end{minipage} \\
\midrule\noalign{}
\endhead
\bottomrule\noalign{}
\endlastfoot
\(0.25\) & \(0\) & \(13.13\) & \(12.82\) & \(1.064\) & \(1.068\) &
\(1.068\) \\
\(0.25\) & \(\tfrac12\) & \(13.49\) & \(13.19\) & \(1.093\) & \(1.099\)
& \(1.095\) \\
\(0.25\) & \(1\) & \(14.53\) & \(14.21\) & \(1.177\) & \(1.185\) &
\(1.165\) \\
\(0.50\) & \(0\) & \(14.87\) & \(14.50\) & \(1.204\) & \(1.209\) &
\(1.209\) \\
\(0.50\) & \(\tfrac12\) & \(15.86\) & \(15.49\) & \(1.285\) & \(1.291\)
& \(1.270\) \\
\(0.50\) & \(1\) & \(19.27\) & \(18.74\) & \(1.560\) & \(1.562\) &
\(1.462\) \\
\end{longtable}

The sample variances are between \(2\%\) and \(3\%\) larger than the
asymptotic ones, for the MLE as well as for the minimum DPD estimators,
which is the effect of the finite sample size. The empirical ratios
agree with the theoretical ones up to the third significant figure, and
for \(c\neq 0\) they are clearly different from \(\omega_{c}(\lambda)\).
For the exponential model with \(\lambda=0.5\) the loss of efficiency
with respect to the MLE is \(56\%\), of which \(46\%\) corresponds to
the scalar factor and the rest to the heterogeneity of the sample, as
established in Theorem~\ref{thm-penalty}.

\subsection{Contamination at equal efficiency}\label{sec-sim-b}

The second experiment considers one population with \(\theta=1\),
\(n=100\) and \(5000\) replications. The three models are compared with
the values of \(\lambda\) which provide \(\omega_{c}(\lambda)=1.2\)
(Table~\ref{tbl-ges}), so that their variances without contamination are
the same. A proportion \(\varepsilon\) of the observations is replaced
by observations from \(\mathcal{E}(10)\) (moderate outliers),
\(\mathcal{E}(50)\) (gross outliers) or \(\mathcal{E}(0.02)\) (inliers).
Table~\ref{tbl-simB} shows the bias and the root mean square error
(RMSE) of the estimators of \(\eta=0\).

\begin{longtable}[]{@{}
  >{\raggedright\arraybackslash}p{(\linewidth - 10\tabcolsep) * \real{0.1667}}
  >{\centering\arraybackslash}p{(\linewidth - 10\tabcolsep) * \real{0.1667}}
  >{\centering\arraybackslash}p{(\linewidth - 10\tabcolsep) * \real{0.1667}}
  >{\centering\arraybackslash}p{(\linewidth - 10\tabcolsep) * \real{0.1667}}
  >{\centering\arraybackslash}p{(\linewidth - 10\tabcolsep) * \real{0.1667}}
  >{\centering\arraybackslash}p{(\linewidth - 10\tabcolsep) * \real{0.1667}}@{}}
\caption{Bias and RMSE (in brackets) of the estimators of \(\eta\) with
\(\omega_{c}(\lambda)=1.2\), \(n=100\).}\label{tbl-simB}\tabularnewline
\toprule\noalign{}
\begin{minipage}[b]{\linewidth}\raggedright
Contamination
\end{minipage} & \begin{minipage}[b]{\linewidth}\centering
\(\varepsilon\)
\end{minipage} & \begin{minipage}[b]{\linewidth}\centering
MLE
\end{minipage} & \begin{minipage}[b]{\linewidth}\centering
\(c=1\)
\end{minipage} & \begin{minipage}[b]{\linewidth}\centering
\(c=\tfrac12\)
\end{minipage} & \begin{minipage}[b]{\linewidth}\centering
\(c=0\)
\end{minipage} \\
\midrule\noalign{}
\endfirsthead
\toprule\noalign{}
\begin{minipage}[b]{\linewidth}\raggedright
Contamination
\end{minipage} & \begin{minipage}[b]{\linewidth}\centering
\(\varepsilon\)
\end{minipage} & \begin{minipage}[b]{\linewidth}\centering
MLE
\end{minipage} & \begin{minipage}[b]{\linewidth}\centering
\(c=1\)
\end{minipage} & \begin{minipage}[b]{\linewidth}\centering
\(c=\tfrac12\)
\end{minipage} & \begin{minipage}[b]{\linewidth}\centering
\(c=0\)
\end{minipage} \\
\midrule\noalign{}
\endhead
\bottomrule\noalign{}
\endlastfoot
None & \(0\) & \(-0.004\) \((0.101)\) & \(-0.002\) \((0.112)\) &
\(-0.003\) \((0.112)\) & \(-0.004\) \((0.112)\) \\
\(\mathcal{E}(10)\) & \(0.05\) & \(0.346\) \((0.405)\) & \(0.058\)
\((0.131)\) & \(0.048\) \((0.127)\) & \(0.044\) \((0.127)\) \\
\(\mathcal{E}(10)\) & \(0.10\) & \(0.616\) \((0.655)\) & \(0.131\)
\((0.185)\) & \(0.110\) \((0.169)\) & \(0.105\) \((0.169)\) \\
\(\mathcal{E}(50)\) & \(0.05\) & \(1.138\) \((1.228)\) & \(0.030\)
\((0.119)\) & \(0.021\) \((0.116)\) & \(0.011\) \((0.116)\) \\
\(\mathcal{E}(50)\) & \(0.10\) & \(1.703\) \((1.744)\) & \(0.068\)
\((0.140)\) & \(0.048\) \((0.130)\) & \(0.025\) \((0.124)\) \\
\(\mathcal{E}(0.02)\) & \(0.05\) & \(-0.052\) \((0.119)\) & \(-0.083\)
\((0.147)\) & \(-0.044\) \((0.126)\) & \(-0.021\) \((0.118)\) \\
\(\mathcal{E}(0.02)\) & \(0.10\) & \(-0.110\) \((0.155)\) & \(-0.177\)
\((0.221)\) & \(-0.099\) \((0.159)\) & \(-0.050\) \((0.129)\) \\
\end{longtable}

Without contamination the three robust estimators have the same RMSE, as
expected. Under contamination the type I extreme value model has the
smallest bias in all the scenarios, and its RMSE is the smallest one, or
equal to the one of the Rayleigh model for moderate outliers. The
advantage increases with the size of the outliers, in agreement with the
redescending shape of its influence function, and it is especially
important for inliers, for which the exponential model is worse than the
MLE. The larger gross error sensitivity of the type I extreme value
model at equal efficiency (Table~\ref{tbl-ges}) is not reflected in a
worse behaviour in any of these scenarios.

\subsection{Tests for the comparison of two
populations}\label{sec-sim-c}

The third experiment studies the statistics
(\ref{eq-wald})-(\ref{eq-lq}) for two populations with
\(n_{1}=n_{2}=m\), nominal size \(0.05\) and \(10000\) replications,
together with the classical likelihood ratio test (LRT). Under the null
hypothesis \(\theta_{1}=\theta_{2}=1\), and under the alternative
hypothesis \(\theta_{1}=1\), \(\theta_{2}=1.5\). In the contaminated
scenarios \(5\%\) of the observations of the second population are
replaced by observations from \(\mathcal{E}(10)\) or
\(\mathcal{E}(0.02)\). Table~\ref{tbl-simC} shows the proportion of
rejections.

\begin{longtable}[]{@{}
  >{\centering\arraybackslash}p{(\linewidth - 16\tabcolsep) * \real{0.1111}}
  >{\raggedright\arraybackslash}p{(\linewidth - 16\tabcolsep) * \real{0.1111}}
  >{\centering\arraybackslash}p{(\linewidth - 16\tabcolsep) * \real{0.1111}}
  >{\centering\arraybackslash}p{(\linewidth - 16\tabcolsep) * \real{0.1111}}
  >{\centering\arraybackslash}p{(\linewidth - 16\tabcolsep) * \real{0.1111}}
  >{\centering\arraybackslash}p{(\linewidth - 16\tabcolsep) * \real{0.1111}}
  >{\centering\arraybackslash}p{(\linewidth - 16\tabcolsep) * \real{0.1111}}
  >{\centering\arraybackslash}p{(\linewidth - 16\tabcolsep) * \real{0.1111}}
  >{\centering\arraybackslash}p{(\linewidth - 16\tabcolsep) * \real{0.1111}}@{}}
\caption{Proportion of rejections at nominal size \(0.05\) for the type
I extreme value model, with
\(\lambda\in\{0.3,0.5\}\).}\label{tbl-simC}\tabularnewline
\toprule\noalign{}
\begin{minipage}[b]{\linewidth}\centering
\(m\)
\end{minipage} & \begin{minipage}[b]{\linewidth}\raggedright
Scenario
\end{minipage} & \begin{minipage}[b]{\linewidth}\centering
LRT
\end{minipage} & \begin{minipage}[b]{\linewidth}\centering
\(W_{0.3}\)
\end{minipage} & \begin{minipage}[b]{\linewidth}\centering
\(R_{0.3}\)
\end{minipage} & \begin{minipage}[b]{\linewidth}\centering
\(S_{0.3}\)
\end{minipage} & \begin{minipage}[b]{\linewidth}\centering
\(W_{0.5}\)
\end{minipage} & \begin{minipage}[b]{\linewidth}\centering
\(R_{0.5}\)
\end{minipage} & \begin{minipage}[b]{\linewidth}\centering
\(S_{0.5}\)
\end{minipage} \\
\midrule\noalign{}
\endfirsthead
\toprule\noalign{}
\begin{minipage}[b]{\linewidth}\centering
\(m\)
\end{minipage} & \begin{minipage}[b]{\linewidth}\raggedright
Scenario
\end{minipage} & \begin{minipage}[b]{\linewidth}\centering
LRT
\end{minipage} & \begin{minipage}[b]{\linewidth}\centering
\(W_{0.3}\)
\end{minipage} & \begin{minipage}[b]{\linewidth}\centering
\(R_{0.3}\)
\end{minipage} & \begin{minipage}[b]{\linewidth}\centering
\(S_{0.3}\)
\end{minipage} & \begin{minipage}[b]{\linewidth}\centering
\(W_{0.5}\)
\end{minipage} & \begin{minipage}[b]{\linewidth}\centering
\(R_{0.5}\)
\end{minipage} & \begin{minipage}[b]{\linewidth}\centering
\(S_{0.5}\)
\end{minipage} \\
\midrule\noalign{}
\endhead
\bottomrule\noalign{}
\endlastfoot
\(50\) & \(H_{0}\) & \(0.048\) & \(0.047\) & \(0.046\) & \(0.046\) &
\(0.048\) & \(0.044\) & \(0.046\) \\
\(50\) & \(H_{0}\), outliers \(\mathcal{E}(10)\) & \(0.423\) & \(0.085\)
& \(0.065\) & \(0.075\) & \(0.071\) & \(0.055\) & \(0.062\) \\
\(50\) & \(H_{1}\) & \(0.524\) & \(0.492\) & \(0.485\) & \(0.491\) &
\(0.453\) & \(0.442\) & \(0.453\) \\
\(50\) & \(H_{1}\), inliers \(\mathcal{E}(0.02)\) & \(0.427\) &
\(0.448\) & \(0.427\) & \(0.440\) & \(0.432\) & \(0.402\) & \(0.418\) \\
\(100\) & \(H_{0}\) & \(0.048\) & \(0.048\) & \(0.048\) & \(0.048\) &
\(0.050\) & \(0.048\) & \(0.048\) \\
\(100\) & \(H_{0}\), outliers \(\mathcal{E}(10)\) & \(0.616\) &
\(0.098\) & \(0.075\) & \(0.087\) & \(0.070\) & \(0.056\) & \(0.063\) \\
\(100\) & \(H_{1}\) & \(0.820\) & \(0.783\) & \(0.780\) & \(0.783\) &
\(0.742\) & \(0.738\) & \(0.742\) \\
\(100\) & \(H_{1}\), inliers \(\mathcal{E}(0.02)\) & \(0.703\) &
\(0.720\) & \(0.704\) & \(0.714\) & \(0.699\) & \(0.677\) & \(0.690\) \\
\end{longtable}

Without contamination the three robust statistics have a size close to
the nominal one, with a moderate loss of power with respect to the LRT.
With \(5\%\) of outliers in one of the populations the LRT rejects a
true null hypothesis in more than \(40\%\) of the samples, while the
size of the robust statistics remains close to the nominal one, the
Rao-type statistic being the most stable.

\subsection{Comparison with an alternative robust
estimator}\label{sec-sim-d}

The last experiment compares the minimum DPD estimators, with the values
of \(\lambda\) of Section~\ref{sec-sim-b}, with the robust estimator
which has been used for Rayleigh regression with SAR data, the weighted
maximum likelihood estimator (WMLE) of \citet{field1994} as applied in
\citet{palm2021}. Let \(F_{i}=1-e^{-w_{i}}\) be the value of the
distribution function of the model at the \(i\)-th observation. The WMLE
solves
\(\sum_{i=1}^{n}\boldsymbol{x}_{i}\,\varpi_{\delta}(\widehat{F}_{i})\left(t_{i}/\theta_{i}-1\right)=\boldsymbol{0}_{k+1}\)
with \begin{equation}\protect\phantomsection\label{eq-wmle}{
\varpi_{\delta}(u)=\begin{cases}u/\delta, & u<\delta,\\ 1, & \delta\leq u\leq 1-\delta,\\ (1-u)/\delta, & u>1-\delta,\end{cases}
}\end{equation} where \(\delta\in(0,\tfrac12)\) is a small constant and
\(\widehat{F}_{i}\) is evaluated at the MLE. Since the weights depend on
the observations only through \(F_{i}\), the WMLE is equivariant under
increasing transformations of the response. The estimating equation
(\ref{eq-estimatingEV}) of the minimum DPD estimator for \(c=0\) has the
same form, with the weight \[
\varpi_{i}=w_{i}^{\lambda}e^{-\lambda w_{i}}=\left[-(1-F_{i})\log(1-F_{i})\right]^{\lambda},
\] and three differences. The weight is a smooth function which modifies
all the observations and not only a proportion \(2\delta\) of them, it
is evaluated at the estimator itself and not at the MLE, and the
estimating equation is unbiased at the model for every \(\lambda\),
which is not the case for the one associated with (\ref{eq-wmle}).

The values \(\delta\in\{0.01,0.001\}\) are considered, in the setting of
Section~\ref{sec-sim-b}. Two contaminations have been added to the ones
considered there, the contamination at zero of Section~\ref{sec-zeros}
and the one used in \citet{frery2022} to illustrate the lack of
robustness of the MLE with SAR data, in which an observation is
multiplied by \(100\) with a very small probability.

\begin{longtable}[]{@{}
  >{\raggedright\arraybackslash}p{(\linewidth - 14\tabcolsep) * \real{0.1481}}
  >{\centering\arraybackslash}p{(\linewidth - 14\tabcolsep) * \real{0.0741}}
  >{\centering\arraybackslash}p{(\linewidth - 14\tabcolsep) * \real{0.1296}}
  >{\centering\arraybackslash}p{(\linewidth - 14\tabcolsep) * \real{0.1296}}
  >{\centering\arraybackslash}p{(\linewidth - 14\tabcolsep) * \real{0.1296}}
  >{\centering\arraybackslash}p{(\linewidth - 14\tabcolsep) * \real{0.1296}}
  >{\centering\arraybackslash}p{(\linewidth - 14\tabcolsep) * \real{0.1296}}
  >{\centering\arraybackslash}p{(\linewidth - 14\tabcolsep) * \real{0.1296}}@{}}
\caption{Bias and RMSE (in brackets) of the estimators of \(\eta\),
\(n=100\), \(5000\) replications. The subindex of WMLE is the value of
\(\delta\). The minimum DPD estimators have
\(\omega_{c}(\lambda)=1.2\).}\label{tbl-simD}\tabularnewline
\toprule\noalign{}
\begin{minipage}[b]{\linewidth}\raggedright
Contam.
\end{minipage} & \begin{minipage}[b]{\linewidth}\centering
\(\varepsilon\)
\end{minipage} & \begin{minipage}[b]{\linewidth}\centering
MLE
\end{minipage} & \begin{minipage}[b]{\linewidth}\centering
WMLE\(_{0.01}\)
\end{minipage} & \begin{minipage}[b]{\linewidth}\centering
WMLE\(_{0.001}\)
\end{minipage} & \begin{minipage}[b]{\linewidth}\centering
\(c=1\)
\end{minipage} & \begin{minipage}[b]{\linewidth}\centering
\(c=\tfrac12\)
\end{minipage} & \begin{minipage}[b]{\linewidth}\centering
\(c=0\)
\end{minipage} \\
\midrule\noalign{}
\endfirsthead
\toprule\noalign{}
\begin{minipage}[b]{\linewidth}\raggedright
Contam.
\end{minipage} & \begin{minipage}[b]{\linewidth}\centering
\(\varepsilon\)
\end{minipage} & \begin{minipage}[b]{\linewidth}\centering
MLE
\end{minipage} & \begin{minipage}[b]{\linewidth}\centering
WMLE\(_{0.01}\)
\end{minipage} & \begin{minipage}[b]{\linewidth}\centering
WMLE\(_{0.001}\)
\end{minipage} & \begin{minipage}[b]{\linewidth}\centering
\(c=1\)
\end{minipage} & \begin{minipage}[b]{\linewidth}\centering
\(c=\tfrac12\)
\end{minipage} & \begin{minipage}[b]{\linewidth}\centering
\(c=0\)
\end{minipage} \\
\midrule\noalign{}
\endhead
\bottomrule\noalign{}
\endlastfoot
None & \(0\) & \(-0.006\) \((0.101)\) & \(-0.025\) \((0.108)\) &
\(-0.008\) \((0.102)\) & \(-0.003\) \((0.111)\) & \(-0.004\) \((0.111)\)
& \(-0.005\) \((0.111)\) \\
\(\mathcal{E}(10)\) & \(0.05\) & \(0.349\) \((0.406)\) & \(0.069\)
\((0.140)\) & \(0.118\) \((0.183)\) & \(0.058\) \((0.132)\) & \(0.048\)
\((0.128)\) & \(0.045\) \((0.128)\) \\
\(\mathcal{E}(10)\) & \(0.10\) & \(0.620\) \((0.658)\) & \(0.221\)
\((0.270)\) & \(0.324\) \((0.373)\) & \(0.136\) \((0.187)\) & \(0.115\)
\((0.172)\) & \(0.110\) \((0.171)\) \\
\(\mathcal{E}(50)\) & \(0.05\) & \(1.142\) \((1.233)\) & \(0.154\)
\((0.242)\) & \(0.235\) \((0.343)\) & \(0.031\) \((0.119)\) & \(0.022\)
\((0.117)\) & \(0.012\) \((0.116)\) \\
\(\mathcal{E}(50)\) & \(0.10\) & \(1.708\) \((1.750)\) & \(0.548\)
\((0.644)\) & \(0.756\) \((0.857)\) & \(0.071\) \((0.140)\) & \(0.051\)
\((0.129)\) & \(0.028\) \((0.123)\) \\
\(\mathcal{E}(0.02)\) & \(0.05\) & \(-0.053\) \((0.119)\) & \(-0.068\)
\((0.132)\) & \(-0.056\) \((0.122)\) & \(-0.083\) \((0.148)\) &
\(-0.045\) \((0.127)\) & \(-0.022\) \((0.119)\) \\
\(\mathcal{E}(0.02)\) & \(0.10\) & \(-0.108\) \((0.155)\) & \(-0.122\)
\((0.170)\) & \(-0.112\) \((0.159)\) & \(-0.176\) \((0.220)\) &
\(-0.097\) \((0.157)\) & \(-0.048\) \((0.127)\) \\
\(\delta_{0}\) & \(0.05\) & \(-0.058\) \((0.120)\) & \(-0.031\)
\((0.112)\) & \(-0.010\) \((0.105)\) & \(-0.088\) \((0.149)\) &
\(0.009\) \((0.113)\) & \(-0.005\) \((0.113)\) \\
\(\delta_{0}\) & \(0.20\) & \(-0.233\) \((0.263)\) & \(-0.061\)
\((0.135)\) & \(-0.019\) \((0.116)\) & \(-0.413\) \((0.448)\) &
\(0.053\) \((0.133)\) & \(-0.006\) \((0.122)\) \\
\(\mathcal{E}(100)\) & \(0.001\) & \(0.044\) \((0.231)\) & \(-0.025\)
\((0.107)\) & \(-0.009\) \((0.101)\) & \(-0.005\) \((0.110)\) &
\(-0.006\) \((0.110)\) & \(-0.007\) \((0.109)\) \\
\(\mathcal{E}(100)\) & \(0.005\) & \(0.263\) \((0.517)\) & \(-0.010\)
\((0.107)\) & \(0.000\) \((0.104)\) & \(0.000\) \((0.111)\) & \(-0.001\)
\((0.112)\) & \(-0.002\) \((0.112)\) \\
\end{longtable}

The WMLE has a very small loss of efficiency without contamination, with
a bias for \(\delta=0.01\) which is due to its estimating equation, and
it behaves very well in the two situations for which its weights are
designed, a very small proportion of gross outliers and observations
equal to zero. When the proportion of outliers is \(5\%\) or \(10\%\) it
is clearly worse than the minimum DPD estimators, with a bias which for
the contamination \(\mathcal{E}(50)\) is more than twelve times the one
of the type I extreme value model. The reason is that its weights are
computed with the MLE, which is itself affected by the outliers, so that
the contaminated observations are not recognized as such. The minimum
DPD estimators do not have this limitation, at the price of a larger
loss of efficiency without contamination, which is the role of the
tuning parameter \(\lambda\).

\section{Application to ground type detection in SAR
imagery}\label{sec-sar}

The CARABAS-II magnitude image used in this application belongs to the
public data set described in \citet{lundberg2006}, and it was available
as an \(8\)-bit JPEG file. Pixel values were imported in \texttt{R}
using the \texttt{imager} package, which represents the JPEG digital
levels on the interval \([0,1]\). This representation is equivalent, up
to the common multiplicative factor \(255\), to the conventional
\(8\)-bit scale \(\{0,\ldots,255\}\). Thus, its use does not constitute
an additional normalization and does not affect comparisons between
regions based on ratios or differences of logarithmic parameter
estimates. For the type I extreme value model this is a consequence of
part c) of Proposition~\ref{prp-invariance} with \(a=255^{2}\), the
factor which multiplies the squared magnitudes. The Wald-type, Rao-type
and divergence-based test statistics take exactly the same values on
both scales, and by part b) they do not depend on the region which is
taken as baseline. The same holds for the statistics with a sandwich
variance used below, since they depend on the data only through
differences of estimates of the linear predictor and through the values
\(t_{i}e^{-\widehat{\eta}_{i,\lambda}}\), which do not change.

Under the fully developed speckle assumption \citep{oliver2004}, the SAR
return within a resolution cell is the coherent sum of many elementary
scatterers with approximately independent random phases. By the central
limit theorem, its real and imaginary components are modelled as
zero-mean Gaussian variables. Consequently, the magnitude of the return
follows a Rayleigh distribution and its square follows an exponential
distribution. Since the observations arise from a JPEG representation
rather than radiometrically calibrated SAR measurements, the models are
interpreted as working approximations for the selected regions.

\begin{figure}[!htb]

\centering{

\pandocbounded{\includegraphics[keepaspectratio]{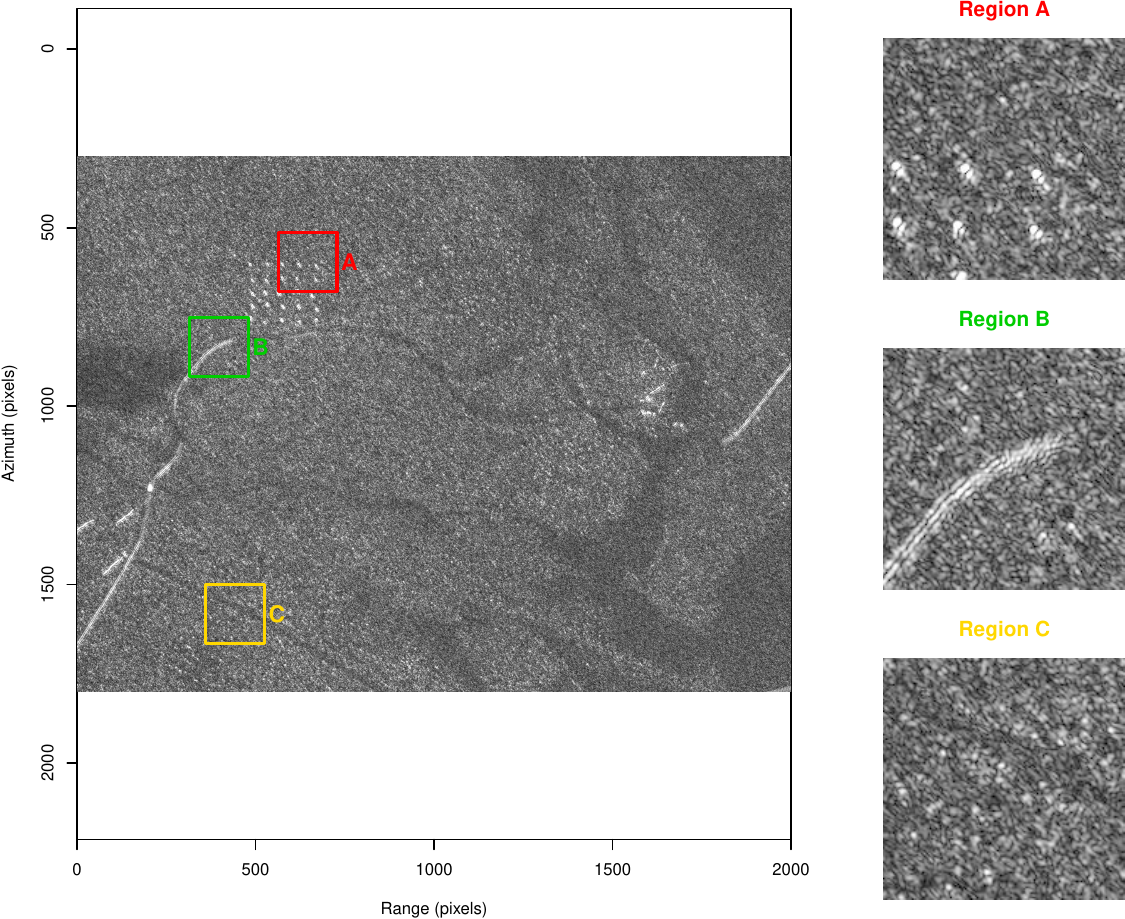}}

}

\caption{\label{fig-sar-map}CARABAS-II magnitude image (azimuth from 300
to 1800) with the regions A (military vehicles), B (power line) and C
(forest background), and the three regions of \(165\times 165\) pixels.
The square root of the magnitude is represented in order to increase the
contrast.}

\end{figure}

The experiment is the ground type detection of \citet{palm2021}, with
the same image. Three regions of interest are compared, A (military
vehicles), B (power lines) and C (forest background), each of them a
square of \(165\times 165\) pixels. In that paper the differences
between the regions are tested with Wald statistics which assume that
the pixels are independent, and the conclusion is that the MLE does not
distinguish the region of the power lines from the one of the vehicles,
while the WMLE does. This is a one-way layout with \(I=3\), for which,
according to Corollary~\ref{cor-penalty}, the three models have the same
matrix factor and the statistics of Section~\ref{sec-tests-oneway}
apply. Regions A and B contain a small proportion of very bright pixels,
the vehicles and the power line respectively, which play the role of
outliers with respect to the background of each region.
Figure~\ref{fig-sar-map} shows their location in the image and the three
regions.

Two features of the data must be taken into account. First, an \(8\)-bit
image contains pixels which are equal to zero (\(61\), \(75\) and \(71\)
in regions A, B and C), for which the logarithm does not exist. Pixels
equal to zero are frequent in magnitude SAR images with a high dynamic
range, and \citet{palm2024} deal with them by adding to the Rayleigh
distribution a point mass at zero, with its own regression structure.
They are not an obstacle for the minimum DPD estimator of the type I
extreme value model, since the estimating equation
(\ref{eq-estimatingEV}) and the statistics of Section~\ref{sec-tests}
are written in terms of \(t_{i}=r_{i}^{2}\), and an observation with
\(t_{i}=0\) has weight zero. Hence the zero-valued pixels do not need to
be replaced or modelled, and they are discarded in the same way as the
very bright ones. Second, neighbouring pixels are strongly correlated,
which is to be expected because the size of a pixel is \(1\) m and the
spatial resolution of the system is \(2.5\) m \citep{palm2021}, so that
the image is oversampled. The sample autocorrelation of \(t_{i}\)
between adjacent pixels is between \(0.74\) and \(0.81\) in the three
regions, and at a distance of \(8\) pixels it is below \(0.1\) in
regions A and C and about \(0.2\) in region B. Hence the assumption of
independence does not hold for the pixels of a region, a feature of SAR
images which has been modelled explicitly in \citet{palm2022arma}. In
addition, the sample value of
\(\operatorname{Var}[T_{i}]/\operatorname{E}^{2}[T_{i}]\), which is
equal to one for the exponential distribution, is \(2.49\), \(1.88\) and
\(1.69\) in regions A, B and C, so that the working model does not fit
exactly even in the forest background. For these reasons the estimators
are computed with all the pixels, as they remain consistent for the
solution of the population estimating equation, but their variances are
not estimated with the expressions of Section~\ref{sec-tests}, which
assume independence and a correct model.

Table~\ref{tbl-sarEst} shows the estimates of \(\eta_{A}-\eta_{B}\) and
\(\eta_{C}-\eta_{B}\) with the \(27225\) pixels of each region. For
\(\lambda=0\) the three models provide the same estimates, the ones of
the MLE.

\begin{longtable}[]{@{}
  >{\centering\arraybackslash}p{(\linewidth - 12\tabcolsep) * \real{0.1429}}
  >{\centering\arraybackslash}p{(\linewidth - 12\tabcolsep) * \real{0.1429}}
  >{\centering\arraybackslash}p{(\linewidth - 12\tabcolsep) * \real{0.1429}}
  >{\centering\arraybackslash}p{(\linewidth - 12\tabcolsep) * \real{0.1429}}
  >{\centering\arraybackslash}p{(\linewidth - 12\tabcolsep) * \real{0.1429}}
  >{\centering\arraybackslash}p{(\linewidth - 12\tabcolsep) * \real{0.1429}}
  >{\centering\arraybackslash}p{(\linewidth - 12\tabcolsep) * \real{0.1429}}@{}}
\caption{Estimates of \(\eta_{A}-\eta_{B}\) (first three columns) and
\(\eta_{C}-\eta_{B}\) (last three columns) with all the
pixels.}\label{tbl-sarEst}\tabularnewline
\toprule\noalign{}
\begin{minipage}[b]{\linewidth}\centering
\(\lambda\)
\end{minipage} & \begin{minipage}[b]{\linewidth}\centering
\(c=0\)
\end{minipage} & \begin{minipage}[b]{\linewidth}\centering
\(c=\tfrac12\)
\end{minipage} & \begin{minipage}[b]{\linewidth}\centering
\(c=1\)
\end{minipage} & \begin{minipage}[b]{\linewidth}\centering
\(c=0\)
\end{minipage} & \begin{minipage}[b]{\linewidth}\centering
\(c=\tfrac12\)
\end{minipage} & \begin{minipage}[b]{\linewidth}\centering
\(c=1\)
\end{minipage} \\
\midrule\noalign{}
\endfirsthead
\toprule\noalign{}
\begin{minipage}[b]{\linewidth}\centering
\(\lambda\)
\end{minipage} & \begin{minipage}[b]{\linewidth}\centering
\(c=0\)
\end{minipage} & \begin{minipage}[b]{\linewidth}\centering
\(c=\tfrac12\)
\end{minipage} & \begin{minipage}[b]{\linewidth}\centering
\(c=1\)
\end{minipage} & \begin{minipage}[b]{\linewidth}\centering
\(c=0\)
\end{minipage} & \begin{minipage}[b]{\linewidth}\centering
\(c=\tfrac12\)
\end{minipage} & \begin{minipage}[b]{\linewidth}\centering
\(c=1\)
\end{minipage} \\
\midrule\noalign{}
\endhead
\bottomrule\noalign{}
\endlastfoot
\(0\) & \(-0.138\) & \(-0.138\) & \(-0.138\) & \(-0.322\) & \(-0.322\) &
\(-0.322\) \\
\(0.1\) & \(-0.181\) & \(-0.181\) & \(-0.180\) & \(-0.310\) & \(-0.306\)
& \(-0.303\) \\
\(0.3\) & \(-0.199\) & \(-0.191\) & \(-0.181\) & \(-0.287\) & \(-0.282\)
& \(-0.278\) \\
\(0.5\) & \(-0.193\) & \(-0.182\) & \(-0.169\) & \(-0.275\) & \(-0.274\)
& \(-0.272\) \\
\(1\) & \(-0.186\) & \(-0.175\) & \(-0.158\) & \(-0.270\) & \(-0.276\) &
\(-0.273\) \\
\end{longtable}

The robust estimates modify the two comparisons in opposite directions.
Region A contains \(117\) saturated pixels, which correspond to the
vehicles, and region B \(35\), which correspond to the power line. When
their influence is reduced, the estimated difference between A and B
increases in absolute value, from \(-0.14\) to values between \(-0.16\)
and \(-0.20\), and the one between C and B decreases, from \(-0.32\) to
about \(-0.27\). That is, part of the difference between the forest and
the region of the power line which is found by the MLE is due to the
power line itself and not to the ground, while the bright pixels of the
vehicles hide part of the difference between the grounds of A and B. For
\(\lambda\geq 0.3\) the estimate of \(\eta_{A}-\eta_{B}\) is more stable
with respect to \(\lambda\) for the type I extreme value model, between
\(-0.199\) and \(-0.186\), than for the exponential model, between
\(-0.181\) and \(-0.158\).

\begin{figure}[!htb]

\centering{

\pandocbounded{\includegraphics[keepaspectratio]{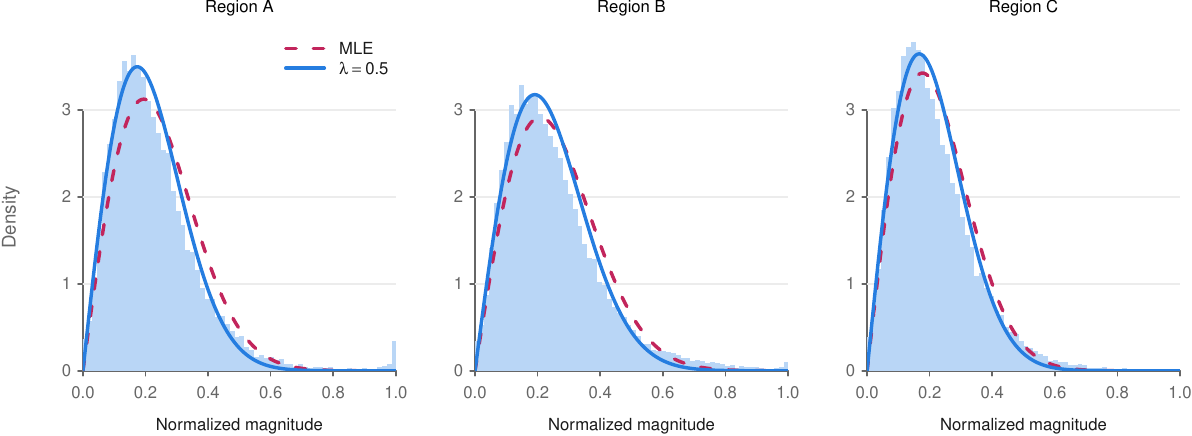}}

}

\caption{\label{fig-sar-fit}Histograms of the magnitudes in the three
regions and Rayleigh density functions fitted with the MLE (red dashed
line) and with the minimum DPD estimator of the type I extreme value
model with \(\lambda=0.5\) (blue solid line). The vertical scale is the
same in the three panels.}

\end{figure}%

\begin{figure}[!htb]

\centering{

\pandocbounded{\includegraphics[keepaspectratio]{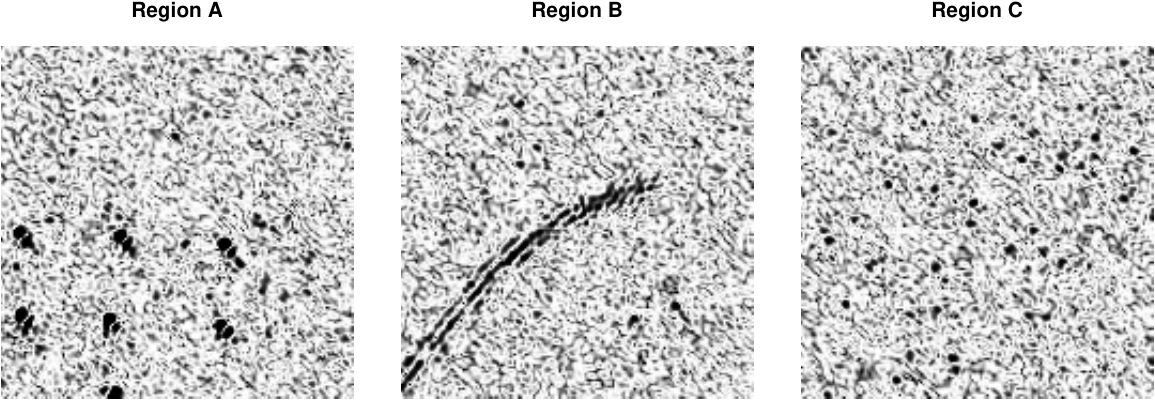}}

}

\caption{\label{fig-sar-weights}Weight
\(\widehat{w}_{i}^{\lambda}e^{-\lambda\widehat{w}_{i}}\) given to each
pixel by the minimum DPD estimator of the type I extreme value model
with \(\lambda=0.5\), relative to its maximum value. White pixels have
the maximum weight and black pixels weight zero.}

\end{figure}%

Figure~\ref{fig-sar-fit} compares the histogram of the magnitudes of
each region with the Rayleigh density functions fitted with the MLE and
with the robust estimator. The MLE is displaced towards the right tail,
and the robust fit follows the bulk of the pixels more closely, although
the right tail of the histograms is heavier than the one of the Rayleigh
distribution in the three regions. Figure~\ref{fig-sar-weights} shows
the weight which the estimating equation (\ref{eq-estimatingEV}) gives
to each pixel. The vehicles in region A and the power line in region B
are recognized as dark areas, that is, they are the pixels which the
robust estimator discards, together with the pixels whose magnitude is
close to zero.

Let \(\widehat{w}_{i}=t_{i}\exp\{-\widehat{\eta}_{j,\lambda}\}\) for the
pixels of the \(j\)-th region and
\(\psi_{\lambda}(w)=(w-1)w^{\lambda}e^{-\lambda w}\). The variance of
\(\widehat{\eta}_{j,\lambda}\) is estimated with the sandwich expression
\begin{equation}\protect\phantomsection\label{eq-sandwich}{
\widehat{\operatorname{Var}}[\widehat{\eta}_{j,\lambda}]=\frac{1}{n_{j}}\frac{\widehat{\sigma}_{j,\lambda}^{2}}{\widehat{J}_{j,\lambda}^{2}},\qquad
\widehat{J}_{j,\lambda}=\frac{1}{n_{j}}\sum_{i\in I_{j}}\left[\widehat{w}_{i}-\lambda(\widehat{w}_{i}-1)^{2}\right]\widehat{w}_{i}^{\lambda}e^{-\lambda\widehat{w}_{i}},
}\end{equation} where \(\widehat{J}_{j,\lambda}\) is the sample mean of
minus the derivative of \(\psi_{\lambda}(t_{i}e^{-\eta})\) with respect
to \(\eta\), and \(\widehat{\sigma}_{j,\lambda}^{2}\) is an estimator of
the long-run variance of the random field
\(\psi_{\lambda}(\widehat{w}_{i})\), the sum of its sample
autocovariances at the spatial lags \((d_{1},d_{2})\) with
\(|d_{1}|,|d_{2}|\leq L\), weighted with the Bartlett kernel
\(\left(1-\tfrac{|d_{1}|}{L+1}\right)\left(1-\tfrac{|d_{2}|}{L+1}\right)\).
This is the heteroskedasticity and autocorrelation consistent estimator
for spatially dependent data of \citet{conley1999}, applied to the
estimating function of the minimum DPD estimator. Under independence and
a correct model, \(\widehat{J}_{j,\lambda}\) and
\(\widehat{\sigma}_{j,\lambda}^{2}\) are consistent estimators of
\(j_{0}(\lambda)\) and \(j_{0}(2\lambda)\), and (\ref{eq-sandwich})
reduces to \(\varphi(\lambda)/n_{j}\). Two regions are compared with the
statistic \begin{equation}\protect\phantomsection\label{eq-Zsandwich}{
Z_{\lambda}=\frac{\widehat{\eta}_{j,\lambda}-\widehat{\eta}_{l,\lambda}}{\left(\widehat{\operatorname{Var}}[\widehat{\eta}_{j,\lambda}]+\widehat{\operatorname{Var}}[\widehat{\eta}_{l,\lambda}]\right)^{1/2}},
}\end{equation} which is the square root of (\ref{eq-two}) with the
estimated variances instead of the ones of the model.

Table~\ref{tbl-sarTest} shows the standard errors and the statistics
(\ref{eq-Zsandwich}) for the type I extreme value model, with \(L=16\).

\begin{longtable}[]{@{}
  >{\centering\arraybackslash}p{(\linewidth - 12\tabcolsep) * \real{0.1429}}
  >{\centering\arraybackslash}p{(\linewidth - 12\tabcolsep) * \real{0.1429}}
  >{\centering\arraybackslash}p{(\linewidth - 12\tabcolsep) * \real{0.1429}}
  >{\centering\arraybackslash}p{(\linewidth - 12\tabcolsep) * \real{0.1429}}
  >{\centering\arraybackslash}p{(\linewidth - 12\tabcolsep) * \real{0.1429}}
  >{\centering\arraybackslash}p{(\linewidth - 12\tabcolsep) * \real{0.1429}}
  >{\centering\arraybackslash}p{(\linewidth - 12\tabcolsep) * \real{0.1429}}@{}}
\caption{Estimated differences, standard errors based on
(\ref{eq-sandwich}) with \(L=16\), and statistics (\ref{eq-Zsandwich}),
type I extreme value model.}\label{tbl-sarTest}\tabularnewline
\toprule\noalign{}
\begin{minipage}[b]{\linewidth}\centering
\(\lambda\)
\end{minipage} & \begin{minipage}[b]{\linewidth}\centering
\(\widehat{\eta}_{A,\lambda}-\widehat{\eta}_{B,\lambda}\)
\end{minipage} & \begin{minipage}[b]{\linewidth}\centering
Standard error
\end{minipage} & \begin{minipage}[b]{\linewidth}\centering
\(Z_{\lambda}\)
\end{minipage} & \begin{minipage}[b]{\linewidth}\centering
\(\widehat{\eta}_{C,\lambda}-\widehat{\eta}_{B,\lambda}\)
\end{minipage} & \begin{minipage}[b]{\linewidth}\centering
Standard error
\end{minipage} & \begin{minipage}[b]{\linewidth}\centering
\(Z_{\lambda}\)
\end{minipage} \\
\midrule\noalign{}
\endfirsthead
\toprule\noalign{}
\begin{minipage}[b]{\linewidth}\centering
\(\lambda\)
\end{minipage} & \begin{minipage}[b]{\linewidth}\centering
\(\widehat{\eta}_{A,\lambda}-\widehat{\eta}_{B,\lambda}\)
\end{minipage} & \begin{minipage}[b]{\linewidth}\centering
Standard error
\end{minipage} & \begin{minipage}[b]{\linewidth}\centering
\(Z_{\lambda}\)
\end{minipage} & \begin{minipage}[b]{\linewidth}\centering
\(\widehat{\eta}_{C,\lambda}-\widehat{\eta}_{B,\lambda}\)
\end{minipage} & \begin{minipage}[b]{\linewidth}\centering
Standard error
\end{minipage} & \begin{minipage}[b]{\linewidth}\centering
\(Z_{\lambda}\)
\end{minipage} \\
\midrule\noalign{}
\endhead
\bottomrule\noalign{}
\endlastfoot
\(0\) & \(-0.138\) & \(0.083\) & \(-1.68\) & \(-0.322\) & \(0.069\) &
\(-4.65\) \\
\(0.1\) & \(-0.181\) & \(0.073\) & \(-2.49\) & \(-0.310\) & \(0.067\) &
\(-4.64\) \\
\(0.3\) & \(-0.199\) & \(0.058\) & \(-3.41\) & \(-0.287\) & \(0.060\) &
\(-4.80\) \\
\(0.5\) & \(-0.193\) & \(0.052\) & \(-3.74\) & \(-0.275\) & \(0.055\) &
\(-5.02\) \\
\(1\) & \(-0.186\) & \(0.045\) & \(-4.08\) & \(-0.270\) & \(0.051\) &
\(-5.33\) \\
\end{longtable}

The standard errors are several times larger than the ones obtained
under independence, which for the MLE are \(0.0086\), and the dependence
is stronger for the MLE than for the robust estimators. In region B the
ratio between \(\widehat{\sigma}_{j,\lambda}^{2}\) and the sample
variance of \(\psi_{\lambda}(\widehat{w}_{i})\) is \(54.5\) for
\(\lambda=0\) and \(14.4\) for \(\lambda=1\), since the pixels of the
power line, which form a long structure, have a bounded contribution for
\(\lambda>0\). As a consequence, in these data the robust estimators
have a smaller standard error than the MLE.

The difference between the forest (C) and the region of the power line
(B) is detected by all the procedures. The conclusion about A and B
depends on the procedure. With the MLE the result is not conclusive,
since the statistic is \(-2.22\), \(-1.68\) and \(-1.49\) for \(L=8\),
\(16\) and \(24\), on both sides of the critical value at the \(5\%\)
significance level. With the robust estimators the difference is
significant for the three values of \(L\), the statistic being
\(-4.50\), \(-3.41\) and \(-2.99\) for \(\lambda=0.3\). The same
conclusion is obtained with the Rayleigh and exponential models, whose
statistics for \(\lambda=0.3\) and \(L=16\) are \(-3.56\) and \(-3.64\).
The pixels of the vehicles, which are not representative of the ground
of region A, reduce the estimated difference and increase its variance,
and both effects hide a difference between the two grounds which the
robust procedure detects.

The Rao-type statistic of Section~\ref{sec-tests-oneway} can be modified
in the same way. Let \(\widetilde{\eta}_{\lambda}\) be the estimator
computed with the pixels of the two regions which are compared,
\(U_{j}=\sum_{i\in I_{j}}\psi_{\lambda}(t_{i}e^{-\widetilde{\eta}_{\lambda}})\),
\(\widetilde{J}_{j}\) and \(\widetilde{\sigma}_{j}^{2}\) the quantities
of (\ref{eq-sandwich}) evaluated at \(\widetilde{\eta}_{\lambda}\)
instead of \(\widehat{\eta}_{j,\lambda}\), and
\(a=n_{j}\widetilde{J}_{j}/(n_{j}\widetilde{J}_{j}+n_{l}\widetilde{J}_{l})\).
Since \(U_{j}+U_{l}=0\), the statistic is
\begin{equation}\protect\phantomsection\label{eq-Zrao}{
Z_{\lambda}^{R}=\frac{(1-a)U_{j}-aU_{l}}{\left((1-a)^{2}n_{j}\widetilde{\sigma}_{j}^{2}+a^{2}n_{l}\widetilde{\sigma}_{l}^{2}\right)^{1/2}},
}\end{equation} which only requires the estimator under the null
hypothesis. The divergence-based statistic is not used with these data,
because under dependence its asymptotic distribution under the null
hypothesis is no longer a chi-square distribution but a weighted
chi-square distribution, while the Wald-type and Rao-type statistics
only need the variance (\ref{eq-sandwich}). For the three regions, the
hypothesis \(\eta_{A}=\eta_{B}=\eta_{C}\) is tested with the Wald-type
statistic
\(\sum_{j}(\widehat{\eta}_{j,\lambda}-\overline{\eta}_{\lambda})^{2}/\widehat{\operatorname{Var}}[\widehat{\eta}_{j,\lambda}]\),
where \(\overline{\eta}_{\lambda}\) is the mean of the three estimates
weighted with the reciprocals of their estimated variances, and with the
corresponding Rao-type statistic, both of them with an asymptotic
\(\chi_{2}^{2}\) distribution under the null hypothesis.
Table~\ref{tbl-sarRao} shows that the Wald-type and the Rao-type
statistics lead to the same conclusions. The global hypothesis is
rejected by all the procedures, since the difference between C and B is
very clear, and the statistics for A and B behave as in
Table~\ref{tbl-sarTest}.

\begin{longtable}[]{@{}
  >{\centering\arraybackslash}p{(\linewidth - 12\tabcolsep) * \real{0.1429}}
  >{\centering\arraybackslash}p{(\linewidth - 12\tabcolsep) * \real{0.1429}}
  >{\centering\arraybackslash}p{(\linewidth - 12\tabcolsep) * \real{0.1429}}
  >{\centering\arraybackslash}p{(\linewidth - 12\tabcolsep) * \real{0.1429}}
  >{\centering\arraybackslash}p{(\linewidth - 12\tabcolsep) * \real{0.1429}}
  >{\centering\arraybackslash}p{(\linewidth - 12\tabcolsep) * \real{0.1429}}
  >{\centering\arraybackslash}p{(\linewidth - 12\tabcolsep) * \real{0.1429}}@{}}
\caption{Wald-type and Rao-type statistics with sandwich variances,
\(L=16\), type I extreme value model. The critical value of the global
statistics at the \(5\%\) significance level is
\(5.99\).}\label{tbl-sarRao}\tabularnewline
\toprule\noalign{}
\begin{minipage}[b]{\linewidth}\centering
\(\lambda\)
\end{minipage} & \begin{minipage}[b]{\linewidth}\centering
Global, Wald
\end{minipage} & \begin{minipage}[b]{\linewidth}\centering
Global, Rao
\end{minipage} & \begin{minipage}[b]{\linewidth}\centering
\(Z_{\lambda}\), A vs.~B
\end{minipage} & \begin{minipage}[b]{\linewidth}\centering
\(Z_{\lambda}^{R}\), A vs.~B
\end{minipage} & \begin{minipage}[b]{\linewidth}\centering
\(Z_{\lambda}\), C vs.~B
\end{minipage} & \begin{minipage}[b]{\linewidth}\centering
\(Z_{\lambda}^{R}\), C vs.~B
\end{minipage} \\
\midrule\noalign{}
\endfirsthead
\toprule\noalign{}
\begin{minipage}[b]{\linewidth}\centering
\(\lambda\)
\end{minipage} & \begin{minipage}[b]{\linewidth}\centering
Global, Wald
\end{minipage} & \begin{minipage}[b]{\linewidth}\centering
Global, Rao
\end{minipage} & \begin{minipage}[b]{\linewidth}\centering
\(Z_{\lambda}\), A vs.~B
\end{minipage} & \begin{minipage}[b]{\linewidth}\centering
\(Z_{\lambda}^{R}\), A vs.~B
\end{minipage} & \begin{minipage}[b]{\linewidth}\centering
\(Z_{\lambda}\), C vs.~B
\end{minipage} & \begin{minipage}[b]{\linewidth}\centering
\(Z_{\lambda}^{R}\), C vs.~B
\end{minipage} \\
\midrule\noalign{}
\endhead
\bottomrule\noalign{}
\endlastfoot
\(0\) & \(24.87\) & \(26.72\) & \(-1.675\) & \(-1.680\) & \(-4.645\) &
\(-4.726\) \\
\(0.1\) & \(22.63\) & \(23.92\) & \(-2.487\) & \(-2.488\) & \(-4.640\) &
\(-4.784\) \\
\(0.3\) & \(23.05\) & \(25.62\) & \(-3.412\) & \(-3.499\) & \(-4.801\) &
\(-5.066\) \\
\(0.5\) & \(25.51\) & \(28.88\) & \(-3.738\) & \(-3.865\) & \(-5.017\) &
\(-5.294\) \\
\(1\) & \(29.90\) & \(33.29\) & \(-4.080\) & \(-4.188\) & \(-5.328\) &
\(-5.527\) \\
\end{longtable}

As a check of the sandwich standard errors, which depend on \(L\), a
moving block bootstrap \citep{kunsch1989, lahiri2003} has been applied.
Each region is rebuilt with \((165/b)^{2}\) square blocks of side \(b\),
drawn with replacement among all the blocks of that size contained in
the region, the three regions being resampled independently, and the
estimators are computed for \(500\) bootstrap samples.
Table~\ref{tbl-sarBoot} shows the bootstrap standard errors of
\(\widehat{\eta}_{A,\lambda}-\widehat{\eta}_{B,\lambda}\) and the ratio
between the estimate and its standard error, for \(b\in\{11,15,33\}\).

\begin{longtable}[]{@{}ccccccc@{}}
\caption{Block bootstrap standard errors of
\(\widehat{\eta}_{A,\lambda}-\widehat{\eta}_{B,\lambda}\) (first three
columns) and ratios between the estimate and its standard error (last
three columns), type I extreme value
model.}\label{tbl-sarBoot}\tabularnewline
\toprule\noalign{}
\(\lambda\) & \(b=11\) & \(b=15\) & \(b=33\) & \(b=11\) & \(b=15\) &
\(b=33\) \\
\midrule\noalign{}
\endfirsthead
\toprule\noalign{}
\(\lambda\) & \(b=11\) & \(b=15\) & \(b=33\) & \(b=11\) & \(b=15\) &
\(b=33\) \\
\midrule\noalign{}
\endhead
\bottomrule\noalign{}
\endlastfoot
\(0\) & \(0.069\) & \(0.077\) & \(0.101\) & \(-2.01\) & \(-1.79\) &
\(-1.37\) \\
\(0.1\) & \(0.061\) & \(0.069\) & \(0.094\) & \(-2.97\) & \(-2.61\) &
\(-1.93\) \\
\(0.3\) & \(0.051\) & \(0.057\) & \(0.079\) & \(-3.93\) & \(-3.49\) &
\(-2.51\) \\
\(0.5\) & \(0.046\) & \(0.051\) & \(0.070\) & \(-4.17\) & \(-3.80\) &
\(-2.75\) \\
\(1\) & \(0.042\) & \(0.045\) & \(0.059\) & \(-4.37\) & \(-4.15\) &
\(-3.14\) \\
\end{longtable}

The bootstrap standard errors with \(b\in\{11,15\}\) are close to the
ones of Table~\ref{tbl-sarTest}, and they are larger with \(b=33\), for
which each region is rebuilt with only \(25\) blocks. The conclusions
are the same with the two methods. For the MLE the ratio is never
clearly beyond the critical value, and for \(\lambda\geq 0.3\) it is for
all the block sizes. The bootstrap percentile confidence intervals at
the \(95\%\) level lead to the same conclusion. For \(b=33\) the
interval for \(\eta_{A}-\eta_{B}\) based on the MLE is
\((-0.381,0.012)\), which contains zero, and the one based on the robust
estimator with \(\lambda=0.3\) is \((-0.405,-0.096)\). For
\(\eta_{C}-\eta_{B}\) none of the intervals contains zero.

A third method, which does not require any assumption about the range of
the dependence, is the one of batch means \citep{lahiri2003}. Each
region is divided into \((165/b)^{2}\) disjoint blocks, the estimator is
computed separately in each block, and two regions are compared with the
Welch test for the means of the two samples of block estimates.
Table~\ref{tbl-sarBatch} shows the results for \(b=55\) (\(9\) blocks)
and \(b=33\) (\(25\) blocks).

\begin{longtable}[]{@{}
  >{\centering\arraybackslash}p{(\linewidth - 16\tabcolsep) * \real{0.1111}}
  >{\centering\arraybackslash}p{(\linewidth - 16\tabcolsep) * \real{0.1111}}
  >{\centering\arraybackslash}p{(\linewidth - 16\tabcolsep) * \real{0.1111}}
  >{\centering\arraybackslash}p{(\linewidth - 16\tabcolsep) * \real{0.1111}}
  >{\centering\arraybackslash}p{(\linewidth - 16\tabcolsep) * \real{0.1111}}
  >{\centering\arraybackslash}p{(\linewidth - 16\tabcolsep) * \real{0.1111}}
  >{\centering\arraybackslash}p{(\linewidth - 16\tabcolsep) * \real{0.1111}}
  >{\centering\arraybackslash}p{(\linewidth - 16\tabcolsep) * \real{0.1111}}
  >{\centering\arraybackslash}p{(\linewidth - 16\tabcolsep) * \real{0.1111}}@{}}
\caption{Batch means. Standard deviations of the block estimates in each
region (\(s_{A}\), \(s_{B}\), \(s_{C}\)) and Welch tests, type I extreme
value model.}\label{tbl-sarBatch}\tabularnewline
\toprule\noalign{}
\begin{minipage}[b]{\linewidth}\centering
\(b\)
\end{minipage} & \begin{minipage}[b]{\linewidth}\centering
\(\lambda\)
\end{minipage} & \begin{minipage}[b]{\linewidth}\centering
\(s_{A}\)
\end{minipage} & \begin{minipage}[b]{\linewidth}\centering
\(s_{B}\)
\end{minipage} & \begin{minipage}[b]{\linewidth}\centering
\(s_{C}\)
\end{minipage} & \begin{minipage}[b]{\linewidth}\centering
A vs.~B, statistic
\end{minipage} & \begin{minipage}[b]{\linewidth}\centering
p-value
\end{minipage} & \begin{minipage}[b]{\linewidth}\centering
C vs.~B, statistic
\end{minipage} & \begin{minipage}[b]{\linewidth}\centering
p-value
\end{minipage} \\
\midrule\noalign{}
\endfirsthead
\toprule\noalign{}
\begin{minipage}[b]{\linewidth}\centering
\(b\)
\end{minipage} & \begin{minipage}[b]{\linewidth}\centering
\(\lambda\)
\end{minipage} & \begin{minipage}[b]{\linewidth}\centering
\(s_{A}\)
\end{minipage} & \begin{minipage}[b]{\linewidth}\centering
\(s_{B}\)
\end{minipage} & \begin{minipage}[b]{\linewidth}\centering
\(s_{C}\)
\end{minipage} & \begin{minipage}[b]{\linewidth}\centering
A vs.~B, statistic
\end{minipage} & \begin{minipage}[b]{\linewidth}\centering
p-value
\end{minipage} & \begin{minipage}[b]{\linewidth}\centering
C vs.~B, statistic
\end{minipage} & \begin{minipage}[b]{\linewidth}\centering
p-value
\end{minipage} \\
\midrule\noalign{}
\endhead
\bottomrule\noalign{}
\endlastfoot
\(55\) & \(0\) & \(0.212\) & \(0.277\) & \(0.123\) & \(-1.04\) &
\(0.315\) & \(-2.88\) & \(0.015\) \\
\(55\) & \(0.3\) & \(0.112\) & \(0.262\) & \(0.141\) & \(-2.20\) &
\(0.050\) & \(-3.08\) & \(0.009\) \\
\(55\) & \(0.5\) & \(0.101\) & \(0.244\) & \(0.150\) & \(-2.45\) &
\(0.033\) & \(-3.19\) & \(0.007\) \\
\(55\) & \(1\) & \(0.097\) & \(0.206\) & \(0.165\) & \(-2.79\) &
\(0.017\) & \(-3.39\) & \(0.004\) \\
\(33\) & \(0\) & \(0.287\) & \(0.363\) & \(0.189\) & \(-1.19\) &
\(0.241\) & \(-3.29\) & \(0.002\) \\
\(33\) & \(0.3\) & \(0.183\) & \(0.346\) & \(0.206\) & \(-2.58\) &
\(0.014\) & \(-3.79\) & \(<0.001\) \\
\(33\) & \(0.5\) & \(0.162\) & \(0.334\) & \(0.209\) & \(-2.96\) &
\(0.006\) & \(-3.93\) & \(<0.001\) \\
\(33\) & \(1\) & \(0.154\) & \(0.312\) & \(0.209\) & \(-3.32\) &
\(0.002\) & \(-4.14\) & \(<0.001\) \\
\end{longtable}

The three methods agree. With the MLE the difference between regions A
and B is not significant (p-values \(0.32\) and \(0.24\)). With the
robust estimators it is significant at the \(5\%\) level for
\(\lambda\geq 0.5\) with the two block sizes and for \(\lambda=0.3\)
with \(b=33\), while for \(\lambda=0.3\) and \(b=55\), with only \(9\)
blocks in each region, the p-value is \(0.050\). The evidence is clearer
the larger \(\lambda\) is. The table also shows the reason. In region A
the standard deviation of the block estimates is halved when the MLE is
replaced by a robust estimator, from \(0.21\) to \(0.10\) for \(b=55\),
because the blocks which contain vehicles are no longer different from
the other ones, while in regions B and C it hardly changes. This is the
conclusion of \citet{palm2021} for these data, the robust procedure
distinguishes the region of the power lines from the one of the vehicles
and the MLE does not, which is now obtained with procedures which take
into account the dependence between pixels.

Finally, a sensitivity analysis is carried out to study the effect of
the zero-valued pixels described in Section~\ref{sec-zeros} on these
data. The proportion of zero-valued pixels in the three regions is
small, about \(0.25\%\), but it can be much larger in images with a high
dynamic range, in which there are not enough digital levels for the
smallest magnitudes \citep{palm2024}. To reproduce that situation in a
controlled way, a proportion \(\varepsilon\) of the pixels of region A,
chosen at random, has been set to zero, and the selection has been
repeated \(20\) times. Table~\ref{tbl-sarZeros} shows the change of the
estimates of \(\eta_{A}\) with respect to the ones of the original
image, for the MLE and for the three models with the values of
\(\lambda\) which provide \(\omega_{c}(\lambda)=1.2\).

\begin{longtable}[]{@{}ccccc@{}}
\caption{Sensitivity analysis. Mean change, over \(20\) random
selections, of the estimate of \(\eta_{A}\) when a proportion
\(\varepsilon\) of the pixels of region A is set to
zero.}\label{tbl-sarZeros}\tabularnewline
\toprule\noalign{}
\(\varepsilon\) & MLE & \(c=1\) & \(c=\tfrac12\) & \(c=0\) \\
\midrule\noalign{}
\endfirsthead
\toprule\noalign{}
\(\varepsilon\) & MLE & \(c=1\) & \(c=\tfrac12\) & \(c=0\) \\
\midrule\noalign{}
\endhead
\bottomrule\noalign{}
\endlastfoot
\(0.01\) & \(-0.011\) & \(-0.018\) & \(0.003\) & \(0.000\) \\
\(0.05\) & \(-0.051\) & \(-0.093\) & \(0.014\) & \(0.000\) \\
\(0.10\) & \(-0.104\) & \(-0.196\) & \(0.029\) & \(0.000\) \\
\(0.20\) & \(-0.222\) & \(-0.444\) & \(0.066\) & \(0.000\) \\
\end{longtable}

\begin{figure}[!htb]

\centering{

\pandocbounded{\includegraphics[keepaspectratio]{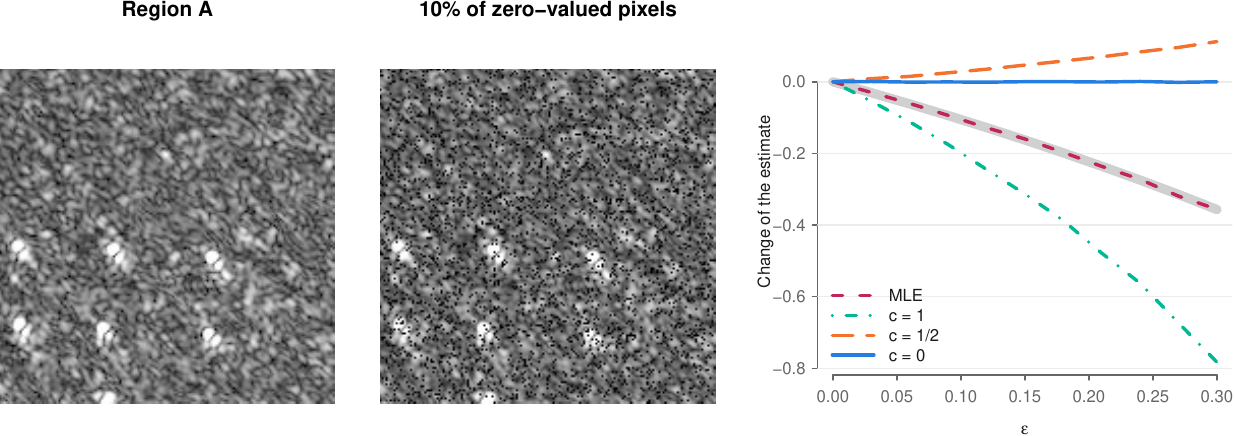}}

}

\caption{\label{fig-sar-zeros}Sensitivity analysis. Region A (left), the
same region with \(10\%\) of its pixels set to zero at random (centre),
and change of the estimates of \(\eta_{A}\) as a function of the
proportion \(\varepsilon\) of zero-valued pixels (right), for the MLE
and the three models with \(\omega_{c}(\lambda)=1.2\). The thick light
grey line is \(\log(1-\varepsilon)\).}

\end{figure}%

Figure~\ref{fig-sar-zeros} shows region A before and after setting to
zero \(10\%\) of its pixels, and the change of the four estimates as a
function of \(\varepsilon\). The changes are very close to the
asymptotic biases of Table~\ref{tbl-zeros}, although the data do not
follow the model exactly, and for the type I extreme value model they
are below \(0.001\) in absolute value. Practically the same values are
obtained for regions B and C. Since the estimated difference between
regions A and B is between \(-0.14\) and \(-0.20\), a proportion of
zero-valued pixels of \(10\%\) in only one of the regions modifies the
comparison in a quantity of the same order when the MLE or the
exponential model are used, and it does not modify it with the type I
extreme value model. The windows around zero-valued pixels used in
\citet{palm2024} have not been considered, because neighbouring pixels
are strongly correlated and such windows are located in the darkest
areas of each region, so that the effect of the zero-valued pixels
cannot be separated from the one of the selection.

\section{Concluding remarks}\label{sec-conclusions}

The minimum DPD estimators of a regression model depend on the scale in
which the response is written, and this paper has quantified that
dependence for the log-linear exponential regression model and the
family of power transformations which contains the exponential, Rayleigh
and type I extreme value formulations. The exponent of the
transformation acts as a second tuning parameter. There is not a scale
which is uniformly the best one, since at equal efficiency the smallest
gross error sensitivity corresponds to exponents between the Rayleigh
and the exponential models. The logarithmic scale is nevertheless
distinguished by four properties which no other member of the family
has. The asymptotic variance-covariance matrix is proportional to the
one of the MLE and does not depend on the unknown parameters, there is
no loss of efficiency due to the heterogeneity of the sample, the
influence function redescends to zero, and the test statistics have
explicit expressions with a chi-square distribution under the null
hypothesis. In the simulation study, at equal efficiency, it also had
the smallest bias under contamination with outliers and with inliers.
Table~\ref{tbl-summary} collects the properties of the three models of
interest.

\begin{longtable}[]{@{}
  >{\raggedright\arraybackslash}p{(\linewidth - 6\tabcolsep) * \real{0.6505}}
  >{\centering\arraybackslash}p{(\linewidth - 6\tabcolsep) * \real{0.1165}}
  >{\centering\arraybackslash}p{(\linewidth - 6\tabcolsep) * \real{0.1165}}
  >{\centering\arraybackslash}p{(\linewidth - 6\tabcolsep) * \real{0.1165}}@{}}
\caption{Summary of the properties of the minimum DPD estimators for the
type I extreme value (\(c=0\)), Rayleigh (\(c=\tfrac12\)) and
exponential (\(c=1\)) models.}\label{tbl-summary}\tabularnewline
\toprule\noalign{}
\begin{minipage}[b]{\linewidth}\raggedright
Property
\end{minipage} & \begin{minipage}[b]{\linewidth}\centering
\(c=0\)
\end{minipage} & \begin{minipage}[b]{\linewidth}\centering
\(c=\tfrac12\)
\end{minipage} & \begin{minipage}[b]{\linewidth}\centering
\(c=1\)
\end{minipage} \\
\midrule\noalign{}
\endfirsthead
\toprule\noalign{}
\begin{minipage}[b]{\linewidth}\raggedright
Property
\end{minipage} & \begin{minipage}[b]{\linewidth}\centering
\(c=0\)
\end{minipage} & \begin{minipage}[b]{\linewidth}\centering
\(c=\tfrac12\)
\end{minipage} & \begin{minipage}[b]{\linewidth}\centering
\(c=1\)
\end{minipage} \\
\midrule\noalign{}
\endhead
\bottomrule\noalign{}
\endlastfoot
Scalar factor \(\omega_{c}(\lambda)\) for \(\lambda=0.5\)
(Table~\ref{tbl-omega}) & \(1.209\) & \(1.270\) & \(1.463\) \\
Matrix factor for a slope, \(\beta_{1}=2\) and \(\lambda=0.5\)
(Table~\ref{tbl-penalty}) & \(1\) & \(1.017\) & \(1.068\) \\
Asymptotic variance free of the unknown parameters & yes & no & no \\
Bounded influence function & yes & yes & yes \\
Limit of the influence function as \(t\rightarrow+\infty\) & zero & not
zero & not zero \\
Gross error sensitivity when \(\omega_{c}(\lambda)=1.2\)
(Table~\ref{tbl-ges}) & \(2.55\) & \(2.28\) & \(2.26\) \\
Asymptotic bias, \(10\%\) of zeros and \(\omega_{c}(\lambda)=1.2\)
(Table~\ref{tbl-zeros}) & \(0\) & \(0.026\) & \(-0.178\) \\
Test statistics without estimated matrices & yes & no & no \\
\end{longtable}

These results lead to a practical recommendation. The minimum DPD
estimator should be computed on the logarithmic scale, that is, with the
type I extreme value model. Its tuning parameter can be chosen from the
loss of efficiency which is accepted, since \(\varphi(\lambda)\) depends
neither on the design nor on the unknown parameters, for example
\(\lambda=0.25\) and \(\lambda=0.5\) for increases of the asymptotic
variance of \(7\%\) and \(21\%\), and the comparisons between
populations are made with statistics which do not require the estimation
of any matrix. The Rayleigh and exponential scales are only preferable
when the criterion is the gross error sensitivity at a given efficiency.
In that case the price is a loss of efficiency which increases with the
heterogeneity of the sample, an asymptotic variance which has to be
estimated, and a bias when there are observations equal to zero, which
is small for the Rayleigh model and larger than the one of the MLE for
the exponential model.

The results can be extended in several directions, among them censored
observations and data-driven procedures for the joint choice of the
exponent and the tuning parameter, in the line of the methods proposed
for the tuning parameter in \citet{warwick2005} and \citet{basak2021}.

\section*{Appendix: Proofs}\label{appendix-proofs}
\addcontentsline{toc}{section}{Appendix: Proofs}

The proofs are given in full detail. The following notation and
preliminary results are used throughout.

\textbf{Notation.} For the \(i\)-th observation,
\(\theta_{i}=\exp\{\boldsymbol{x}_{i}^{\top}\boldsymbol{\beta}\}\),
\(t\) is the value of the exponential response (\(t=z^{1/c}\) for
\(c\neq 0\) and \(t=e^{z}\) for \(c=0\)), \(w=t/\theta_{i}\), and \[
a=(1-c)\lambda,\qquad \varkappa_{i}=\begin{cases}|c|^{-\lambda}\theta_{i}^{-c\lambda}, & c\neq 0,\\ 1, & c=0.\end{cases}
\] Since
\(\theta_{i}=\exp\{\boldsymbol{x}_{i}^{\top}\boldsymbol{\beta}\}\) and
\(w=t\exp\{-\boldsymbol{x}_{i}^{\top}\boldsymbol{\beta}\}\),
\begin{equation}\protect\phantomsection\label{eq-A-derivatives}{
\frac{\partial}{\partial\boldsymbol{\beta}}\theta_{i}=\theta_{i}\boldsymbol{x}_{i},\qquad \frac{\partial}{\partial\boldsymbol{\beta}}\theta_{i}^{-c\lambda}=-c\lambda\,\theta_{i}^{-c\lambda}\boldsymbol{x}_{i},\qquad \frac{\partial}{\partial\boldsymbol{\beta}}w=-w\,\boldsymbol{x}_{i}.
}\end{equation}

\textbf{(P1) A gamma integral.} For \(\lambda\geq 0\) and \(a+m>-1\),
the change of variable \(u=(1+\lambda)w\) gives \[
I_{m}=\int_{0}^{+\infty}w^{a+m}e^{-(1+\lambda)w}dw=\frac{\Gamma(a+m+1)}{(1+\lambda)^{a+m+1}}.
\] From \(\Gamma(x+1)=x\Gamma(x)\),
\begin{equation}\protect\phantomsection\label{eq-A-I}{
I_{1}=\frac{a+1}{1+\lambda}I_{0},\qquad I_{2}=\frac{(a+2)(a+1)}{(1+\lambda)^{2}}I_{0},\qquad I_{0}=\frac{\Gamma(a+1)}{(1+\lambda)^{a+1}}.
}\end{equation} Hence the functions (\ref{eq-je}) are
\begin{equation}\protect\phantomsection\label{eq-A-ej}{
e_{c}(\lambda)=-\frac{c\lambda}{1+\lambda}I_{0},\qquad j_{c}(\lambda)=\frac{c^{2}\lambda^{2}+(1-c)\lambda+1}{(1+\lambda)^{2}}I_{0}.
}\end{equation}

\textbf{(P2) The density function in terms of \(w\).} For \(c\neq 0\),
\(z=t^{c}=(\theta_{i}w)^{c}\), so that \(z^{1/c}=\theta_{i}w\) and
\(z^{\frac{1}{c}-1}=(\theta_{i}w)^{1-c}\). Replacing in
(\ref{eq-density}), \[
f_{Z_{i}^{(c)}}(z;\boldsymbol{\beta})=\frac{1}{|c|\,\theta_{i}}(\theta_{i}w)^{1-c}e^{-w}=|c|^{-1}\theta_{i}^{-c}w^{1-c}e^{-w},
\] and therefore, for every \(\lambda>0\),
\begin{equation}\protect\phantomsection\label{eq-A-flambda}{
f_{Z_{i}^{(c)}}^{\lambda}(z;\boldsymbol{\beta})=\varkappa_{i}\,w^{a}e^{-\lambda w}.
}\end{equation} For \(c=0\), \(z=\eta_{i}+\log w\) and
\(f_{Y_{i}}(z;\boldsymbol{\beta})=\exp\{z-\eta_{i}-e^{z-\eta_{i}}\}=we^{-w}\),
so that (\ref{eq-A-flambda}) also holds, with \(a=\lambda\) and
\(\varkappa_{i}=1\).

\textbf{(P3) Change of variable.} For \(c\neq 0\),
\(dz=|c|\,\theta_{i}^{c}w^{c-1}dw\) in absolute value, and \(w\) ranges
over \((0,+\infty)\) when \(z\) does. For any function \(g\) for which
the integrals exist,
\begin{equation}\protect\phantomsection\label{eq-A-change}{
\begin{aligned}
\int_{0}^{+\infty}g(w)f_{Z_{i}^{(c)}}^{1+\lambda}(z;\boldsymbol{\beta})dz
&=|c|^{-(1+\lambda)}\theta_{i}^{-c(1+\lambda)}|c|\,\theta_{i}^{c}\int_{0}^{+\infty}g(w)w^{(1-c)(1+\lambda)+c-1}e^{-(1+\lambda)w}dw\\
&=\varkappa_{i}\int_{0}^{+\infty}g(w)w^{a}e^{-(1+\lambda)w}dw,
\end{aligned}
}\end{equation} because \((1-c)(1+\lambda)+c-1=(1-c)\lambda=a\). For
\(c=0\), \(dz=w^{-1}dw\) and
\(f^{1+\lambda}=w^{1+\lambda}e^{-(1+\lambda)w}\), so that
(\ref{eq-A-change}) holds with \(a=\lambda\) and \(\varkappa_{i}=1\),
the integral in \(z\) being over \(\mathbb{R}\).

\textbf{(P4) Three integrals.} From (\ref{eq-A-change}) with \(g(w)=1\),
\(g(w)=w-1\) and \(g(w)=(w-1)^{2}\), and (\ref{eq-A-I}),
\begin{equation}\protect\phantomsection\label{eq-A-int0}{
\int f_{Z_{i}^{(c)}}^{1+\lambda}dz=\varkappa_{i}I_{0},
}\end{equation}
\begin{equation}\protect\phantomsection\label{eq-A-int1}{
\int (w-1)f_{Z_{i}^{(c)}}^{1+\lambda}dz=\varkappa_{i}(I_{1}-I_{0})=\varkappa_{i}I_{0}\frac{(a+1)-(1+\lambda)}{1+\lambda}=-\varkappa_{i}I_{0}\frac{c\lambda}{1+\lambda}=\varkappa_{i}\,e_{c}(\lambda),
}\end{equation} since \(a-\lambda=-c\lambda\), and
\begin{equation}\protect\phantomsection\label{eq-A-int2}{
\begin{aligned}
\int (w-1)^{2}f_{Z_{i}^{(c)}}^{1+\lambda}dz&=\varkappa_{i}(I_{2}-2I_{1}+I_{0})\\
&=\frac{\varkappa_{i}I_{0}}{(1+\lambda)^{2}}\left[(a+2)(a+1)-2(a+1)(1+\lambda)+(1+\lambda)^{2}\right]=\varkappa_{i}\,j_{c}(\lambda).
\end{aligned}
}\end{equation} The last equality holds because, with \(u=a+1\) and
\(v=1+\lambda\),
\((u+1)u-2uv+v^{2}=(u-v)^{2}+u=(a-\lambda)^{2}+a+1=c^{2}\lambda^{2}+(1-c)\lambda+1\).
The three integrals are finite if and only if \(a>-1\), that is,
\(c<1+\tfrac{1}{\lambda}\).

\emph{Proof of Proposition~\ref{prp-equations}.} The function minimized
in (\ref{eq-mdpde}) is
\(\frac{1}{n}\sum_{i=1}^{n}V_{i}(\boldsymbol{\beta})\), with \[
V_{i}(\boldsymbol{\beta})=\int f_{Z_{i}^{(c)}}^{1+\lambda}(z;\boldsymbol{\beta})dz-\frac{1+\lambda}{\lambda}f_{Z_{i}^{(c)}}^{\lambda}(z_{i};\boldsymbol{\beta}).
\] By (\ref{eq-A-int0}), the first term of \(V_{i}(\boldsymbol{\beta})\)
is \(|c|^{-\lambda}\theta_{i}^{-c\lambda}I_{0}\), in which only
\(\theta_{i}^{-c\lambda}\) depends on \(\boldsymbol{\beta}\). By
(\ref{eq-A-derivatives}) and (\ref{eq-A-ej}), \[
\frac{\partial}{\partial\boldsymbol{\beta}}\int f_{Z_{i}^{(c)}}^{1+\lambda}dz=-c\lambda\,\varkappa_{i}I_{0}\,\boldsymbol{x}_{i}=(1+\lambda)\varkappa_{i}\,e_{c}(\lambda)\boldsymbol{x}_{i}.
\] For the second term, by (\ref{eq-A-flambda}),
\(f_{Z_{i}^{(c)}}^{\lambda}(z_{i};\boldsymbol{\beta})=|c|^{-\lambda}\theta_{i}^{-c\lambda}w_{i}^{a}e^{-\lambda w_{i}}\).
By the product rule and (\ref{eq-A-derivatives}),
\begin{equation}\protect\phantomsection\label{eq-A-dflambda}{
\frac{\partial}{\partial\boldsymbol{\beta}}\left(\theta_{i}^{-c\lambda}w_{i}^{a}e^{-\lambda w_{i}}\right)
=\theta_{i}^{-c\lambda}w_{i}^{a}e^{-\lambda w_{i}}\left(-c\lambda-a+\lambda w_{i}\right)\boldsymbol{x}_{i}
=\lambda\,\theta_{i}^{-c\lambda}(w_{i}-1)w_{i}^{a}e^{-\lambda w_{i}}\boldsymbol{x}_{i},
}\end{equation} where the three terms of the bracket come, respectively,
from \(\theta_{i}^{-c\lambda}\), from \(w_{i}^{a}\) and from
\(e^{-\lambda w_{i}}\), and the last equality uses
\(c\lambda+a=\lambda\).

Combining the two terms,
\begin{equation}\protect\phantomsection\label{eq-A-dV}{
\frac{\partial}{\partial\boldsymbol{\beta}}V_{i}(\boldsymbol{\beta})=(1+\lambda)\varkappa_{i}\,e_{c}(\lambda)\boldsymbol{x}_{i}-\frac{1+\lambda}{\lambda}\lambda\,\varkappa_{i}(w_{i}-1)w_{i}^{a}e^{-\lambda w_{i}}\boldsymbol{x}_{i}
=-(1+\lambda)|c|^{-\lambda}\,\theta_{i}^{-c\lambda}\boldsymbol{x}_{i}\,h_{c,\lambda}(w_{i}),
}\end{equation} with
\(h_{c,\lambda}(w)=(w-1)w^{a}e^{-\lambda w}-e_{c}(\lambda)\). Equating
to the zero vector the sum over \(i\) and dividing by the constant
\(-(1+\lambda)|c|^{-\lambda}\), (\ref{eq-estimating}) is obtained. For
\(c=0\) the same derivation holds with \(|c|^{-\lambda}\) replaced by
\(1\). \(\square\)

The asymptotic results are proven under the conditions (C1) and (C2) of
Section~\ref{sec-asymptotic}. In what follows \(\|\cdot\|\) denotes the
Euclidean norm of a vector and the spectral norm of a matrix, and a
sequence of random vectors \(\boldsymbol{V}_{n}\) is said to be bounded
in probability if for every \(\varepsilon>0\) there is a constant \(K\)
such that \(P(\|\boldsymbol{V}_{n}\|>K)\leq\varepsilon\) for \(n\) large
enough.

\begin{lemma}[]\protect\hypertarget{lem-bounds}{}\label{lem-bounds}

Let \(\lambda>0\), \(c\leq 1\), \(a=(1-c)\lambda\), and, for \(w>0\), \[
g_{0}(w)=h_{c,\lambda}(w),\qquad g_{1}(w)=c\lambda\,g_{0}(w)+w\,g_{0}'(w),\qquad g_{2}(w)=c\lambda\,g_{1}(w)+w\,g_{1}'(w).
\]

\begin{enumerate}
\def\labelenumi{\alph{enumi})}
\item
  The three functions are bounded,
  \(B_{m}=\sup_{w>0}|g_{m}(w)|<+\infty\), \(m=0,1,2\).
\item
  If \(W\sim\mathcal{E}(1)\), then \(\operatorname{E}[g_{0}(W)]=0\),
  \(\operatorname{E}[g_{0}^{2}(W)]=j_{c}(2\lambda)-e_{c}^{2}(\lambda)\)
  and \(\operatorname{E}[g_{1}(W)]=j_{c}(\lambda)\).
\item
  With
  \(\theta_{i}=\exp\{\boldsymbol{x}_{i}^{\top}\boldsymbol{\beta}\}\) and
  \(w_{i}=t_{i}/\theta_{i}\), for \(m=0,1\), \[
  \frac{\partial}{\partial\boldsymbol{\beta}}\left[\theta_{i}^{-c\lambda}g_{m}(w_{i})\right]=-\theta_{i}^{-c\lambda}g_{m+1}(w_{i})\boldsymbol{x}_{i}.
  \]
\end{enumerate}

\end{lemma}

\emph{Proof.} a) For a polynomial \(q\), \[
w\frac{\partial}{\partial w}\left[q(w)w^{a}e^{-\lambda w}\right]=\left[(a-\lambda w)q(w)+w\,q'(w)\right]w^{a}e^{-\lambda w},
\] which is again a polynomial multiplied by \(w^{a}e^{-\lambda w}\).
Since \(g_{0}(w)=(w-1)w^{a}e^{-\lambda w}-e_{c}(\lambda)\), the three
functions are the sum of a constant and a polynomial multiplied by
\(w^{a}e^{-\lambda w}\). For every integer \(m\geq 0\) the function
\(w^{a+m}e^{-\lambda w}\) is continuous on \((0,+\infty)\), it tends to
zero as \(w\rightarrow+\infty\) because \(\lambda>0\), and it has a
finite limit as \(w\rightarrow 0^{+}\) because \(a\geq 0\), so that it
is bounded.

\begin{enumerate}
\def\labelenumi{\alph{enumi})}
\setcounter{enumi}{1}
\item
  By (\ref{eq-A-I}),
  \(\operatorname{E}[(W-1)W^{a}e^{-\lambda W}]=I_{1}-I_{0}=e_{c}(\lambda)\),
  as in (\ref{eq-A-int1}), which gives the first expectation. For the
  second one,
  \(\operatorname{E}[(W-1)^{2}W^{2a}e^{-2\lambda W}]=\int_{0}^{+\infty}(w-1)^{2}w^{2a}e^{-(1+2\lambda)w}dw\)
  is the integral computed in (\ref{eq-A-int2}) with \(2\lambda\)
  instead of \(\lambda\), that is, \(j_{c}(2\lambda)\), and therefore
  \(\operatorname{E}[g_{0}^{2}(W)]=j_{c}(2\lambda)-2e_{c}^{2}(\lambda)+e_{c}^{2}(\lambda)\).
  For the third one, integrating by parts, \[
  \begin{aligned}
  \operatorname{E}[W g_{0}'(W)]&=\int_{0}^{+\infty}w\,g_{0}'(w)e^{-w}dw=\left[w e^{-w}g_{0}(w)\right]_{0}^{+\infty}-\int_{0}^{+\infty}g_{0}(w)(1-w)e^{-w}dw\\
  &=\operatorname{E}[(W-1)g_{0}(W)],
  \end{aligned}
  \] where the first term is zero because \(g_{0}\) is bounded. Since
  \(\operatorname{E}[W-1]=0\),
  \(\operatorname{E}[(W-1)g_{0}(W)]=\operatorname{E}[(W-1)^{2}W^{a}e^{-\lambda W}]=I_{2}-2I_{1}+I_{0}=j_{c}(\lambda)\),
  as in (\ref{eq-A-int2}), and
  \(\operatorname{E}[g_{1}(W)]=c\lambda\operatorname{E}[g_{0}(W)]+j_{c}(\lambda)=j_{c}(\lambda)\).
\item
  By (\ref{eq-A-derivatives}),
  \(\frac{\partial}{\partial\boldsymbol{\beta}}\left[\theta_{i}^{-c\lambda}g_{m}(w_{i})\right]=-c\lambda\,\theta_{i}^{-c\lambda}g_{m}(w_{i})\boldsymbol{x}_{i}-\theta_{i}^{-c\lambda}g_{m}'(w_{i})w_{i}\boldsymbol{x}_{i}\).
  \(\square\)
\end{enumerate}

\begin{lemma}[]\protect\hypertarget{lem-linear}{}\label{lem-linear}

Let \(\lambda>0\) and \(c\leq 1\), assume that (C1) and (C2) hold, and
let \(\boldsymbol{\beta}\) be the true value of the parameter. Let
\(\boldsymbol{z}_{n}=n^{-1/2}\boldsymbol{u}_{\lambda}^{(c)}(\boldsymbol{\beta})\)
and, for \(\boldsymbol{b}\in\mathbb{R}^{k+1}\), with
\(\theta_{i}(\boldsymbol{b})=\exp\{\boldsymbol{x}_{i}^{\top}\boldsymbol{b}\}\)
and \(w_{i}(\boldsymbol{b})=t_{i}/\theta_{i}(\boldsymbol{b})\),
\begin{equation}\protect\phantomsection\label{eq-A-Dn}{
\boldsymbol{D}_{n}(\boldsymbol{b})=-\frac{1}{n}\frac{\partial}{\partial\boldsymbol{b}^{\top}}\boldsymbol{u}_{\lambda}^{(c)}(\boldsymbol{b})=\frac{1}{n}\sum_{i=1}^{n}\theta_{i}^{-c\lambda}(\boldsymbol{b})\,g_{1}(w_{i}(\boldsymbol{b}))\,\boldsymbol{x}_{i}\boldsymbol{x}_{i}^{\top}.
}\end{equation}

\begin{enumerate}
\def\labelenumi{\alph{enumi})}
\item
  \(\boldsymbol{z}_{n}\overset{\mathcal{L}}{\underset{n\rightarrow\infty}{\longrightarrow}}\mathcal{N}\left(\boldsymbol{0}_{k+1},\left[j_{c}(2\lambda)-e_{c}^{2}(\lambda)\right]\boldsymbol{G}_{2c\lambda}(\boldsymbol{\theta})\right)\).
\item
  \(\boldsymbol{D}_{n}(\boldsymbol{\beta})\overset{P}{\underset{n\rightarrow\infty}{\longrightarrow}}j_{c}(\lambda)\boldsymbol{G}_{c\lambda}(\boldsymbol{\theta})\).
\item
  If \(\|\boldsymbol{b}-\boldsymbol{\beta}\|\leq 1\) and
  \(\|\boldsymbol{b}'-\boldsymbol{\beta}\|\leq 1\), then
  \(\|\boldsymbol{D}_{n}(\boldsymbol{b})-\boldsymbol{D}_{n}(\boldsymbol{b}')\|\leq L\|\boldsymbol{b}-\boldsymbol{b}'\|\),
  with \(L=B_{2}M^{3}\exp\{|c|\lambda M(\|\boldsymbol{\beta}\|+1)\}\).
\item
  If \(\check{\boldsymbol{\beta}}_{n}\) is a sequence of random vectors
  such that
  \(\sqrt{n}(\check{\boldsymbol{\beta}}_{n}-\boldsymbol{\beta})\) is
  bounded in probability, then
  \begin{equation}\protect\phantomsection\label{eq-A-expansion}{
  n^{-1/2}\boldsymbol{u}_{\lambda}^{(c)}(\check{\boldsymbol{\beta}}_{n})=\boldsymbol{z}_{n}-j_{c}(\lambda)\boldsymbol{G}_{c\lambda}(\boldsymbol{\theta})\sqrt{n}(\check{\boldsymbol{\beta}}_{n}-\boldsymbol{\beta})+\boldsymbol{R}_{n},\qquad \boldsymbol{R}_{n}\overset{P}{\underset{n\rightarrow\infty}{\longrightarrow}}\boldsymbol{0}_{k+1}.
  }\end{equation}
\item
  With probability tending to one, the equation
  \(\boldsymbol{u}_{\lambda}^{(c)}(\boldsymbol{b})=\boldsymbol{0}_{k+1}\)
  has a solution \(\widehat{\boldsymbol{\beta}}_{n}\) such that
  \(\sqrt{n}(\widehat{\boldsymbol{\beta}}_{n}-\boldsymbol{\beta})\) is
  bounded in probability, and for every sequence of solutions with this
  property \begin{equation}\protect\phantomsection\label{eq-A-root}{
  \sqrt{n}(\widehat{\boldsymbol{\beta}}_{n}-\boldsymbol{\beta})=\frac{1}{j_{c}(\lambda)}\boldsymbol{G}_{c\lambda}^{-1}(\boldsymbol{\theta})\boldsymbol{z}_{n}+\boldsymbol{R}_{n},\qquad \boldsymbol{R}_{n}\overset{P}{\underset{n\rightarrow\infty}{\longrightarrow}}\boldsymbol{0}_{k+1}.
  }\end{equation}
\end{enumerate}

\end{lemma}

\emph{Proof.} The second expression of (\ref{eq-A-Dn}) follows from part
c) of Lemma~\ref{lem-bounds} with \(m=0\), and it shows that
\(\boldsymbol{D}_{n}(\boldsymbol{b})\) is a symmetric matrix. Under
(C1),
\(e^{-M\|\boldsymbol{\beta}\|}\leq\theta_{i}\leq e^{M\|\boldsymbol{\beta}\|}\)
for every \(i\).

\begin{enumerate}
\def\labelenumi{\alph{enumi})}
\item
  At the true value,
  \(\boldsymbol{u}_{\lambda}^{(c)}(\boldsymbol{\beta})=\sum_{i=1}^{n}\theta_{i}^{-c\lambda}\boldsymbol{x}_{i}g_{0}(W_{i})\),
  where \(W_{i}=T_{i}/\theta_{i}\), \(i=1,\ldots,n\), are independent
  with distribution \(\mathcal{E}(1)\). By part b) of
  Lemma~\ref{lem-bounds} the terms of the sum are independent random
  vectors with zero mean vector and variance-covariance matrices
  \(\theta_{i}^{-2c\lambda}[j_{c}(2\lambda)-e_{c}^{2}(\lambda)]\boldsymbol{x}_{i}\boldsymbol{x}_{i}^{\top}\),
  whose average converges to
  \([j_{c}(2\lambda)-e_{c}^{2}(\lambda)]\boldsymbol{G}_{2c\lambda}(\boldsymbol{\theta})\),
  a positive definite matrix because
  \(j_{c}(2\lambda)-e_{c}^{2}(\lambda)\) is the variance of
  \(g_{0}(W_{i})\), which is not a degenerate random variable. For a
  fixed vector \(\boldsymbol{t}\neq\boldsymbol{0}_{k+1}\) the random
  variables
  \(\boldsymbol{t}^{\top}\boldsymbol{x}_{i}\theta_{i}^{-c\lambda}g_{0}(W_{i})\)
  are bounded in absolute value by
  \(\|\boldsymbol{t}\|MB_{0}e^{|c|\lambda M\|\boldsymbol{\beta}\|}\),
  while the variance of their sum is of order \(n\), so that the
  Lindeberg condition holds, the indicator functions which appear in it
  being zero for \(n\) large enough. The result follows from the
  Lindeberg-Feller central limit theorem and the Cramér-Wold device.
\item
  By part b) of Lemma~\ref{lem-bounds},
  \(\operatorname{E}[\boldsymbol{D}_{n}(\boldsymbol{\beta})]=j_{c}(\lambda)\frac{1}{n}\mathbb{X}^{\top}diag^{-c\lambda}\{\boldsymbol{\theta}\}\mathbb{X}\),
  which converges to
  \(j_{c}(\lambda)\boldsymbol{G}_{c\lambda}(\boldsymbol{\theta})\). Each
  element of \(\boldsymbol{D}_{n}(\boldsymbol{\beta})\) is the mean of
  \(n\) independent random variables bounded in absolute value by
  \(B_{1}M^{2}e^{|c|\lambda M\|\boldsymbol{\beta}\|}\), so that its
  variance is not greater than the square of this constant divided by
  \(n\), and the convergence in probability follows from the Chebyshev
  inequality.
\item
  For fixed \(t_{i}\), let
  \(f_{i}(s)=e^{-c\lambda s}g_{1}(t_{i}e^{-s})\), so that the \(i\)-th
  coefficient in (\ref{eq-A-Dn}) is
  \(f_{i}(\boldsymbol{x}_{i}^{\top}\boldsymbol{b})\) and, as in part c)
  of Lemma~\ref{lem-bounds},
  \(f_{i}'(s)=-e^{-c\lambda s}g_{2}(t_{i}e^{-s})\). If
  \(\|\boldsymbol{b}-\boldsymbol{\beta}\|\leq 1\), then
  \(|\boldsymbol{x}_{i}^{\top}\boldsymbol{b}|\leq M(\|\boldsymbol{\beta}\|+1)\),
  and the same holds for \(\boldsymbol{b}'\) and for the points between
  them. By the mean value theorem,
  \(|f_{i}(\boldsymbol{x}_{i}^{\top}\boldsymbol{b})-f_{i}(\boldsymbol{x}_{i}^{\top}\boldsymbol{b}')|\leq B_{2}e^{|c|\lambda M(\|\boldsymbol{\beta}\|+1)}M\|\boldsymbol{b}-\boldsymbol{b}'\|\),
  and the inequality follows because
  \(\|\boldsymbol{x}_{i}\boldsymbol{x}_{i}^{\top}\|=\|\boldsymbol{x}_{i}\|^{2}\leq M^{2}\).
\item
  Let
  \(\boldsymbol{\delta}_{n}=\check{\boldsymbol{\beta}}_{n}-\boldsymbol{\beta}\).
  Since \(\boldsymbol{u}_{\lambda}^{(c)}\) is continuously
  differentiable with matrix of derivatives
  \(-n\boldsymbol{D}_{n}(\boldsymbol{b})\),
  \begin{equation}\protect\phantomsection\label{eq-A-mean}{
  \boldsymbol{u}_{\lambda}^{(c)}(\check{\boldsymbol{\beta}}_{n})-\boldsymbol{u}_{\lambda}^{(c)}(\boldsymbol{\beta})=-n\,\overline{\boldsymbol{D}}_{n}\boldsymbol{\delta}_{n},\qquad \overline{\boldsymbol{D}}_{n}=\int_{0}^{1}\boldsymbol{D}_{n}(\boldsymbol{\beta}+s\boldsymbol{\delta}_{n})ds,
  }\end{equation} so that (\ref{eq-A-expansion}) holds with the
  remainder \[
  \boldsymbol{R}_{n}=-\left[\overline{\boldsymbol{D}}_{n}-j_{c}(\lambda)\boldsymbol{G}_{c\lambda}(\boldsymbol{\theta})\right]\sqrt{n}\,\boldsymbol{\delta}_{n}.
  \] On the event \(\|\boldsymbol{\delta}_{n}\|\leq 1\), whose
  probability tends to one, part c) gives
  \(\|\overline{\boldsymbol{D}}_{n}-\boldsymbol{D}_{n}(\boldsymbol{\beta})\|\leq L\|\boldsymbol{\delta}_{n}\|\int_{0}^{1}s\,ds=\tfrac{1}{2}L\|\boldsymbol{\delta}_{n}\|\),
  and hence \[
  \|\boldsymbol{R}_{n}\|\leq\left(\tfrac{1}{2}L\|\boldsymbol{\delta}_{n}\|+\|\boldsymbol{D}_{n}(\boldsymbol{\beta})-j_{c}(\lambda)\boldsymbol{G}_{c\lambda}(\boldsymbol{\theta})\|\right)\|\sqrt{n}\,\boldsymbol{\delta}_{n}\|.
  \] The first factor converges in probability to zero, by part b) and
  because \(\boldsymbol{\delta}_{n}\) does, and the second one is
  bounded in probability.
\item
  Let \(\gamma>0\) be the smallest eigenvalue of
  \(j_{c}(\lambda)\boldsymbol{G}_{c\lambda}(\boldsymbol{\theta})\) and
  \(\varepsilon>0\). By part a) the sequence \(\boldsymbol{z}_{n}\) is
  bounded in probability, so that there is a constant \(K>0\) such that
  \(P(\|\boldsymbol{z}_{n}\|>\tfrac{1}{4}K\gamma)\leq\tfrac{\varepsilon}{2}\)
  for \(n\) large enough, and by part b) the probability that the
  smallest eigenvalue of \(\boldsymbol{D}_{n}(\boldsymbol{\beta})\) is
  less than \(\tfrac{1}{2}\gamma\) is not greater than
  \(\tfrac{\varepsilon}{2}\) for \(n\) large enough. Let \(n\geq K^{2}\)
  and \(\boldsymbol{v}\in\mathbb{R}^{k+1}\) with
  \(\|\boldsymbol{v}\|=K\), so that the points
  \(\boldsymbol{\beta}+sn^{-1/2}\boldsymbol{v}\), \(s\in[0,1]\), belong
  to the ball of part c). By (\ref{eq-A-mean}) with the fixed point
  \(\boldsymbol{\beta}+n^{-1/2}\boldsymbol{v}\), and by part c), \[
  \boldsymbol{v}^{\top}n^{-1/2}\boldsymbol{u}_{\lambda}^{(c)}(\boldsymbol{\beta}+n^{-1/2}\boldsymbol{v})=\boldsymbol{v}^{\top}\boldsymbol{z}_{n}-\boldsymbol{v}^{\top}\overline{\boldsymbol{D}}_{n}\boldsymbol{v}\leq K\|\boldsymbol{z}_{n}\|-\boldsymbol{v}^{\top}\boldsymbol{D}_{n}(\boldsymbol{\beta})\boldsymbol{v}+\frac{LK^{3}}{2\sqrt{n}}.
  \] On the intersection of the two events above the right-hand side is
  not greater than
  \(\tfrac{1}{4}K^{2}\gamma-\tfrac{1}{2}K^{2}\gamma+\tfrac{1}{2}LK^{3}n^{-1/2}\),
  which is negative when \(\sqrt{n}>2LK/\gamma\). Then the continuous
  function
  \(\boldsymbol{F}(\boldsymbol{v})=-n^{-1/2}\boldsymbol{u}_{\lambda}^{(c)}(\boldsymbol{\beta}+n^{-1/2}\boldsymbol{v})\)
  verifies \(\boldsymbol{v}^{\top}\boldsymbol{F}(\boldsymbol{v})>0\) for
  every \(\boldsymbol{v}\) with \(\|\boldsymbol{v}\|=K\), and by a
  consequence of the Brouwer fixed point theorem \citep[Theorem
  6.3.4]{ortega1970} it has a zero in the ball
  \(\|\boldsymbol{v}\|\leq K\). Therefore, for \(n\) large enough, with
  probability at least \(1-\varepsilon\) the estimating equation has a
  solution at a distance of \(\boldsymbol{\beta}\) not greater than
  \(Kn^{-1/2}\). Let \(\widehat{\boldsymbol{\beta}}_{n}\) be a solution
  of minimum distance to \(\boldsymbol{\beta}\) when the equation has
  some solution, which exists because the set of solutions is closed.
  Since \(\varepsilon\) is arbitrary, the probability that a solution
  exists tends to one and
  \(\sqrt{n}(\widehat{\boldsymbol{\beta}}_{n}-\boldsymbol{\beta})\) is
  bounded in probability.
\end{enumerate}

Finally, for a sequence of solutions such that
\(\sqrt{n}(\widehat{\boldsymbol{\beta}}_{n}-\boldsymbol{\beta})\) is
bounded in probability, (\ref{eq-A-expansion}) with
\(\check{\boldsymbol{\beta}}_{n}=\widehat{\boldsymbol{\beta}}_{n}\) is
\(\boldsymbol{0}_{k+1}=\boldsymbol{z}_{n}-j_{c}(\lambda)\boldsymbol{G}_{c\lambda}(\boldsymbol{\theta})\sqrt{n}(\widehat{\boldsymbol{\beta}}_{n}-\boldsymbol{\beta})+\boldsymbol{R}_{n}\),
and multiplying by the inverse of
\(j_{c}(\lambda)\boldsymbol{G}_{c\lambda}(\boldsymbol{\theta})\),
(\ref{eq-A-root}) is obtained, its remainder being that inverse
multiplied by the one of (\ref{eq-A-expansion}). \(\square\)

\begin{lemma}[]\protect\hypertarget{lem-local}{}\label{lem-local}

Let \(c=0\) and \(\lambda>0\), assume that (C1) holds and that
\(\boldsymbol{G}_{0}\) exists and is positive definite, and let the true
value of the parameter be
\(\boldsymbol{\beta}_{n}=(\boldsymbol{\beta}_{\ast}^{\top},n^{-1/2}\boldsymbol{d}^{\top})^{\top}\),
with \(\boldsymbol{d}\in\mathbb{R}^{r}\) fixed.

\begin{enumerate}
\def\labelenumi{\alph{enumi})}
\item
  Parts a), b), c) and d) of Lemma~\ref{lem-linear} hold with
  \(\boldsymbol{\beta}_{n}\) instead of \(\boldsymbol{\beta}\), the
  limits being
  \(\mathcal{N}(\boldsymbol{0}_{k+1},j_{0}(2\lambda)\boldsymbol{G}_{0})\)
  in a) and \(j_{0}(\lambda)\boldsymbol{G}_{0}\) in b) and d), and the
  inequality of part c) holds for every \(\boldsymbol{b}\) and
  \(\boldsymbol{b}'\) with \(L=B_{2}M^{3}\).
\item
  With probability tending to one, the estimating equation
  \(\boldsymbol{u}_{\lambda}(\boldsymbol{b})=\boldsymbol{0}_{k+1}\) has
  a solution \(\widehat{\boldsymbol{\beta}}_{\lambda}\) such that
  \(\sqrt{n}(\widehat{\boldsymbol{\beta}}_{\lambda}-\boldsymbol{\beta}_{n})\)
  is bounded in probability, and the estimating equation of the model
  with design matrix \(\mathbb{X}_{\ast}\) has a solution
  \(\widetilde{\boldsymbol{\beta}}_{\ast,\lambda}\) such that
  \(\sqrt{n}(\widetilde{\boldsymbol{\beta}}_{\ast,\lambda}-\boldsymbol{\beta}_{\ast})\)
  is bounded in probability.
\end{enumerate}

\end{lemma}

\emph{Proof.} a) For \(c=0\) the factors \(\theta_{i}^{-c\lambda}\) are
equal to one,
\(\boldsymbol{z}_{n}=n^{-1/2}\sum_{i=1}^{n}\boldsymbol{x}_{i}g_{0}(W_{i})\)
and
\(\boldsymbol{D}_{n}(\boldsymbol{\beta}_{n})=\frac{1}{n}\sum_{i=1}^{n}g_{1}(W_{i})\boldsymbol{x}_{i}\boldsymbol{x}_{i}^{\top}\),
where
\(W_{i}=T_{i}\exp\{-\boldsymbol{x}_{i}^{\top}\boldsymbol{\beta}_{n}\}\),
\(i=1,\ldots,n\), are independent with distribution \(\mathcal{E}(1)\)
for every \(n\). Hence the distributions of \(\boldsymbol{z}_{n}\) and
\(\boldsymbol{D}_{n}(\boldsymbol{\beta}_{n})\) do not depend on
\(\boldsymbol{\beta}_{n}\), and the proofs of parts a) and b) of
Lemma~\ref{lem-linear} apply. In the proof of part c) the exponential
factor is not present, so that the bound holds for every
\(\boldsymbol{b}\) and \(\boldsymbol{b}'\), and the proof of part d) is
the same.

\begin{enumerate}
\def\labelenumi{\alph{enumi})}
\setcounter{enumi}{1}
\tightlist
\item
  The first statement is proven as part e) of Lemma~\ref{lem-linear}.
  For the second one, the estimating function of the model with design
  matrix \(\mathbb{X}_{\ast}\) at \(\boldsymbol{b}_{\ast}\) is the first
  component, \(\boldsymbol{u}_{1}\), of
  \(\boldsymbol{u}_{\lambda}((\boldsymbol{b}_{\ast}^{\top},\boldsymbol{0}_{r}^{\top})^{\top})\).
  Let \(\boldsymbol{v}\in\mathbb{R}^{k+1-r}\) with
  \(\|\boldsymbol{v}\|=K\) and
  \(\boldsymbol{h}=(\boldsymbol{v}^{\top},-\boldsymbol{d}^{\top})^{\top}\),
  so that
  \(\boldsymbol{\beta}_{n}+n^{-1/2}\boldsymbol{h}=((\boldsymbol{\beta}_{\ast}+n^{-1/2}\boldsymbol{v})^{\top},\boldsymbol{0}_{r}^{\top})^{\top}\).
  By (\ref{eq-A-mean}) at this point, with the matrix
  \(\overline{\boldsymbol{D}}_{n}\) partitioned as
  \(\boldsymbol{G}_{0}\), \[
  n^{-1/2}\boldsymbol{u}_{1}(\boldsymbol{\beta}_{n}+n^{-1/2}\boldsymbol{h})=\boldsymbol{z}_{n,1}-\overline{\boldsymbol{D}}_{n,11}\boldsymbol{v}+\overline{\boldsymbol{D}}_{n,12}\boldsymbol{d}.
  \] By part a),
  \(\|\overline{\boldsymbol{D}}_{n}-\boldsymbol{D}_{n}(\boldsymbol{\beta}_{n})\|\leq\tfrac{1}{2}L\|\boldsymbol{h}\|n^{-1/2}\),
  with \(\|\boldsymbol{h}\|\leq K+\|\boldsymbol{d}\|\), and
  \(\|\boldsymbol{D}_{n}(\boldsymbol{\beta}_{n})\|\leq B_{1}M^{2}\). Let
  \(\gamma>0\) be the smallest eigenvalue of
  \(j_{0}(\lambda)\boldsymbol{G}_{0}\). On the event in which the
  smallest eigenvalue of \(\boldsymbol{D}_{n}(\boldsymbol{\beta}_{n})\)
  is at least \(\tfrac{1}{2}\gamma\), the same holds for its first
  diagonal block, and \[
  \boldsymbol{v}^{\top}n^{-1/2}\boldsymbol{u}_{1}(\boldsymbol{\beta}_{n}+n^{-1/2}\boldsymbol{h})\leq K\left(\|\boldsymbol{z}_{n}\|+B_{1}M^{2}\|\boldsymbol{d}\|\right)-\tfrac{1}{2}\gamma K^{2}+\frac{L(K+\|\boldsymbol{d}\|)^{3}}{2\sqrt{n}}.
  \] If \(K\geq 8B_{1}M^{2}\|\boldsymbol{d}\|/\gamma\) and
  \(\|\boldsymbol{z}_{n}\|\leq\tfrac{1}{8}K\gamma\), the right-hand side
  is not greater than
  \(-\tfrac{1}{4}\gamma K^{2}+\tfrac{1}{2}L(K+\|\boldsymbol{d}\|)^{3}n^{-1/2}\),
  which is negative for \(n\) large enough, and the argument of part e)
  of Lemma~\ref{lem-linear} applies. \(\square\)
\end{enumerate}

\emph{Proof of Theorem~\ref{thm-asymptotic}.} For \(c\leq 1\), the
existence of the sequence of solutions is part e) of
Lemma~\ref{lem-linear}, and by (\ref{eq-A-root}) and part a) of the same
lemma the limit distribution of
\(\sqrt{n}(\widehat{\boldsymbol{\beta}}_{\lambda}^{(c)}-\boldsymbol{\beta})\)
is normal with zero mean vector and variance-covariance matrix \[
\frac{j_{c}(2\lambda)-e_{c}^{2}(\lambda)}{j_{c}^{2}(\lambda)}\boldsymbol{G}_{c\lambda}^{-1}(\boldsymbol{\theta})\boldsymbol{G}_{2c\lambda}(\boldsymbol{\theta})\boldsymbol{G}_{c\lambda}^{-1}(\boldsymbol{\theta}),
\] which is (\ref{eq-asymptotic}). For \(1<c<1+\tfrac{1}{2\lambda}\) the
functions of Lemma~\ref{lem-bounds} are not bounded, and the result is
obtained from Theorem 3.1 of \citet{ghosh2013}, according to which,
under the regularity conditions established there, \[
\sqrt{n}(\widehat{\boldsymbol{\beta}}_{\lambda}^{(c)}-\boldsymbol{\beta})\overset{\mathcal{L}}{\underset{n\rightarrow\infty}{\longrightarrow}}\mathcal{N}\left(\boldsymbol{0}_{k+1},\boldsymbol{J}_{\lambda}^{-1}(\boldsymbol{\beta})\boldsymbol{K}_{\lambda}(\boldsymbol{\beta})\boldsymbol{J}_{\lambda}^{-1}(\boldsymbol{\beta})\right),
\] where
\(\boldsymbol{J}_{\lambda}(\boldsymbol{\beta})=\lim_{n\rightarrow\infty}\frac{1}{n}\sum_{i=1}^{n}\boldsymbol{J}_{\lambda}^{(i)}(\boldsymbol{\beta})\),
\(\boldsymbol{K}_{\lambda}(\boldsymbol{\beta})=\lim_{n\rightarrow\infty}\frac{1}{n}\sum_{i=1}^{n}\boldsymbol{K}_{\lambda}^{(i)}(\boldsymbol{\beta})\)
and, when the model is correct, \[
\begin{aligned}
\boldsymbol{J}_{\lambda}^{(i)}(\boldsymbol{\beta})&=\int\boldsymbol{s}_{\boldsymbol{\beta}}^{(i)}\boldsymbol{s}_{\boldsymbol{\beta}}^{(i)\top}f_{Z_{i}^{(c)}}^{1+\lambda}dz,\qquad
\boldsymbol{\xi}_{\lambda}^{(i)}(\boldsymbol{\beta})=\int\boldsymbol{s}_{\boldsymbol{\beta}}^{(i)}f_{Z_{i}^{(c)}}^{1+\lambda}dz,\\
\boldsymbol{K}_{\lambda}^{(i)}(\boldsymbol{\beta})&=\int\boldsymbol{s}_{\boldsymbol{\beta}}^{(i)}\boldsymbol{s}_{\boldsymbol{\beta}}^{(i)\top}f_{Z_{i}^{(c)}}^{1+2\lambda}dz-\boldsymbol{\xi}_{\lambda}^{(i)}(\boldsymbol{\beta})\boldsymbol{\xi}_{\lambda}^{(i)\top}(\boldsymbol{\beta}).
\end{aligned}
\] The three quantities are computed with the score function
(\ref{eq-score}),
\(\boldsymbol{s}_{\boldsymbol{\beta}}^{(i)}=(w-1)\boldsymbol{x}_{i}\),
for every \(c<1+\tfrac{1}{2\lambda}\), since they are also used in the
proof of Proposition~\ref{prp-influence}, and for \(c\leq 1\) they lead
again to the expression obtained above.

By (\ref{eq-A-int1}) and (\ref{eq-A-int2}),
\begin{equation}\protect\phantomsection\label{eq-A-Ji}{
\boldsymbol{\xi}_{\lambda}^{(i)}(\boldsymbol{\beta})=\varkappa_{i}\,e_{c}(\lambda)\boldsymbol{x}_{i},\qquad \boldsymbol{J}_{\lambda}^{(i)}(\boldsymbol{\beta})=\varkappa_{i}\,j_{c}(\lambda)\boldsymbol{x}_{i}\boldsymbol{x}_{i}^{\top}.
}\end{equation} The first integral of
\(\boldsymbol{K}_{\lambda}^{(i)}(\boldsymbol{\beta})\) is the one of
\(\boldsymbol{J}_{\lambda}^{(i)}(\boldsymbol{\beta})\) with \(2\lambda\)
instead of \(\lambda\), for which \(\varkappa_{i}\) becomes
\(|c|^{-2\lambda}\theta_{i}^{-2c\lambda}=\varkappa_{i}^{2}\) and \(a\)
becomes \(2a\), so that it is equal to
\(\varkappa_{i}^{2}j_{c}(2\lambda)\boldsymbol{x}_{i}\boldsymbol{x}_{i}^{\top}\),
and it is finite if and only if \(2a>-1\), that is,
\(c<1+\tfrac{1}{2\lambda}\). Subtracting
\(\boldsymbol{\xi}_{\lambda}^{(i)}(\boldsymbol{\beta})\boldsymbol{\xi}_{\lambda}^{(i)\top}(\boldsymbol{\beta})\),
\[
\boldsymbol{K}_{\lambda}^{(i)}(\boldsymbol{\beta})=\varkappa_{i}^{2}\left[j_{c}(2\lambda)-e_{c}^{2}(\lambda)\right]\boldsymbol{x}_{i}\boldsymbol{x}_{i}^{\top}.
\]

With respect to the limits, since
\(\frac{1}{n}\sum_{i=1}^{n}\theta_{i}^{-c\lambda}\boldsymbol{x}_{i}\boldsymbol{x}_{i}^{\top}=\frac{1}{n}\mathbb{X}^{\top}diag^{-c\lambda}\{\boldsymbol{\theta}\}\mathbb{X}\),
by (\ref{eq-G}), \[
\boldsymbol{J}_{\lambda}(\boldsymbol{\beta})=|c|^{-\lambda}j_{c}(\lambda)\boldsymbol{G}_{c\lambda}(\boldsymbol{\theta}),\qquad
\boldsymbol{K}_{\lambda}(\boldsymbol{\beta})=|c|^{-2\lambda}\left[j_{c}(2\lambda)-e_{c}^{2}(\lambda)\right]\boldsymbol{G}_{2c\lambda}(\boldsymbol{\theta}).
\]

Finally, in
\(\boldsymbol{J}_{\lambda}^{-1}\boldsymbol{K}_{\lambda}\boldsymbol{J}_{\lambda}^{-1}\)
the constants are \(|c|^{\lambda}|c|^{-2\lambda}|c|^{\lambda}=1\) and
\([j_{c}(2\lambda)-e_{c}^{2}(\lambda)]/j_{c}^{2}(\lambda)=\omega_{c}(\lambda)\),
which gives (\ref{eq-asymptotic}).

The matrix used in the algorithm (\ref{eq-algorithm}) is justified as
follows. Differentiating
\(\boldsymbol{\xi}_{\lambda}^{(i)}=\int\boldsymbol{s}_{\boldsymbol{\beta}}^{(i)}f^{1+\lambda}dz\)
under the integral sign,
\(\frac{\partial}{\partial\boldsymbol{\beta}^{\top}}\boldsymbol{\xi}_{\lambda}^{(i)}=\int\frac{\partial\boldsymbol{s}_{\boldsymbol{\beta}}^{(i)}}{\partial\boldsymbol{\beta}^{\top}}f^{1+\lambda}dz+(1+\lambda)\boldsymbol{J}_{\lambda}^{(i)}\),
and
\(\operatorname{E}\left[\frac{\partial}{\partial\boldsymbol{\beta}^{\top}}\left(\boldsymbol{s}_{\boldsymbol{\beta}}^{(i)}f^{\lambda}\right)\right]=\int\frac{\partial\boldsymbol{s}_{\boldsymbol{\beta}}^{(i)}}{\partial\boldsymbol{\beta}^{\top}}f^{1+\lambda}dz+\lambda\boldsymbol{J}_{\lambda}^{(i)}\).
Subtracting,
\begin{equation}\protect\phantomsection\label{eq-A-identity}{
\operatorname{E}\left[\frac{\partial}{\partial\boldsymbol{\beta}^{\top}}\left(\boldsymbol{s}_{\boldsymbol{\beta}}^{(i)}f_{Z_{i}^{(c)}}^{\lambda}-\boldsymbol{\xi}_{\lambda}^{(i)}\right)\right]=-\boldsymbol{J}_{\lambda}^{(i)}(\boldsymbol{\beta}).
}\end{equation} Since
\(\boldsymbol{s}_{\boldsymbol{\beta}}^{(i)}f^{\lambda}-\boldsymbol{\xi}_{\lambda}^{(i)}=|c|^{-\lambda}\theta_{i}^{-c\lambda}\boldsymbol{x}_{i}h_{c,\lambda}(w_{i})\),
the expectation of minus the matrix of derivatives of
\(\boldsymbol{u}_{\lambda}^{(c)}(\boldsymbol{\beta})\) is
\(j_{c}(\lambda)\mathbb{X}^{\top}diag^{-c\lambda}\{\boldsymbol{\theta}\}\mathbb{X}\),
the matrix used in (\ref{eq-algorithm}). \(\square\)

\emph{Proof of Corollary~\ref{cor-three}.} The three functions are
obtained from (\ref{eq-omega}) and (\ref{eq-je}).

\emph{Case \(c=0\).} Here \(e_{0}(\lambda)=0\) and
\(j_{0}(\lambda)=\frac{\Gamma(\lambda+1)}{(1+\lambda)^{\lambda+3}}(\lambda+1)=\frac{\Gamma(\lambda+1)}{(1+\lambda)^{\lambda+2}}\),
so that \[
\omega_{0}(\lambda)=\frac{j_{0}(2\lambda)}{j_{0}^{2}(\lambda)}=\frac{\Gamma(2\lambda+1)}{(1+2\lambda)^{2\lambda+2}}\frac{(1+\lambda)^{2\lambda+4}}{\Gamma^{2}(\lambda+1)}.
\] Since
\(\mathcal{B}(1+\lambda,1+\lambda)=\frac{\Gamma^{2}(\lambda+1)}{\Gamma(2\lambda+2)}=\frac{\Gamma^{2}(\lambda+1)}{(2\lambda+1)\Gamma(2\lambda+1)}\),
it holds
\(\frac{\Gamma(2\lambda+1)}{\Gamma^{2}(\lambda+1)}=\frac{1}{(1+2\lambda)\mathcal{B}(1+\lambda,1+\lambda)}\),
and \(\omega_{0}(\lambda)=\varphi(\lambda)\).

\emph{Case \(c=1\).} Here \(a=0\),
\(j_{1}(\lambda)=\frac{\lambda^{2}+1}{(1+\lambda)^{3}}\),
\(j_{1}(2\lambda)=\frac{4\lambda^{2}+1}{(1+2\lambda)^{3}}\) and
\(e_{1}(\lambda)=-\frac{\lambda}{(1+\lambda)^{2}}\), so that \[
\omega_{1}(\lambda)=\frac{(1+\lambda)^{6}}{(\lambda^{2}+1)^{2}}\left(\frac{4\lambda^{2}+1}{(1+2\lambda)^{3}}-\frac{\lambda^{2}}{(1+\lambda)^{4}}\right)=\phi(\lambda).
\]

\emph{Case \(c=\tfrac12\).} Here \(a=\tfrac{\lambda}{2}\), and \[
\begin{aligned}
j_{1/2}(\lambda)&=\frac{\Gamma(\tfrac{\lambda}{2}+1)}{(1+\lambda)^{\frac{\lambda}{2}+3}}\frac{\lambda^{2}+2\lambda+4}{4},\qquad
e_{1/2}(\lambda)=-\frac{\lambda}{2}\frac{\Gamma(\tfrac{\lambda}{2}+1)}{(1+\lambda)^{\frac{\lambda}{2}+2}},\\
j_{1/2}(2\lambda)&=\frac{\Gamma(\lambda+1)}{(1+2\lambda)^{\lambda+3}}(\lambda^{2}+\lambda+1).
\end{aligned}
\] Then \[
\frac{e_{1/2}^{2}(\lambda)}{j_{1/2}^{2}(\lambda)}=\frac{\lambda^{2}}{4}\frac{16(1+\lambda)^{\lambda+6}}{(1+\lambda)^{\lambda+4}(\lambda^{2}+2\lambda+4)^{2}}=\frac{4\lambda^{2}(1+\lambda)^{2}}{(\lambda^{2}+2\lambda+4)^{2}},
\] \[
\frac{j_{1/2}(2\lambda)}{j_{1/2}^{2}(\lambda)}=\frac{16(1+\lambda)^{\lambda+6}(\lambda^{2}+\lambda+1)}{(1+2\lambda)^{\lambda+3}(\lambda^{2}+2\lambda+4)^{2}}\frac{\Gamma(\lambda+1)}{\Gamma^{2}(\tfrac{\lambda}{2}+1)}
=\frac{16(1+\lambda)^{\lambda+5}(\lambda^{2}+\lambda+1)}{(1+2\lambda)^{\lambda+3}(\lambda^{2}+2\lambda+4)^{2}}\frac{1}{\mathcal{B}(\tfrac{\lambda}{2}+1,\tfrac{\lambda}{2}+1)},
\] because
\(\mathcal{B}(\tfrac{\lambda}{2}+1,\tfrac{\lambda}{2}+1)=\frac{\Gamma^{2}(\frac{\lambda}{2}+1)}{\Gamma(\lambda+2)}=\frac{\Gamma^{2}(\frac{\lambda}{2}+1)}{(\lambda+1)\Gamma(\lambda+1)}\).
Subtracting the two expressions and taking the common factor
\(4/(\lambda^{2}+2\lambda+4)^{2}\),
\(\omega_{1/2}(\lambda)=\psi(\lambda)\). \(\square\)

\emph{Proof of Theorem~\ref{thm-penalty}.} Let
\(\boldsymbol{A}=(\mathbb{X}^{\top}\boldsymbol{D}\mathbb{X})^{-1}\mathbb{X}^{\top}\boldsymbol{D}\),
\(\boldsymbol{B}=(\mathbb{X}^{\top}\mathbb{X})^{-1}\mathbb{X}^{\top}\)
and \(\boldsymbol{M}=\boldsymbol{A}-\boldsymbol{B}\). The matrix
\(\mathbb{X}^{\top}\boldsymbol{D}\mathbb{X}\) is positive definite,
because \(\boldsymbol{D}\) is diagonal with positive elements and
\(\mathbb{X}\) has full column rank. Since \(\boldsymbol{D}\) is
symmetric, \[
\boldsymbol{A}\boldsymbol{A}^{\top}=(\mathbb{X}^{\top}\boldsymbol{D}\mathbb{X})^{-1}\mathbb{X}^{\top}\boldsymbol{D}^{2}\mathbb{X}(\mathbb{X}^{\top}\boldsymbol{D}\mathbb{X})^{-1},\qquad
\boldsymbol{B}\boldsymbol{B}^{\top}=(\mathbb{X}^{\top}\mathbb{X})^{-1},
\] \[
\boldsymbol{A}\boldsymbol{B}^{\top}=(\mathbb{X}^{\top}\boldsymbol{D}\mathbb{X})^{-1}(\mathbb{X}^{\top}\boldsymbol{D}\mathbb{X})(\mathbb{X}^{\top}\mathbb{X})^{-1}=(\mathbb{X}^{\top}\mathbb{X})^{-1}=\boldsymbol{B}\boldsymbol{A}^{\top}.
\] Therefore
\(\boldsymbol{M}\boldsymbol{M}^{\top}=\boldsymbol{A}\boldsymbol{A}^{\top}-\boldsymbol{A}\boldsymbol{B}^{\top}-\boldsymbol{B}\boldsymbol{A}^{\top}+\boldsymbol{B}\boldsymbol{B}^{\top}=\boldsymbol{A}\boldsymbol{A}^{\top}-(\mathbb{X}^{\top}\mathbb{X})^{-1}\),
which is (\ref{eq-penalty}), and a matrix of the form
\(\boldsymbol{M}\boldsymbol{M}^{\top}\) is positive semidefinite.

It is the zero matrix if and only if \(\boldsymbol{M}\) is the zero
matrix, that is,
\((\mathbb{X}^{\top}\boldsymbol{D}\mathbb{X})^{-1}\mathbb{X}^{\top}\boldsymbol{D}=(\mathbb{X}^{\top}\mathbb{X})^{-1}\mathbb{X}^{\top}\).
Multiplying on the left by \(\mathbb{X}^{\top}\boldsymbol{D}\mathbb{X}\)
and transposing, this is
\(\boldsymbol{D}\mathbb{X}=\boldsymbol{H}\boldsymbol{D}\mathbb{X}\),
with
\(\boldsymbol{H}=\mathbb{X}(\mathbb{X}^{\top}\mathbb{X})^{-1}\mathbb{X}^{\top}\)
the Hat matrix, which means that the columns of
\(\boldsymbol{D}\mathbb{X}\) belong to the column space of
\(\mathbb{X}\). If this holds,
\(\boldsymbol{D}\mathbb{X}=\mathbb{X}\boldsymbol{\Delta}\) with
\(\boldsymbol{\Delta}=(\mathbb{X}^{\top}\mathbb{X})^{-1}\mathbb{X}^{\top}\boldsymbol{D}\mathbb{X}\),
which is nonsingular. Conversely, if
\(\boldsymbol{D}\mathbb{X}=\mathbb{X}\boldsymbol{\Delta}\) for some
matrix \(\boldsymbol{\Delta}\), then
\(\boldsymbol{H}\boldsymbol{D}\mathbb{X}=\boldsymbol{H}\mathbb{X}\boldsymbol{\Delta}=\mathbb{X}\boldsymbol{\Delta}=\boldsymbol{D}\mathbb{X}\).

Finally, \(n\) times (\ref{eq-penalty}) is
\(\left(\tfrac{1}{n}\mathbb{X}^{\top}\boldsymbol{D}\mathbb{X}\right)^{-1}\tfrac{1}{n}\mathbb{X}^{\top}\boldsymbol{D}^{2}\mathbb{X}\left(\tfrac{1}{n}\mathbb{X}^{\top}\boldsymbol{D}\mathbb{X}\right)^{-1}-\left(\tfrac{1}{n}\mathbb{X}^{\top}\mathbb{X}\right)^{-1}\),
with \(\boldsymbol{D}^{2}=diag^{-2a}\{\boldsymbol{\theta}\}\), which
converges to
\(\boldsymbol{G}_{a}^{-1}(\boldsymbol{\theta})\boldsymbol{G}_{2a}(\boldsymbol{\theta})\boldsymbol{G}_{a}^{-1}(\boldsymbol{\theta})-\boldsymbol{G}_{0}^{-1}\),
and the limit of a sequence of positive semidefinite matrices is
positive semidefinite. \(\square\)

\emph{Proof of Corollary~\ref{cor-penalty}.} By
Theorem~\ref{thm-asymptotic} the asymptotic variance-covariance matrix
is
\(\omega_{c}(\lambda)\boldsymbol{G}_{c\lambda}^{-1}\boldsymbol{G}_{2c\lambda}\boldsymbol{G}_{c\lambda}^{-1}\),
and by Theorem~\ref{thm-penalty} with \(a=c\lambda\) its difference with
\(\omega_{c}(\lambda)\boldsymbol{G}_{0}^{-1}\) is
\(\omega_{c}(\lambda)>0\) times a positive semidefinite matrix. a) For
\(c=0\), \(a=0\) and \(\boldsymbol{D}\) is the identity matrix. b) For
identically distributed observations
\(\boldsymbol{D}=\theta^{-a}\boldsymbol{I}_{n}\) and
\(\boldsymbol{D}\mathbb{X}=\mathbb{X}(\theta^{-a}\boldsymbol{I}_{k+1})\).
c) In a one-way layout with \(I\) populations, let \(\boldsymbol{Z}\) be
the \(n\times I\) matrix whose \((i,j)\)-th element is one if the
\(i\)-th observation belongs to the \(j\)-th population and zero
otherwise. Any coding of the populations gives
\(\mathbb{X}=\boldsymbol{Z}\boldsymbol{C}\) for a nonsingular
\(I\times I\) matrix \(\boldsymbol{C}\). Since \(\theta_{i}\) is
constant inside each population,
\(\boldsymbol{D}\boldsymbol{Z}=\boldsymbol{Z}\boldsymbol{\Lambda}\),
with \(\boldsymbol{\Lambda}\) the diagonal matrix of the values of
\(\theta^{-a}\) in the \(I\) populations, and
\(\boldsymbol{D}\mathbb{X}=\boldsymbol{Z}\boldsymbol{\Lambda}\boldsymbol{C}=\mathbb{X}\boldsymbol{C}^{-1}\boldsymbol{\Lambda}\boldsymbol{C}\).
In the three cases the condition of Theorem~\ref{thm-penalty} holds.
\(\square\)

\emph{Proof of Proposition~\ref{prp-small}.} Let \(\alpha=1-c\), so that
\(a=\alpha\lambda\), and let \(\gamma\) be the Euler constant. Three
expansions as \(\lambda\rightarrow 0^{+}\) are used, in which
\(O(\lambda^{3})\) denotes a function whose quotient by \(\lambda^{3}\)
is bounded in a neighbourhood of zero.

\begin{enumerate}
\def\labelenumi{\roman{enumi})}
\item
  Since the first two derivatives of \(\log\Gamma(x)\) at \(x=1\) are
  \(-\gamma\) and \(\pi^{2}/6\),
  \(\log\Gamma(1+a)=-\gamma a+\frac{\pi^{2}}{12}a^{2}+O(a^{3})=-\gamma\alpha\lambda+\frac{\pi^{2}}{12}\alpha^{2}\lambda^{2}+O(\lambda^{3})\).
\item
  Since
  \(\log(1+\lambda)=\lambda-\frac{\lambda^{2}}{2}+O(\lambda^{3})\),
  \((a+3)\log(1+\lambda)=(\alpha\lambda+3)\left(\lambda-\frac{\lambda^{2}}{2}\right)+O(\lambda^{3})=3\lambda+\left(\alpha-\frac{3}{2}\right)\lambda^{2}+O(\lambda^{3})\).
\item
  Since \(\log(1+x)=x-\frac{x^{2}}{2}+O(x^{3})\) with
  \(x=a+c^{2}\lambda^{2}=\alpha\lambda+c^{2}\lambda^{2}\),
  \(\log(1+a+c^{2}\lambda^{2})=\alpha\lambda+\left(c^{2}-\frac{\alpha^{2}}{2}\right)\lambda^{2}+O(\lambda^{3})\).
\end{enumerate}

By (\ref{eq-je}), \(\log j_{c}(\lambda)\) is the expansion i), minus
ii), plus iii), \[
\log j_{c}(\lambda)=A\lambda+B\lambda^{2}+O(\lambda^{3}),\qquad A=\alpha(1-\gamma)-3,\qquad B=\left(\frac{\pi^{2}}{12}-\frac{1}{2}\right)\alpha^{2}-\alpha+c^{2}+\frac{3}{2},
\] and replacing \(\lambda\) by \(2\lambda\),
\(\log j_{c}(2\lambda)=2A\lambda+4B\lambda^{2}+O(\lambda^{3})\).

For the numerator of (\ref{eq-omega}),
\(e_{c}^{2}(\lambda)=c^{2}\lambda^{2}\Gamma^{2}(a+1)(1+\lambda)^{-2a-4}=c^{2}\lambda^{2}+O(\lambda^{3})\)
and \(j_{c}(2\lambda)=1+O(\lambda)\), so that
\(e_{c}^{2}(\lambda)/j_{c}(2\lambda)=c^{2}\lambda^{2}+O(\lambda^{3})\)
and \[
\log\left[j_{c}(2\lambda)-e_{c}^{2}(\lambda)\right]=\log j_{c}(2\lambda)+\log\left[1-\frac{e_{c}^{2}(\lambda)}{j_{c}(2\lambda)}\right]=2A\lambda+(4B-c^{2})\lambda^{2}+O(\lambda^{3}).
\] Subtracting
\(2\log j_{c}(\lambda)=2A\lambda+2B\lambda^{2}+O(\lambda^{3})\), the
terms in \(\lambda\) cancel and \[
\log\omega_{c}(\lambda)=(2B-c^{2})\lambda^{2}+O(\lambda^{3}),\qquad 2B-c^{2}=\left(\frac{\pi^{2}}{6}-1\right)\alpha^{2}-2\alpha+c^{2}+3.
\] With \(\alpha=1-c\), \(-2\alpha+c^{2}+3=c^{2}+2c+1=(1+c)^{2}\), which
gives \(\kappa(c)\).

Finally,
\(\frac{\partial}{\partial c}\kappa(c)=-2\left(\frac{\pi^{2}}{6}-1\right)(1-c)+2(1+c)\)
and
\(\frac{\partial^{2}}{\partial c^{2}}\kappa(c)=2\left(\frac{\pi^{2}}{6}-1\right)+2=\frac{\pi^{2}}{3}>0\),
so that \(\kappa\) is strictly convex. The first derivative is zero when
\((1+c)=\left(\frac{\pi^{2}}{6}-1\right)(1-c)\), that is,
\(c\frac{\pi^{2}}{6}=\frac{\pi^{2}}{6}-2\),
\(c^{\ast}=1-\frac{12}{\pi^{2}}\), and it is positive for
\(c>c^{\ast}\). The three values are
\(\kappa(0)=\frac{\pi^{2}}{6}-1+1\),
\(\kappa(\tfrac12)=\frac{1}{4}\left(\frac{\pi^{2}}{6}-1\right)+\frac{9}{4}\)
and \(\kappa(1)=4\). \(\square\)

\emph{Proof of Proposition~\ref{prp-influence}.} By (\ref{eq-A-dV}), the
minimum DPD estimator solves
\(\sum_{l=1}^{n}\boldsymbol{\Psi}_{l}(z_{l};\boldsymbol{\beta})=\boldsymbol{0}_{k+1}\),
with
\(\boldsymbol{\Psi}_{l}(z;\boldsymbol{\beta})=\boldsymbol{s}_{\boldsymbol{\beta}}^{(l)}(z)f_{Z_{l}^{(c)}}^{\lambda}(z;\boldsymbol{\beta})-\boldsymbol{\xi}_{\lambda}^{(l)}(\boldsymbol{\beta})=\varkappa_{l}\boldsymbol{x}_{l}h_{c,\lambda}(w)\).
For distribution functions \(G_{1},\ldots,G_{n}\) of the observations,
the functional \(\boldsymbol{U}_{\lambda}^{(c)}(G_{1},\ldots,G_{n})\) is
the solution in \(\boldsymbol{\beta}\) of
\(\sum_{l=1}^{n}\int\boldsymbol{\Psi}_{l}(z;\boldsymbol{\beta})dG_{l}(z)=\boldsymbol{0}_{k+1}\),
and at the model, \(G_{l}=F_{l}\), it is the true value, since
\(\int\boldsymbol{\Psi}_{l}dF_{l}=\boldsymbol{0}_{k+1}\) by the
definition of \(\boldsymbol{\xi}_{\lambda}^{(l)}\).

Let now \(G_{i}=(1-\varepsilon)F_{i}+\varepsilon\delta_{t}\), with
\(\delta_{t}\) the degenerate distribution at the point which
corresponds to \(t\), and \(G_{l}=F_{l}\) for \(l\neq i\), and let
\(\boldsymbol{\beta}_{\varepsilon}\) be the value of the functional.
Then \[
\sum_{l=1}^{n}\int\boldsymbol{\Psi}_{l}(z;\boldsymbol{\beta}_{\varepsilon})dF_{l}(z)+\varepsilon\left[\boldsymbol{\Psi}_{i}(t;\boldsymbol{\beta}_{\varepsilon})-\int\boldsymbol{\Psi}_{i}(z;\boldsymbol{\beta}_{\varepsilon})dF_{i}(z)\right]=\boldsymbol{0}_{k+1}.
\] Differentiating with respect to \(\varepsilon\) at \(\varepsilon=0\),
where \(\boldsymbol{\beta}_{0}=\boldsymbol{\beta}\) and
\(\int\boldsymbol{\Psi}_{i}dF_{i}=\boldsymbol{0}_{k+1}\), and using
(\ref{eq-A-identity}), \[
-\left(\sum_{l=1}^{n}\boldsymbol{J}_{\lambda}^{(l)}(\boldsymbol{\beta})\right)\left.\frac{\partial\boldsymbol{\beta}_{\varepsilon}}{\partial\varepsilon}\right|_{\varepsilon=0}+\boldsymbol{\Psi}_{i}(t;\boldsymbol{\beta})=\boldsymbol{0}_{k+1},
\] so that the influence function is
\(\left(\sum_{l=1}^{n}\boldsymbol{J}_{\lambda}^{(l)}(\boldsymbol{\beta})\right)^{-1}\boldsymbol{\Psi}_{i}(t;\boldsymbol{\beta})\).

By (\ref{eq-A-Ji}),
\(\sum_{l=1}^{n}\boldsymbol{J}_{\lambda}^{(l)}=|c|^{-\lambda}j_{c}(\lambda)\mathbb{X}^{\top}diag^{-c\lambda}\{\boldsymbol{\theta}\}\mathbb{X}\),
and
\(\boldsymbol{\Psi}_{i}(t;\boldsymbol{\beta})=|c|^{-\lambda}\theta_{i}^{-c\lambda}\boldsymbol{x}_{i}h_{c,\lambda}(t/\theta_{i})\).
The factors \(|c|^{-\lambda}\) cancel and, multiplying and dividing by
\(n\), (\ref{eq-if}) is obtained.

With respect to the boundedness, the function
\(h_{c,\lambda}(w)=(w-1)w^{a}e^{-\lambda w}-e_{c}(\lambda)\) is
continuous on \((0,+\infty)\). As \(w\rightarrow+\infty\),
\((w-1)w^{a}e^{-\lambda w}\rightarrow 0\) for every \(a\), because
\(\lambda>0\). As \(w\rightarrow 0^{+}\), \((w-1)w^{a}\) tends to \(0\)
if \(a>0\) (\(c<1\)), to \(-1\) if \(a=0\) (\(c=1\)), and it is not
bounded if \(a<0\) (\(c>1\)). Hence \(h_{c,\lambda}\) is bounded if and
only if \(c\leq 1\), with the limits of the statement. Its limit as
\(w\rightarrow+\infty\) is \(-e_{c}(\lambda)\), which is zero if and
only if \(c=0\). \(\square\)

\emph{Proof of Theorem~\ref{thm-tests}.} Let \(c=0\), so that the
functions of Lemma~\ref{lem-bounds} are
\(g_{0}(w)=\psi_{\lambda}(w)=(w-1)w^{\lambda}e^{-\lambda w}\) and
\(g_{1}(w)=w\psi_{\lambda}'(w)=\left[w-\lambda(w-1)^{2}\right]w^{\lambda}e^{-\lambda w}\),
with \(\operatorname{E}[g_{0}^{2}(W)]=j_{0}(2\lambda)\) and
\(\operatorname{E}[g_{1}(W)]=j_{0}(\lambda)\). The two parts are proven
at the same time, the true value of the parameter being
\(\boldsymbol{\beta}_{n}=(\boldsymbol{\beta}_{\ast}^{\top},n^{-1/2}\boldsymbol{d}^{\top})^{\top}\),
and the null hypothesis (\ref{eq-H0}) the case
\(\boldsymbol{d}=\boldsymbol{0}_{r}\). Let \[
\boldsymbol{z}_{n}=n^{-1/2}\boldsymbol{u}_{\lambda}(\boldsymbol{\beta}_{n}),\qquad \boldsymbol{a}_{n}=\sqrt{n}(\widehat{\boldsymbol{\beta}}_{\lambda}-\boldsymbol{\beta}_{n}),\qquad \boldsymbol{b}_{n}=\sqrt{n}(\widetilde{\boldsymbol{\beta}}_{\lambda}-\boldsymbol{\beta}_{n}),
\] with components \(\boldsymbol{z}_{n,1}\), \(\boldsymbol{a}_{n,1}\),
\(\boldsymbol{b}_{n,1}\) associated with \(\boldsymbol{\beta}_{\ast}\)
and \(\boldsymbol{z}_{n,2}\), \(\boldsymbol{a}_{n,2}\),
\(\boldsymbol{b}_{n,2}\) associated with \(\boldsymbol{\beta}^{\ast}\).
The sequences \(\boldsymbol{a}_{n}\) and \(\boldsymbol{b}_{n}\) are
bounded in probability, and
\begin{equation}\protect\phantomsection\label{eq-A-ab}{
\boldsymbol{a}_{n,2}=\sqrt{n}\,\widehat{\boldsymbol{\beta}}_{\lambda}^{\ast}-\boldsymbol{d},\qquad \boldsymbol{b}_{n,2}=-\boldsymbol{d}.
}\end{equation} In what follows \(\boldsymbol{R}_{n}\) denotes a random
vector, and \(R_{n}\) a random variable, not always the same ones, which
converge in probability to zero. All of them are obtained from the
remainders of part d) of Lemma~\ref{lem-linear}, which holds here by
Lemma~\ref{lem-local}, through products with matrices which converge and
with random vectors which are bounded in probability.

By (\ref{eq-A-expansion}) with
\(\check{\boldsymbol{\beta}}_{n}=\widehat{\boldsymbol{\beta}}_{\lambda}\),
for which
\(\boldsymbol{u}_{\lambda}(\widehat{\boldsymbol{\beta}}_{\lambda})=\boldsymbol{0}_{k+1}\),
and with
\(\check{\boldsymbol{\beta}}_{n}=\widetilde{\boldsymbol{\beta}}_{\lambda}\),
\begin{equation}\protect\phantomsection\label{eq-A-linear}{
\boldsymbol{0}_{k+1}=\boldsymbol{z}_{n}-j_{0}(\lambda)\boldsymbol{G}_{0}\boldsymbol{a}_{n}+\boldsymbol{R}_{n},
}\end{equation}
\begin{equation}\protect\phantomsection\label{eq-A-linearR}{
n^{-1/2}\boldsymbol{u}_{\lambda}(\widetilde{\boldsymbol{\beta}}_{\lambda})=\boldsymbol{z}_{n}-j_{0}(\lambda)\boldsymbol{G}_{0}\boldsymbol{b}_{n}+\boldsymbol{R}_{n}.
}\end{equation}

For the Wald-type statistic, the blocks of \(\boldsymbol{G}_{0}^{-1}\)
in the rows associated with \(\boldsymbol{\beta}^{\ast}\) are
\(-\boldsymbol{G}_{22\cdot 1}^{-1}\boldsymbol{G}_{21}\boldsymbol{G}_{11}^{-1}\)
and \(\boldsymbol{G}_{22\cdot 1}^{-1}\), by the formula of the inverse
of a partitioned matrix, so that from (\ref{eq-A-linear}) and
(\ref{eq-A-ab})
\begin{equation}\protect\phantomsection\label{eq-A-wald}{
\sqrt{n}\,\widehat{\boldsymbol{\beta}}_{\lambda}^{\ast}=\boldsymbol{d}+\frac{1}{j_{0}(\lambda)}\boldsymbol{G}_{22\cdot 1}^{-1}\left(\boldsymbol{z}_{n,2}-\boldsymbol{G}_{21}\boldsymbol{G}_{11}^{-1}\boldsymbol{z}_{n,1}\right)+\boldsymbol{R}_{n}.
}\end{equation} By Lemma~\ref{lem-local}, \(\boldsymbol{z}_{n}\)
converges in distribution to
\(\mathcal{N}(\boldsymbol{0}_{k+1},j_{0}(2\lambda)\boldsymbol{G}_{0})\),
and the asymptotic variance-covariance matrix of
\(\boldsymbol{z}_{n,2}-\boldsymbol{G}_{21}\boldsymbol{G}_{11}^{-1}\boldsymbol{z}_{n,1}\)
is \[
j_{0}(2\lambda)\left[\boldsymbol{G}_{22}-\boldsymbol{G}_{21}\boldsymbol{G}_{11}^{-1}\boldsymbol{G}_{12}-\boldsymbol{G}_{21}\boldsymbol{G}_{11}^{-1}\boldsymbol{G}_{12}+\boldsymbol{G}_{21}\boldsymbol{G}_{11}^{-1}\boldsymbol{G}_{11}\boldsymbol{G}_{11}^{-1}\boldsymbol{G}_{12}\right]=j_{0}(2\lambda)\boldsymbol{G}_{22\cdot 1}.
\] Therefore \(\sqrt{n}\,\widehat{\boldsymbol{\beta}}_{\lambda}^{\ast}\)
converges in distribution to
\(\mathcal{N}(\boldsymbol{d},\varphi(\lambda)\boldsymbol{G}_{22\cdot 1}^{-1})\),
since \(j_{0}(2\lambda)/j_{0}^{2}(\lambda)=\varphi(\lambda)\). Since
\(\tfrac{1}{n}\mathbb{X}^{\top}\mathbb{X}\rightarrow\boldsymbol{G}_{0}\),
the matrix (\ref{eq-S221}) verifies
\(\tfrac{1}{n}\boldsymbol{S}_{22\cdot 1}\rightarrow\boldsymbol{G}_{22\cdot 1}\),
and \begin{equation}\protect\phantomsection\label{eq-A-W}{
W_{n,\lambda}=\frac{1}{\varphi(\lambda)}\sqrt{n}\,\widehat{\boldsymbol{\beta}}_{\lambda}^{\ast\top}\boldsymbol{G}_{22\cdot 1}\sqrt{n}\,\widehat{\boldsymbol{\beta}}_{\lambda}^{\ast}+R_{n},
}\end{equation} which is the quadratic form of an asymptotically normal
vector with the inverse of its asymptotic variance-covariance matrix.
Its limit distribution is the non-central chi-square distribution with
\(r\) degrees of freedom and non-centrality parameter
\(\boldsymbol{d}^{\top}\boldsymbol{G}_{22\cdot 1}\boldsymbol{d}/\varphi(\lambda)\),
which is the \(\chi_{r}^{2}\) distribution when
\(\boldsymbol{d}=\boldsymbol{0}_{r}\).

For the Rao-type statistic, the first component of
\(\boldsymbol{u}_{\lambda}(\widetilde{\boldsymbol{\beta}}_{\lambda})\)
is the zero vector, by the definition of the restricted estimator. With
(\ref{eq-A-ab}), the first component of (\ref{eq-A-linearR}) is
\(\boldsymbol{0}_{k+1-r}=\boldsymbol{z}_{n,1}-j_{0}(\lambda)\left(\boldsymbol{G}_{11}\boldsymbol{b}_{n,1}-\boldsymbol{G}_{12}\boldsymbol{d}\right)+\boldsymbol{R}_{n}\),
that is, \[
\boldsymbol{b}_{n,1}=\boldsymbol{G}_{11}^{-1}\left(\frac{1}{j_{0}(\lambda)}\boldsymbol{z}_{n,1}+\boldsymbol{G}_{12}\boldsymbol{d}\right)+\boldsymbol{R}_{n},
\] and replacing in the second component, \[
n^{-1/2}\boldsymbol{u}_{2}(\widetilde{\boldsymbol{\beta}}_{\lambda})=\boldsymbol{z}_{n,2}-j_{0}(\lambda)\left(\boldsymbol{G}_{21}\boldsymbol{b}_{n,1}-\boldsymbol{G}_{22}\boldsymbol{d}\right)+\boldsymbol{R}_{n}
=\boldsymbol{z}_{n,2}-\boldsymbol{G}_{21}\boldsymbol{G}_{11}^{-1}\boldsymbol{z}_{n,1}+j_{0}(\lambda)\boldsymbol{G}_{22\cdot 1}\boldsymbol{d}+\boldsymbol{R}_{n}.
\] By (\ref{eq-A-wald}), the right-hand side is
\begin{equation}\protect\phantomsection\label{eq-A-score}{
n^{-1/2}\boldsymbol{u}_{2}(\widetilde{\boldsymbol{\beta}}_{\lambda})=j_{0}(\lambda)\boldsymbol{G}_{22\cdot 1}\sqrt{n}\,\widehat{\boldsymbol{\beta}}_{\lambda}^{\ast}+\boldsymbol{R}_{n}.
}\end{equation} On the other hand, the quadratic form of
\((\mathbb{X}^{\top}\mathbb{X})^{-1}\) on a vector whose first component
is zero only involves the block of that matrix associated with
\(\boldsymbol{\beta}^{\ast}\), which is
\(\boldsymbol{S}_{22\cdot 1}^{-1}\), and hence \[
R_{n,\lambda}=\frac{1}{j_{0}(2\lambda)}n^{-1/2}\boldsymbol{u}_{2}^{\top}(\widetilde{\boldsymbol{\beta}}_{\lambda})\left(\tfrac{1}{n}\boldsymbol{S}_{22\cdot 1}\right)^{-1}n^{-1/2}\boldsymbol{u}_{2}(\widetilde{\boldsymbol{\beta}}_{\lambda}),
\] with
\(\left(\tfrac{1}{n}\boldsymbol{S}_{22\cdot 1}\right)^{-1}\rightarrow\boldsymbol{G}_{22\cdot 1}^{-1}\).
Replacing (\ref{eq-A-score}), and since
\(j_{0}^{2}(\lambda)/j_{0}(2\lambda)=1/\varphi(\lambda)\), by
(\ref{eq-A-W}) \begin{equation}\protect\phantomsection\label{eq-A-RW}{
R_{n,\lambda}=W_{n,\lambda}+R_{n}.
}\end{equation}

For the divergence-based statistic, by (\ref{eq-A-int0}) the integral in
(\ref{eq-mdpde}) is \(I_{0}\), which does not depend on
\(\boldsymbol{\beta}\), and by (\ref{eq-A-flambda}) the function which
is minimized is \begin{equation}\protect\phantomsection\label{eq-A-H}{
H_{n}(\boldsymbol{\beta})=I_{0}-\frac{1+\lambda}{\lambda n}L_{n,\lambda}(\boldsymbol{\beta}),\quad\text{so that}\quad 2n\left[H_{n}(\widetilde{\boldsymbol{\beta}}_{\lambda})-H_{n}(\widehat{\boldsymbol{\beta}}_{\lambda})\right]=\frac{2(1+\lambda)}{\lambda}\left[L_{n,\lambda}(\widehat{\boldsymbol{\beta}}_{\lambda})-L_{n,\lambda}(\widetilde{\boldsymbol{\beta}}_{\lambda})\right].
}\end{equation} By (\ref{eq-A-dV}),
\(\frac{\partial}{\partial\boldsymbol{\beta}}H_{n}(\boldsymbol{\beta})=-\frac{1+\lambda}{n}\boldsymbol{u}_{\lambda}(\boldsymbol{\beta})\),
which is zero at \(\widehat{\boldsymbol{\beta}}_{\lambda}\), and by
(\ref{eq-A-Dn}) the matrix of second derivatives of \(H_{n}\) is
\((1+\lambda)\boldsymbol{D}_{n}(\boldsymbol{\beta})\). The second order
Taylor expansion of \(H_{n}(\widetilde{\boldsymbol{\beta}}_{\lambda})\)
around \(\widehat{\boldsymbol{\beta}}_{\lambda}\) with the remainder in
integral form is \[
H_{n}(\widetilde{\boldsymbol{\beta}}_{\lambda})-H_{n}(\widehat{\boldsymbol{\beta}}_{\lambda})=(1+\lambda)(\widetilde{\boldsymbol{\beta}}_{\lambda}-\widehat{\boldsymbol{\beta}}_{\lambda})^{\top}\left[\int_{0}^{1}(1-s)\boldsymbol{D}_{n}\left(\widehat{\boldsymbol{\beta}}_{\lambda}+s(\widetilde{\boldsymbol{\beta}}_{\lambda}-\widehat{\boldsymbol{\beta}}_{\lambda})\right)ds\right](\widetilde{\boldsymbol{\beta}}_{\lambda}-\widehat{\boldsymbol{\beta}}_{\lambda}).
\] The distance to \(\boldsymbol{\beta}_{n}\) of the points at which
\(\boldsymbol{D}_{n}\) is evaluated is not greater than
\(n^{-1/2}(\|\boldsymbol{a}_{n}\|+\|\boldsymbol{b}_{n}\|)\), and
\(2\int_{0}^{1}(1-s)ds=1\), so that by Lemma~\ref{lem-local} the norm of
the difference between twice the matrix in square brackets and
\(j_{0}(\lambda)\boldsymbol{G}_{0}\) is not greater than
\(\|\boldsymbol{D}_{n}(\boldsymbol{\beta}_{n})-j_{0}(\lambda)\boldsymbol{G}_{0}\|+Ln^{-1/2}(\|\boldsymbol{a}_{n}\|+\|\boldsymbol{b}_{n}\|)\),
which converges in probability to zero. Since
\(\sqrt{n}(\widetilde{\boldsymbol{\beta}}_{\lambda}-\widehat{\boldsymbol{\beta}}_{\lambda})=\boldsymbol{b}_{n}-\boldsymbol{a}_{n}\)
is bounded in probability,
\begin{equation}\protect\phantomsection\label{eq-A-taylor2}{
2n\left[H_{n}(\widetilde{\boldsymbol{\beta}}_{\lambda})-H_{n}(\widehat{\boldsymbol{\beta}}_{\lambda})\right]=(1+\lambda)j_{0}(\lambda)(\boldsymbol{b}_{n}-\boldsymbol{a}_{n})^{\top}\boldsymbol{G}_{0}(\boldsymbol{b}_{n}-\boldsymbol{a}_{n})+R_{n}.
}\end{equation} The difference \(\boldsymbol{b}_{n}-\boldsymbol{a}_{n}\)
is obtained by subtracting the first components of (\ref{eq-A-linear})
and (\ref{eq-A-linearR}), which gives
\(\boldsymbol{G}_{11}(\boldsymbol{b}_{n,1}-\boldsymbol{a}_{n,1})+\boldsymbol{G}_{12}(\boldsymbol{b}_{n,2}-\boldsymbol{a}_{n,2})=\boldsymbol{R}_{n}\),
and from (\ref{eq-A-ab}),
\(\boldsymbol{b}_{n,2}-\boldsymbol{a}_{n,2}=-\sqrt{n}\,\widehat{\boldsymbol{\beta}}_{\lambda}^{\ast}\).
Hence \[
\boldsymbol{b}_{n,1}-\boldsymbol{a}_{n,1}=\boldsymbol{G}_{11}^{-1}\boldsymbol{G}_{12}\sqrt{n}\,\widehat{\boldsymbol{\beta}}_{\lambda}^{\ast}+\boldsymbol{R}_{n},\qquad \boldsymbol{b}_{n,2}-\boldsymbol{a}_{n,2}=-\sqrt{n}\,\widehat{\boldsymbol{\beta}}_{\lambda}^{\ast}.
\] Writing
\(\boldsymbol{q}=\sqrt{n}\,\widehat{\boldsymbol{\beta}}_{\lambda}^{\ast}\),
the quadratic form in (\ref{eq-A-taylor2}) is \[
\boldsymbol{q}^{\top}\boldsymbol{G}_{21}\boldsymbol{G}_{11}^{-1}\boldsymbol{G}_{11}\boldsymbol{G}_{11}^{-1}\boldsymbol{G}_{12}\boldsymbol{q}-2\boldsymbol{q}^{\top}\boldsymbol{G}_{21}\boldsymbol{G}_{11}^{-1}\boldsymbol{G}_{12}\boldsymbol{q}+\boldsymbol{q}^{\top}\boldsymbol{G}_{22}\boldsymbol{q}+R_{n}=\boldsymbol{q}^{\top}\boldsymbol{G}_{22\cdot 1}\boldsymbol{q}+R_{n},
\] and, by (\ref{eq-A-W}),
\(2n[H_{n}(\widetilde{\boldsymbol{\beta}}_{\lambda})-H_{n}(\widehat{\boldsymbol{\beta}}_{\lambda})]=(1+\lambda)j_{0}(\lambda)\varphi(\lambda)W_{n,\lambda}+R_{n}\).
Dividing by the constant
\((1+\lambda)j_{0}(\lambda)\varphi(\lambda)=(1+\lambda)j_{0}(2\lambda)/j_{0}(\lambda)\)
and using (\ref{eq-A-H}), the left-hand side becomes (\ref{eq-lq}), so
that \begin{equation}\protect\phantomsection\label{eq-A-SW}{
S_{n,\lambda}=W_{n,\lambda}+R_{n}.
}\end{equation} By (\ref{eq-A-RW}) and (\ref{eq-A-SW}), the statistics
\(R_{n,\lambda}\) and \(S_{n,\lambda}\) have the same limit distribution
as \(W_{n,\lambda}\), for \(\boldsymbol{d}=\boldsymbol{0}_{r}\), which
is part a), and for any \(\boldsymbol{d}\), which is part b).
\(\square\)

\emph{Proof of Proposition~\ref{prp-invariance}.} a) \emph{Wald-type
statistic.} Since \(\boldsymbol{H}\mathbb{X}=\mathbb{X}\),
\(\boldsymbol{H}_{\ast}\mathbb{X}_{\ast}=\mathbb{X}_{\ast}\) and the
column space of \(\mathbb{X}_{\ast}\) is contained in the one of
\(\mathbb{X}\),
\(\boldsymbol{H}\boldsymbol{H}_{\ast}=\boldsymbol{H}_{\ast}\boldsymbol{H}=\boldsymbol{H}_{\ast}\)
and \(\boldsymbol{H}-\boldsymbol{H}_{\ast}\) is symmetric and
idempotent. Writing
\(\widehat{\boldsymbol{\eta}}_{\lambda}=\mathbb{X}_{\ast}\widehat{\boldsymbol{\beta}}_{\ast,\lambda}+\mathbb{X}^{\ast}\widehat{\boldsymbol{\beta}}_{\lambda}^{\ast}\),
it holds
\((\boldsymbol{H}-\boldsymbol{H}_{\ast})\mathbb{X}_{\ast}=\boldsymbol{0}\)
and
\((\boldsymbol{H}-\boldsymbol{H}_{\ast})\mathbb{X}^{\ast}=(\boldsymbol{I}_{n}-\boldsymbol{H}_{\ast})\mathbb{X}^{\ast}\),
so that \[
\widehat{\boldsymbol{\eta}}_{\lambda}^{\top}(\boldsymbol{H}-\boldsymbol{H}_{\ast})\widehat{\boldsymbol{\eta}}_{\lambda}=\widehat{\boldsymbol{\beta}}_{\lambda}^{\ast\top}\mathbb{X}^{\ast\top}(\boldsymbol{I}_{n}-\boldsymbol{H}_{\ast})\mathbb{X}^{\ast}\widehat{\boldsymbol{\beta}}_{\lambda}^{\ast},
\] and
\(\mathbb{X}^{\ast\top}(\boldsymbol{I}_{n}-\boldsymbol{H}_{\ast})\mathbb{X}^{\ast}=\mathbb{X}^{\ast\top}\mathbb{X}^{\ast}-\mathbb{X}^{\ast\top}\mathbb{X}_{\ast}(\mathbb{X}_{\ast}^{\top}\mathbb{X}_{\ast})^{-1}\mathbb{X}_{\ast}^{\top}\mathbb{X}^{\ast}\)
is the matrix \(\boldsymbol{S}_{22\cdot 1}\) of (\ref{eq-S221}).
\emph{Rao-type statistic.} Since
\(\boldsymbol{u}_{\lambda}(\widetilde{\boldsymbol{\beta}}_{\lambda})=\mathbb{X}^{\top}\widetilde{\boldsymbol{\psi}}_{\lambda}\),
the quadratic form in (\ref{eq-rao}) is
\(\widetilde{\boldsymbol{\psi}}_{\lambda}^{\top}\boldsymbol{H}\widetilde{\boldsymbol{\psi}}_{\lambda}\),
and the estimating equation of the restricted estimator,
\(\mathbb{X}_{\ast}^{\top}\widetilde{\boldsymbol{\psi}}_{\lambda}=\boldsymbol{0}\),
implies
\(\boldsymbol{H}_{\ast}\widetilde{\boldsymbol{\psi}}_{\lambda}=\boldsymbol{0}_{n}\).
\emph{Divergence-based statistic.} By its definition,
\(L_{n,\lambda}(\boldsymbol{\beta})\) depends on \(\boldsymbol{\beta}\)
and on the data only through \(w_{i}=t_{i}e^{-\eta_{i}}\).

\begin{enumerate}
\def\labelenumi{\alph{enumi})}
\setcounter{enumi}{1}
\item
  The estimating equation (\ref{eq-estimatingEV}) is
  \(\mathbb{X}^{\top}\boldsymbol{\psi}=\boldsymbol{0}_{k+1}\), with
  \(\boldsymbol{\psi}\) the vector of components
  \(\psi_{\lambda}(t_{i}e^{-\eta_{i}})\) and
  \(\boldsymbol{\eta}=\mathbb{X}\boldsymbol{\beta}\). It states that
  \(\boldsymbol{\psi}\) is orthogonal to the column space of
  \(\mathbb{X}\), and \(\boldsymbol{\eta}\) ranges over that column
  space, so that the vector of fitted linear predictors depends on
  \(\mathbb{X}\) only through its column space. The same holds for
  \(\widetilde{\boldsymbol{\eta}}_{\lambda}\) and \(\mathbb{X}_{\ast}\),
  and for the matrices \(\boldsymbol{H}\) and \(\boldsymbol{H}_{\ast}\).
  By a), the three statistics only depend on these quantities.
\item
  If \(\boldsymbol{\eta}\) solves the estimating equation for the
  observations \(t_{i}\), then
  \(\boldsymbol{\eta}+(\log a)\boldsymbol{1}_{n}\), which belongs to the
  column space of \(\mathbb{X}\) because it contains
  \(\boldsymbol{1}_{n}\), solves it for the observations \(a\,t_{i}\),
  since \(a\,t_{i}e^{-\eta_{i}-\log a}=t_{i}e^{-\eta_{i}}\). Hence the
  values \(t_{i}e^{-\widehat{\eta}_{i,\lambda}}\) and
  \(t_{i}e^{-\widetilde{\eta}_{i,\lambda}}\) do not change, and neither
  do \(\widetilde{\boldsymbol{\psi}}_{\lambda}\), \(R_{n,\lambda}\) and
  \(S_{n,\lambda}\). The vector
  \(\widehat{\boldsymbol{\eta}}_{\lambda}\) changes in
  \((\log a)\boldsymbol{1}_{n}\), and
  \((\boldsymbol{H}-\boldsymbol{H}_{\ast})\boldsymbol{1}_{n}=\boldsymbol{1}_{n}-\boldsymbol{1}_{n}=\boldsymbol{0}_{n}\),
  so that \(W_{n,\lambda}\) does not change. \(\square\)
\end{enumerate}

\bibliography{referencesSAR}

\end{document}